\pdfoutput = 1

\documentclass[12pt,a4paper]{article}

\usepackage{lscape}
\usepackage{rotating}
\usepackage{float}
\usepackage{amsmath}
\usepackage{amssymb}
\usepackage{cite}
\usepackage{hyperref}
\usepackage{graphicx}
\usepackage{booktabs}
\usepackage{multirow}
\usepackage[caption=false]{subfig}

\newcommand{\lsim}{\mathrel{\hbox{\rlap{\hbox{\lower4pt\hbox{$\sim$}}}\hbox{$<$}}}}

\usepackage{color}

\usepackage{xspace}

\newcommand{\Bbar}{\kern 0.18em\overline{\kern -0.18em B}{}\xspace}

\newcommand{\Kbar}{\kern 0.18em\overline{\kern -0.18em K}{}\xspace}

\makeatletter

\newcommand{\Rmnum}[1]{\expandafter\@slowromancap\romannumeral #1@}
\makeatother

\begin{document}
\begin{titlepage}
\vspace*{-0.3truecm}

\begin{flushright}
Nikhef-2017-038 
\end{flushright}

\vspace{0.9truecm}

\begin{center}
\boldmath
{\Large{\bf CP Violation in Leptonic Rare $B^0_s$ Decays\\

\vspace*{0.2truecm}

as a Probe of New Physics}
}
\unboldmath
\end{center}

\vspace{1.8truecm}

\begin{center}
{\bf Robert Fleischer\,${}^{a,b}$, Daniela Gal\' arraga Espinosa\,${}^{a,c}$, Ruben Jaarsma\,${}^{a}$\\ 
and Gilberto~Tetlalmatzi-Xolocotzi\,${}^{a}$}

\vspace{0.5truecm}

${}^a${\sl Nikhef, Science Park 105, NL-1098 XG Amsterdam, Netherlands}

${}^b${\sl  Department of Physics and Astronomy, Vrije Universiteit Amsterdam,\\
NL-1081 HV Amsterdam, Netherlands}

${}^c${\sl Physics Department, Universit\'e  Paris-Sud, F-91405 Orsay, France}

\end{center}

\vspace*{1.7cm}


\begin{abstract}
\noindent
The decay $B^0_s\to\mu^+\mu^-$ is a key probe for the search of physics beyond the Standard Model.
While the current measurements of the corresponding branching ratio agree with the Standard Model
within the uncertainties, significant New-Physics effects may still be hiding in $B^0_s\to\mu^+\mu^-$.
In order to reveal them, the observable $\mathcal{A}^{\mu\mu}_{\Delta \Gamma_s}$, which is provided
by the decay width difference $\Delta\Gamma_s$ of the $B^0_s$-meson system, plays a central role. 
We point out that a measurement of a CP-violating observable ${\cal S}_{\rm \mu\mu}$, which is induced
through interference between $B^0_s$--$\bar B^0_s$ mixing and $B_s\to\mu^+\mu^-$ decay processes,
is essential to obtain the full picture, in particular to establish new scalar contributions and CP-violating phases. 
We illustrate these findings with future scenarios for the upgrade(s) of the LHC, exploiting also relations which 
emerge within an effective field theory description of the Standard Model, complemented with New Physics entering 
significantly beyond the electroweak scale. 
\end{abstract}


\vspace*{2.1truecm}

\vfill

\noindent
September 2017

\end{titlepage}

\thispagestyle{empty}
\vbox{}
\newpage

\setcounter{page}{1}

\section{Introduction}
The decay $B^0_s\to \mu^+\mu^-$ is one of the most interesting processes offered by Nature, allowing
us to test the Standard Model (SM) and probe New Physics (NP). In the SM, this channel has
no contributions at the tree level and shows a helicity suppression \cite{Bsmumu-SM}. Consequently, 
the SM branching ratio is enormously suppressed, and only about three out of one billion $B^0_s$ 
mesons decay into the $\mu^+\mu^-$ final state. Another key feature of $B^0_s\to \mu^+\mu^-$ 
is related to the impact of strong interactions. As gluons do not couple to the leptonic final state, only 
the $B^0_s$ decay constant $f_{B_s}$ enters the theoretical description, which can be calculated by 
means of lattice QCD \cite{lattice-rev}.

As NP effects may enhance the branching ratio of $B^0_s\to \mu^+\mu^-$ significantly, experiments
have searched for this channel for decades \cite{rev}. It has been a highlight of the results of the
Large Hadron Collider (LHC) that $B^0_s\to \mu^+\mu^-$ could eventually be observed by the CMS 
and LHCb collaborations and is now experimentally well established \cite{CMS-LHCb}, with a 
measured branching ratio in the ballpark of the SM prediction. In addition to the branching ratio, 
$B^0_s\to \mu^+\mu^-$ offers another observable, $\mathcal{A}^{\mu\mu}_{\Delta \Gamma_s}$, 
which is accessible thanks to the sizeable decay width difference $\Delta\Gamma_s$ of the mass 
eigenstates of the $B^0_s$-meson system \cite{Bsmumu-ADG}. This observable is theoretically clean 
and plays an important role in the search for NP effects \cite{BFGK,ANS,FJTX}. A pioneering measurement 
of $\mathcal{A}^{\mu\mu}_{\Delta \Gamma_s}$ has recently been reported by the LHCb collaboration
\cite{LHCb-2017}. This analysis requires, in contrast to the measurement of the branching ratio,
time information for untagged $B_s$ data samples.

If also tagging information is available, a CP-violating observable $\mathcal{S}_{\mu\mu}$ can be
measured which arises from the interference between $B^0_s$--$\bar B^0_s$ mixing and decay processes.
Should it be possible to determine the helicity of the final-state muons, yet another CP asymmetry
${\cal C}_{\mu\mu}$ can be measured, as discussed in detail in Refs.~\cite{Bsmumu-ADG,BFGK}.
It is not independent from $\mathcal{A}^{\mu\mu}_{\Delta \Gamma_s}$ and $\mathcal{S}_{\mu\mu}$, 
as the observables satisfy the following relation:
\begin{equation}\label{CP-rel}
\left(\mathcal{A}^{\mu\mu}_{\Delta \Gamma_s}\right)^2+
\left(\mathcal{S}_{\mu\mu}\right)^2
+\left({\cal C}_{\mu\mu}\right)^2=1\,.
\end{equation}
In these observables, as in the case of $\mathcal{A}^{\mu\mu}_{\Delta \Gamma_s}$, the decay 
constant $f_{B_s}$ cancels. Consequently, they are theoretically clean. Within the SM, the CP 
asymmetries vanish. However, in the presence of physics beyond the SM, we may in general encounter 
new sources of CP violation, generating non-vanishing CP asymmetries and affecting also the
observable $\mathcal{A}^{\mu\mu}_{\Delta \Gamma_s}$. 

In analyses of rare $B_{(s)}$ decays, it is usually -- for simplicity -- assumed that CP-violating NP phases vanish. 
Within specific models, such assumptions can be made, where an important example is given by 
scenarios with ``Minimal Flavour Violation" \cite{MFV}. However, we would rather like to learn from experimental
data whether new CP-violating phases enter the dynamics of the decay $B^0_s\to \mu^+\mu^-$. 

In this paper, we explore this question. Interestingly, we find that $\mathcal{S}_{\mu\mu}$ is an essential
observable to reveal the nature of possible NP effects. The sign of the CP asymmetry ${\cal C}_{\mu\mu}$ 
would allow us to resolve certain ambiguities. We shall illustrate these findings with various examples, 
showing in particular how we may establish new (pseudo)-scalar contributions to $B^0_s\to \mu^+\mu^-$
and further resolve their structure and dynamics. These considerations are completely general and can also 
be applied to the rare $B^0_s\to \tau^+\tau^-$ and $B^0_s\to e^+e^-$ decays \cite{FJTX}. 

The outline of this paper is as follows: in Section~\ref{sec:TH}, we discuss the theoretical description 
of $B^0_s\to \mu^+\mu^-$ and introduce the corresponding observables. In Section~\ref{sec:CP-gen},
we explore then the situation with general CP-violating NP contributions. Assuming relations between
short-distance coefficients, which are motivated by considerations within effective field theory, we analyze
the interplay between the $B^0_s\to \mu^+\mu^-$ observables in Section~\ref{sec:CP-rel}. In Section~\ref{sec:Exp},
we shall address experimental aspects by discussing scenarios and illustrating their physics reach by making 
assumptions about the experimental precision. Finally, we summarize our key results and give a brief outlook 
in Section~\ref{sec:concl}.

\boldmath
\section{Theoretical Description and Observables}\label{sec:TH}
\unboldmath
\subsection{Decay Amplitude}
The theoretical framework to describe the decay $\bar B^0_s\to \mu^+\mu^-$ is given by effective quantum
field theory, which allows the calculation of a low-energy effective Hamiltonian of the following
general structure \cite{Bsmumu-SM,Bsmumu-ADG,ANS}:
\begin{equation}\label{Heff}
{\cal H}_{\rm eff}=-\frac{G_{\rm F}}{\sqrt{2}\pi} V_{ts}^\ast V_{tb} \alpha
\bigl[C_{10} O_{10} + C_{S} O_S + C_P O_P+ C_{10}' O_{10}' + C_{S}' O_S' + C_P' O_P' \bigr].
\end{equation}
Here $G_{\rm F}$ is Fermi's constant, $V_{ts}^\ast V_{tb}$ is a factor with elements of the 
Cabibbo--Kobayashi--Maskawa (CKM) matrix, and $\alpha$ denotes the QED fine structure constant.
The Wilson coefficients $C_{10}^{(')}$, $C_{P}^{(')}$ and $C_{S}^{(')}$ describe heavy
degrees of freedom, which have been integrated out from appearing as explicit fields, and are
associated with the four-fermion operators
\begin{equation}\label{ops}
\begin{array}{rclcrcl}
O_{10}&=&(\bar s \gamma_\mu P_L b) (\bar\mu\gamma^\mu \gamma_5\mu), & \mbox{}\qquad &
 O_{10}'&=&(\bar s \gamma_\mu P_R b) (\bar\mu\gamma^\mu \gamma_5\mu),  \\
O_S&=&m_b (\bar s P_R b)(\bar \mu \mu), & \mbox{}\qquad &  O_S'&=&m_b (\bar s P_L b)(\bar \mu \mu),\\
O_P&=&m_b (\bar s P_R b)(\bar \mu \gamma_5 \mu), & \mbox{}\qquad  
& O_P'&=&m_b (\bar s P_L b)(\bar \mu \gamma_5 \mu),
\end{array}
\end{equation}
with $m_b$ denoting the $b$-quark mass, and 
\begin{equation}
P_{L}\equiv\frac{1}{2}\left(1-\gamma_5\right), \quad P_{R}\equiv\frac{1}{2}\left(1+\gamma_5\right).
\end{equation}
In general, the Wilson coefficients are different for $b \to s$ and $b \to d$ transitions, 
and depend on the flavour of the final-state leptons \cite{FJTX}. For simplicity, we do not give the corresponding
labels explicitly in the following discussion. In the SM, we have only to deal with the $O_{10}$ operator, having 
a real coefficient $C_{10}^{\rm SM}$. 

Introducing the combinations of Wilson coefficients
\begin{eqnarray}
	P &\equiv &\frac{C_{10} - C'_{10}}{C_{10}^{\rm SM}} + \frac{M_{B_s}^2}{2m_\mu}
	\left(\frac{m_b}{m_b + m_s}\right) \left( \frac{C_{P} - C'_{P}}{C_{10}^{\rm SM}}\right)\equiv 
	|P|e^{i\varphi_P}, \label{P-def} \\
	S &\equiv &\sqrt{1-\frac{4m_\mu^2}{M_{B_s}^2}}\frac{M_{B_s}^2}{2m_\mu}\left(\frac{m_b}{m_b + m_s}\right) 
	\left( \frac{C_{S} - C'_{S}}{C_{10}^{\rm SM}}\right)\equiv |S|e^{i\varphi_S},
	\label{S-def}
\end{eqnarray}
where $M_{B_s}$, $m_\mu$, $m_b$, $m_s$ are the corresponding particle masses
and $\varphi_P$, $\varphi_S$ denote CP-violating phases, we obtain the following
expression for the decay amplitude \cite{Bsmumu-ADG}:
\begin{equation}
A(\bar B^0_s\to\mu_\lambda^+\mu_\lambda^-)\propto V_{ts}^\ast V_{tb} f_{B_s} M_{B_s}   
m_\mu C_{10}^{\rm SM} \left [\eta_\lambda P +  S  \right].
\end{equation}
Here $\lambda={\rm L, R}$ describes the helicity of the final-state leptons with $\eta_{\rm L}=+1$ and  
$\eta_{\rm R}=-1$. 

In the SM, we have
\begin{equation}\label{SM_PS}
P|_{\rm SM}=1, \quad S|_{\rm SM}=0,
\end{equation}
and the relevant Wilson coefficient is given as \cite{BFGK}
\begin{equation}\label{C10-SM}
C^{\rm SM}_{10}=-\eta_Y \sin^{-2}\theta_W Y_{0}(x_t) = -4.134,
\end{equation}
where $\eta_Y$ describes QCD corrections, $\theta_W$ is the weak mixing angle, $Y(x_t)$ represents one of the 
Inami--Lim functions, and $x_t\equiv m_t^2/M_W^2$ parametrizes the top-quark and $W$ mass dependence 
\cite{Buras:1996}. We would like to emphasize that, by convention, $C^{\rm SM}_{10}$ does not have a complex phase. However, it takes a negative value, such that
\begin{equation}\label{C10-neg}
C^{\rm SM}_{10}= - |C^{\rm SM}_{10}|.
\end{equation}

In the following discussion, the CP-violating phases $\varphi_P$ and $\varphi_S$ play a central role. While the
latter is directly related to the phase of the short-distance coefficient $C_{S} - C'_{S}$ of new scalar contributions, 
the former may get contributions both from $C_{10} - C'_{10}$ and from the coefficient $C_{P} - C'_{P}$, which arises 
from new pseudo-scalar operators.

\subsection{Branching Ratio and Effective Lifetime}
Due to $B^0_s$--$\bar B^0_s$ mixing, an initially, i.e.\ at time $t=0$, present $B^0_s$ meson evolves into
a time-dependent linear combination of $B^0_s$ and $\bar B^0_s$ states. For the ``untagged" rate
\begin{displaymath}
\langle \Gamma(B_s(t)\to \mu_\lambda^+\mu_\lambda^-)\rangle\equiv
\Gamma(B^0_s(t)\to \mu_\lambda^+\mu_\lambda^-)+ \Gamma(\bar B^0_s(t)\to 
\mu_\lambda^+\mu_\lambda^-)
\end{displaymath}
\vspace*{-0.5truecm}
\begin{equation}\label{untagged}
\propto e^{-t/\tau_{B_s}}\bigl[\cosh(y_st/ \tau_{B_s})+  {\cal A}_{\Delta\Gamma_s}^\lambda
\sinh(y_st/ \tau_{B_s})\bigr]=R_{\rm H}^\lambda e^{-\Gamma_{\rm H}^{(s)}t}+
R_{\rm L}^\lambda e^{-\Gamma_{\rm L}^{(s)}t},
\end{equation}
no ``tagging" of the initially present $B_s$ meson is needed. This quantity depends only on two exponentials and
involves the parameter
\begin{equation}\label{ys-exp}
y_s\equiv \frac{\Delta\Gamma_s}{2\Gamma_s}= 0.0645\pm0.0045,
\end{equation}
which characterizes the decay width difference of the $B_s$ mass eigenstates, with $\tau_{B_s}\equiv1/\Gamma_s$ 
denoting the $B_s$ mean lifetime \cite{DFN,BR-Bs}; for the experimental value, see Ref.~\cite{Amh16}.
The decay dynamics enters through the following observable  \cite{Bsmumu-ADG,BFGK}:
\begin{equation}\label{ADG-expr}
 {\cal A}^{\lambda}_{\Delta\Gamma_s}=\frac{R_{\rm H}^\lambda-R_{\rm L}^\lambda}{R_{\rm H}^\lambda
 +R_{\rm L}^\lambda}
 = \frac{|P|^2\cos(2\varphi_P-\phi_s^{\rm NP}) - |S|^2\cos(2\varphi_S-\phi_s^{\rm NP})}{|P|^2 + |S|^2}\equiv 
  {\cal A}^{\mu\mu}_{\Delta\Gamma_s},
\end{equation}
which is independent of the muon helicity, as reflected by the definition of 
${\cal A}^{\mu\mu}_{\Delta\Gamma_s} $. Within the SM, we have
\begin{equation}
{\cal A}^{\mu\mu}_{\Delta\Gamma_s}|_{\rm SM}=+1.
\end{equation}

The phase $\phi_s^{\rm NP}$ originates from possible CP-violating NP contributions to the 
$B_s^0$--$\bar B_s^0$ mixing phase
\begin{equation}
\phi_s = -2\beta_s + \phi_s^{\rm NP},
\end{equation}
which is already strongly constrained by experimental data for CP-violating effects
in $B^0_s\to J/\psi \phi$ and decays with similar dynamics, yielding the following results 
\cite{peng-anat,Amh16,CKMFitter:2016}:
\begin{equation}\label{phi_s}
\phi_s = -0.030\pm 0.033 = -(1.72\pm 1.89)^{\circ}
\end{equation}
\begin{equation}\label{phis-NP}
\phi_s^{\rm NP}=0.007 \pm 0.033 =(0.4\pm 1.9)^{\circ},
\end{equation}
where we have used the SM value $\phi_s^{\rm SM}=-2\beta_s=-(2.12\pm0.04)^\circ$ in Eq.~(\ref{phis-NP}). 

Since it is challenging to measure the muon helicity, we consider the helicity-summed rates
\begin{equation}
\Gamma({B}_s^0(t)\to \mu^+\mu^-)\equiv \sum_{\lambda={\rm L,R}}
\Gamma({B}_s^0(t)\to \mu^+_\lambda \mu^-_\lambda)
\end{equation}
\begin{equation}
\Gamma(\bar{B}_s^0(t)\to \mu^+\mu^-)\equiv \sum_{\lambda={\rm L,R}}
\Gamma(\bar{B}_s^0(t)\to \mu^+_\lambda \mu^-_\lambda),
\end{equation}
and use them to define an untagged rate $\langle\Gamma(B_s(t)\to \ell^+\ell^-)\rangle$ in analogy to
Eq.~(\ref{untagged}). The branching ratio reported by experiments actually corresponds to the following 
time-integrated untagged rate \cite{Bsmumu-ADG,DFN}:
\begin{equation}
	\overline{\mathcal{B}}(B_s\to\ell^+\ell^-) \equiv \frac{1}{2}\int_0^\infty \langle
	\Gamma(B_s(t)\to \ell^+\ell^-)
	\rangle\,dt.\label{BRexp}
\end{equation}
Combining the CMS result from 2013 \cite{CMSmumu} with the most recent LHCb analysis 
\cite{LHCb-2017} yields 
\begin{equation} \label{eq:BsmumuExpComb}
\overline{\mathcal{B}}(B_s \to \mu^+\mu^-)_\text{LHCb'17+CMS} = (3.0 \pm 0.5 ) \times 10^{-9}.
\end{equation}
This average was calculated by means of the Particle Data Group (PDG) procedure \cite{PDG:2016}.
For comparison, we give also the constraint 
$\overline{\mathcal{B}}(B_s \to \mu^+\mu^-)_\text{ATLAS'16}=(0.9^{+1.1}_{-0.8})
\times 10^{-9}$ reported by the ATLAS collaboration \cite{ATLAS:2016}.

In the SM, we have the following expression \cite{Bsmumu-SM}:
\begin{equation}
\overline{\mathcal{B}}(B_s\rightarrow \mu^+\mu^-)_{\rm SM}=
\frac{\tau_{B_s}G^4_F M^4_W \sin^4\theta_W}{8\pi^5}
\frac{\left|C^{\rm SM}_{10}V_{ts} V^{*}_{tb} \right|^2 }{(1-y_s)}f^2_{B_s}M_{B_s}m^2_{\mu}
\sqrt{1-4\frac{m^2_{\mu}}{M^2_{B_s}}},
\end{equation}
where special care has to be taken concerning the use of renormalization schemes to properly include
next-to-leading-order electroweak corrections (for details, see Ref.~\cite{Bsmumu-SM}). Using current 
state-of-the-art input parameters yields the following result \cite{FJTX}:
\begin{equation} \label{BsmumuSM}
\overline{\mathcal{B}}(B_s \to \mu^+\mu^-)_\text{SM} = (3.57 \pm 0.16) \times 10^{-9}.
\end{equation}
In a very recent analysis \cite{BBS}, QED corrections from dynamics below the renormalization scale
$\mu = m_b$ were calculated, affecting the branching ratio by almost $1\%$. 
 
In order to search for NP effects by means of the branching ratio of $B^0_s\to\mu^+\mu^-$, the 
following ratio plays the key role \cite{Bsmumu-ADG,BFGK}:
\begin{equation}\label{Rbar}
 \overline{R} \equiv \frac{\overline{\mathcal{B}}(B_s\to\mu^+\mu^-)}{\overline{\mathcal{B}}
   (B_s\to\mu^+\mu^-)_{\rm SM}},
\end{equation}
taking by definition the SM value 
\begin{equation}
\overline{R}|_{\rm SM}=1.
\end{equation}
Using the expressions given above yields
\begin{equation}\label{R-expr} 
   \overline{R} = \left[\frac{1+{\cal A}^{\mu\mu}_{\Delta\Gamma_s}\,y_s}{1+y_s} \right] 
	 (|P|^2 + |S|^2)= \Upsilon_{P} |P|^2 + \Upsilon_{S} |S|^2
\end{equation}
with
\begin{equation} \label{y-pm}
\Upsilon_{P}\equiv \left[\frac{1 + y_s\cos(2\varphi_P - \phi_s^{\rm NP})}{1+y_s} \right],
\quad
\Upsilon_{S}\equiv \left[\frac{1 - y_s\cos(2\varphi_S - \phi_s^{\rm NP})}{1+y_s} \right].
\end{equation}
The numerical results in Eqs.~(\ref{eq:BsmumuExpComb}) and (\ref{BsmumuSM}) give
\begin{equation}
\left.\overline{R}\right|_\text{LHCb'17+CMS} = 0.84 \pm 0.16.
\label{eq:Rbar}
\end{equation}

The effective lifetime of the decay $B^0_s\to\mu^+\mu^-$, which is defined through 
\begin{equation}
	\tau^s_{\mu\mu} \equiv \frac{\int^\infty_0 t\,
	\langle\Gamma(B_s(t)\to \mu^+ \mu^-)\rangle\, dt}{\int_0^\infty \langle
	\Gamma(B_s(t)\to \mu^+ \mu^-)\rangle\, dt},
\end{equation}
contains the same physics information as the observable 
${\cal A}^{\mu\mu}_{\Delta\Gamma_s}$ \cite{Bsmumu-ADG}:
\begin{equation} \label{eq:aDGLifetime}
 {\cal A}^{\mu\mu}_{\Delta\Gamma_s}  = \frac{1}{y_s}\left[\frac{(1-y_s^2)\tau^s_{\mu\mu}-(1+
 y_s^2)\tau_{B_s}}{2\tau_{B_s}-(1-y_s^2)\tau^s_{\mu\mu}}\right].
\end{equation} 
A pioneering measurement of the effective lifetime of $B^0_s\to\mu^+\mu^-$ was recently 
reported by the LHCb collaboration \cite{LHCb-2017}:
\begin{equation} \label{eq:tauEffmumu}
\tau^s_{\mu\mu} = \left[2.04 \pm 0.44 ({\rm stat}) \pm 0.05 ({\rm syst}) \right]\hbox{ps}.
\end{equation}
Using Eq.~(\ref{eq:aDGLifetime}), this result can be converted into
\begin{equation}\label{ADG-exp}
\mathcal{A}_{\Delta\Gamma_s}^{\mu\mu} = 8.24 \pm 10.72,
\end{equation}
where the error is fully dominated by the uncertainty of $\tau^s_{\mu\mu}$. In view of the general 
model-independent range
\begin{equation}\label{ADG-range}
-1\leq\mathcal{A}_{\Delta\Gamma_s}^{\mu\mu}\leq+1, 
\end{equation}
it will be crucial to improve the experimental precision for this observable at the LHC upgrade(s) 
in order to use this quantity for testing the flavour sector of the SM.

\subsection{CP Asymmetries}
In contrast to the untagged $B_s$ rate in Eq.~(\ref{untagged}), the tagged, time-dependent rates involve oscillatory 
$\sin(\Delta M_st)$ and $\cos(\Delta M_st)$ terms, where $\Delta M_s$ is the mass difference between the heavy 
and light $B_s$ mass eigenstates. We obtain a CP-violating rate asymmetry of the following form
\cite{Bsmumu-ADG,BFGK}:
\begin{equation}
\frac{\Gamma(B^0_s(t)\to \mu_\lambda^+\mu^-_\lambda)-
\Gamma(\bar B^0_s(t)\to \mu_\lambda^+
\mu^-_\lambda)}{\Gamma(B^0_s(t)\to \mu_\lambda^+\mu^-_\lambda)+
\Gamma(\bar B^0_s(t)\to \mu_\lambda^+\mu^-_\lambda)}
=\frac{{\cal C}_{\mu\mu}^\lambda\cos(\Delta M_st)+{\cal S}_{\mu\mu}^\lambda
\sin(\Delta M_st)}{\cosh(y_st/\tau_{B_s}) + 
{\cal A}_{\Delta\Gamma_s}^\lambda \sinh(y_st/\tau_{B_s})},
\end{equation}
with the observables 
\begin{eqnarray}
	{\cal C}_{\mu\mu}^\lambda &= & -\eta_\lambda\left[\frac{2|PS|\cos(\varphi_P-\varphi_S)}{|P|^2+|S|^2} 
	\right] \equiv -\eta_\lambda {\cal C}_{\rm \mu\mu} \overset{\text{SM}}{\longrightarrow} 0 ,\label{Cobs}\\
	{\cal S}_{\mu\mu}^\lambda
	&=&\frac{|P|^2\sin(2\varphi_P-\phi_s^{\rm NP})-|S|^2\sin(2\varphi_S-\phi_s^{\rm NP})}{|P|^2+|S|^2}
	\equiv {\cal S}_{\mu\mu} \overset{\text{SM}}{\longrightarrow} 0 ,\label{S-expr}
\end{eqnarray}
where $\eta_{\rm L}=+1$ and $\eta_{\rm R}=-1$ for left- and right-handed muon helicity, respectively. 
It should be noted that the CP asymmetry ${\cal S}_{\mu\mu}^\lambda$, which is caused by interference
between $B^0_s$--$\bar B^0_s$ mixing and $B_s\to\mu^+\mu^-$ decay processes, does actually not depend 
on the muon helicity, just as the observable $\mathcal{A}_{\Delta\Gamma_s}^{\mu\mu} \equiv
{\cal A}_{\Delta\Gamma_s}^\lambda$. Using the helicity-summed 
rates introduced above yields
\begin{equation}\label{CP-asym}
\frac{\Gamma(B^0_s(t)\to \mu^+\mu^-)-
\Gamma(\bar B^0_s(t)\to \mu^+\mu^-)}{\Gamma(B^0_s(t)\to \mu^+\mu^-)+
\Gamma(\bar B^0_s(t)\to \mu^+\mu^-)}
=\frac{{\cal S}_{\mu\mu}\sin(\Delta M_st)}{\cosh(y_st/ \tau_{B_s}) + 
{\cal A}^{\mu\mu}_{\Delta\Gamma_s} \sinh(y_st/ \tau_{B_s})},
\end{equation}
where the ${\cal C}^\lambda_{\mu\mu}$ terms cancel because of the $\eta_\lambda$ factor. It should be noted that
a non-vanishing ${\cal C}_{\rm \mu\mu}$ would be a smoking-gun signal for a new scalar contribution $S$.
CP-violating asymmetries of this kind in $B_{s,d}\to\ell^+\ell^-$ decays were also considered for various 
NP scenarios in Refs.~\cite{HL,DP,CKWW}, neglecting the effects of $\Delta\Gamma_s$ and the associated 
observable $\mathcal{A}^{\mu\mu}_{\Delta \Gamma_s}$. For a more recent study, including the
untagged observable, see Ref.~\cite{BFGK}.

It should be stressed that the non-perturbative decay constant $f_{B_s}$ cancels in 
$\mathcal{A}^{\mu\mu}_{\Delta \Gamma_s}$ as well as in ${\cal S}_{\mu\mu}$ and ${\cal C}_{\mu\mu}$, 
thereby making these observables theoretically clean probes for the search of NP signals \cite{Bsmumu-ADG,BFGK}. 
In the SM, a tiny residual uncertainty arises from QED corrections, which lead to effects at the 
$10^{-5}$ and $10^{-3}$ levels for $\mathcal{A}^{\mu\mu}_{\Delta \Gamma_s}$
and the CP asymmetries ${\cal S}_{\mu\mu}$, ${\cal C}_{\mu\mu}$, respectively \cite{BBS}. 

In the following discussion, we will explore the interplay of ${\cal A}^{\mu\mu}_{\Delta\Gamma_s}$ and
${\cal S}_{\mu\mu}$ with the observable $\overline{R}$ to search for NP and reveal its nature, in particular
whether it involves new (pseudo)-scalar contributions. Experimental feasibility studies of measurements of
the CP asymmetry in Eq.~(\ref{CP-asym}) have not yet been performed to the best of our knowledge. However,
we envision that an effort should be made to perform such a measurement at the LHC upgrade(s). In view of the
relation in Eq.~(\ref{CP-rel}), a measurement of ${\cal C}_{\rm \mu\mu}$ would not provide independent 
information. As such an analysis would require the reconstruction of the muon helicity, it is much more challenging
than the asymmetry in Eq.~(\ref{CP-asym}) involving the helicity-averaged rates. However, we will show that 
already information on just the sign of ${\cal C}_{\rm \mu\mu}$ would be sufficient to resolve certain ambiguities 
affecting the determination of $P$ and $S$. We encourage experimentalists to explore avenues to eventually 
measure the sign of the ${\cal C}_{\rm \mu\mu}$ observable.

\boldmath
\section{General CP-Violating New Physics}\label{sec:CP-gen}
\unboldmath
\subsection{Theoretical Description}\label{ssec:theo-des}
\begin{center}
\begin{figure}
\begin{center}
\includegraphics[width=0.5\textwidth]{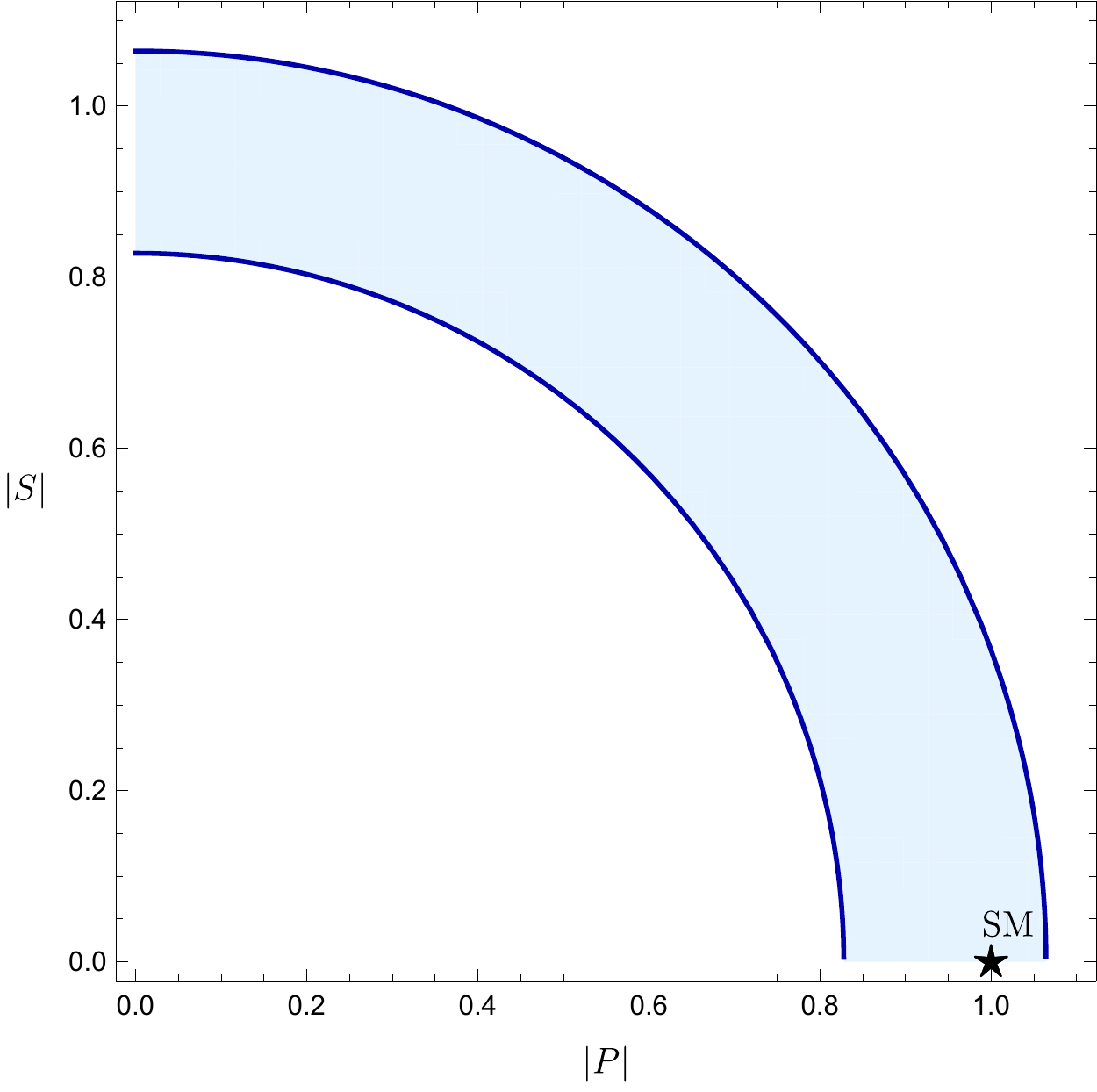}
\caption{Allowed region in the $|P|$--$|S|$ plane following from $r$, which is obtained by varying 
${\cal A}^{\mu\mu}_{\Delta\Gamma_s}$ between $-1$ and $+1$ and taking the $1\,\sigma$ uncertainty 
of the current $\overline{R}$ measurement into account. The black star indicates the SM values given in
Eq.~(\ref{SM_PS}).\label{fig:absSabsP_ex}}
\end{center}
\end{figure}
\end{center}

Let us start the general discussion of the CP-violating coefficients $P$ and $S$ in Eqs.~(\ref{P-def}) and
(\ref{S-def}), respectively, with the ratio $\overline{R}$ in Eq.~(\ref{Rbar}). Using the expression in 
Eq.~(\ref{R-expr}), we obtain
\begin{equation}\label{rr-def}
 r\equiv  \left[\frac{1+y_s}{1+{\cal A}^{\mu\mu}_{\Delta\Gamma_s}\,y_s} \right]\overline{R} =  |P|^2 + |S|^2.
\end{equation}
If we had a precise measurement of ${\cal A}^{\mu\mu}_{\Delta\Gamma_s}$, we could straightforwardly convert
$\overline{R}$ into $r$. In view of the large uncertainty in Eq.~(\ref{ADG-exp}), we use the general range in 
Eq.~(\ref{ADG-range}) to calculate
\begin{equation}\label{r-val}
0.69 \leq r \leq 1.13 ,
\end{equation}
where we have also taken into account the $1\sigma$ uncertainty of $\overline{R}$, given in Eq.~(\ref{eq:Rbar}).
This observable fixes a circular band with radius $\sqrt{r}$ in the $|P|$--$|S|$ plane, which we show in Fig.~\ref{fig:absSabsP_ex}. Using the observable 
${\cal A}^{\mu\mu}_{\Delta\Gamma_s}$, we can calculate a straight line in this plane through
\begin{equation}
\frac{|S|}{|P|}=\sqrt{\frac{\cos\Phi_P-{\cal A}^{\mu\mu}_{\Delta\Gamma_s}}{\cos\Phi_S+
{\cal A}^{\mu\mu}_{\Delta\Gamma_s}}},
\end{equation}
where we have introduced the abbreviations
\begin{equation}
\Phi_P\equiv 2\varphi_P-\phi_s^{\rm NP}, \quad
\Phi_S\equiv 2\varphi_S-\phi_s^{\rm NP}.
\end{equation}
If we assume that the CP-violating phases $\varphi_P$ and $\varphi_S$ take trivial values, i.e. $0^\circ$ or
$180^\circ$, $\overline{R}$ allows us to fix a circle in the $|P|$--$|S|$ plane through Eq.~(\ref{R-expr}), and 
the intersection with the straight line following from
\begin{equation}
\frac{|S|}{|P|}=\sqrt{\frac{\cos\phi_s^{\rm NP}-{\cal A}^{\mu\mu}_{\Delta\Gamma_s}}{\cos\phi_s^{\rm NP}
+{\cal A}^{\mu\mu}_{\Delta\Gamma_s}}}=\sqrt{\frac{1-{\cal A}^{\mu\mu}_{\Delta\Gamma_s}}{1+
{\cal A}^{\mu\mu}_{\Delta\Gamma_s}}}
\end{equation}
fixes $|P|$ and $|S|$, as discussed in detail in Refs.~\cite{Bsmumu-ADG,BFGK,ANS,FJTX}; note that we 
use the result for $\phi_s^{\rm NP}$ in Eq.~(\ref{phis-NP}). However, if we
allow for general CP-violating phases, any point on the circle with radius $\sqrt{r}$ is allowed since we obtain $|S|=0$ 
for $\cos\Phi_P={\cal A}^{\mu\mu}_{\Delta\Gamma_s}$ and $|P|=0$ for 
$\cos\Phi_S=-{\cal A}^{\mu\mu}_{\Delta\Gamma_s}$.

The measurement of a non-vanishing CP asymmetry ${\cal S}_{\mu\mu}$ would immediately establish the 
presence of non-trivial CP-violating phases. This observable fixes another straight line in the $|P|$--$|S|$ plane:
\begin{equation}
\frac{|S|}{|P|}=\sqrt{\frac{\sin\Phi_P-{\cal S}_{\mu\mu}}{\sin\Phi_S+{\cal S}_{\mu\mu}}}.
\end{equation}
However, as the CP-violating phases are in general unknown, the slope of this straight line is not determined, 
in analogy to the constraint following from ${\cal A}^{\mu\mu}_{\Delta\Gamma_s}$.

We have three independent observables at our disposal, $r$ as well as ${\cal A}^{\mu\mu}_{\Delta\Gamma_s}$ and 
${\cal S}_{\mu\mu}$, which depend on the four unknown parameters $|P|$, $\Phi_P$ and $|S|$, $\Phi_S$. Using the
general expressions for ${\cal A}^{\mu\mu}_{\Delta\Gamma_s}$ and ${\cal S}_{\mu\mu}$ in Eqs.~(\ref{ADG-expr}) 
and (\ref{S-expr}), respectively, yields
\begin{equation}
A\cos\Phi_P - B\sin\Phi_P  = C
\end{equation}
with
\begin{eqnarray}
A & \equiv & {\cal S}_{\mu\mu} + \sin\Phi_S \\
B & \equiv & {\cal A}^{\mu\mu}_{\Delta\Gamma_s} + \cos\Phi_S \\
C & \equiv & {\cal A}^{\mu\mu}_{\Delta\Gamma_s}\sin\Phi_S -  {\cal S}_{\mu\mu} \cos\Phi_S.
\end{eqnarray}
This equation allows us to determine $\Phi_P$ as a function of $\Phi_S$ with the help of
\begin{eqnarray}
\sin\Phi_P&=&-\left(\frac{BC}{A^2+B^2}\right)\pm\sqrt{\left(\frac{BC}{A^2+B^2}\right)^2+
\left(\frac{A^2-C^2}{A^2+B^2}\right)}\\
          &=&\frac{-BC\pm |A|\sqrt{A^2+B^2-C^2}}{A^2+B^2}.
\label{eq:sinPhiP}
\end{eqnarray}
The expression under the square root is actually factorizable, thereby yielding
\begin{eqnarray}
\sqrt{A^2+B^2-C^2}&=&|1+{\cal A}^{\mu\mu}_{\Delta\Gamma_s}\cos\Phi_S+{\cal S}_{\mu\mu}\sin\Phi_S|.
\end{eqnarray}

\begin{figure}[t] 
   \centering
   \includegraphics[width=2.8in]{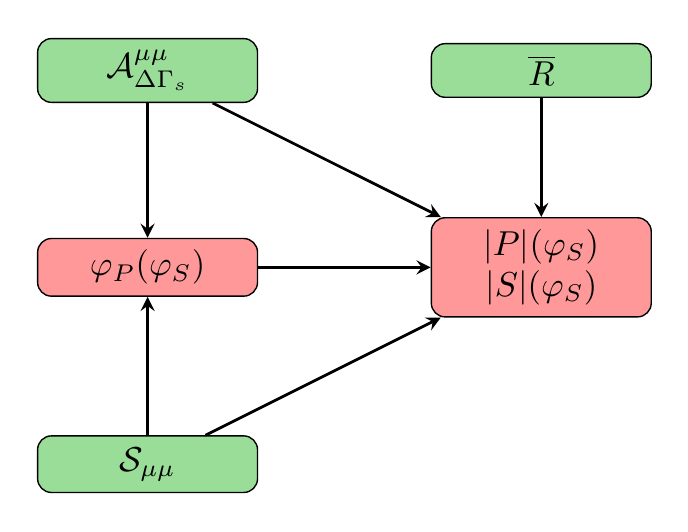} 
   \caption{Flowchart to illustrate the general strategy to determine $|P|$ and $|S|$ as functions of the
   the CP-violating phase $\varphi_S$ from the $B^0_s\to\mu^+\mu^-$ observables.}\label{fig:strategy_general}
\end{figure}

Using then the observables $r$ and ${\cal A}^{\mu\mu}_{\Delta\Gamma_s}$, we may determine
\begin{equation}
|P|=\sqrt{\left(\frac{\cos\Phi_S+{\cal A}^{\mu\mu}_{\Delta\Gamma_s}}{\cos\Phi_S+\cos\Phi_P}\right)r},
\quad
|S|=\sqrt{\left(\frac{\cos\Phi_P-{\cal A}^{\mu\mu}_{\Delta\Gamma_s}}{\cos\Phi_P+\cos\Phi_S}\right)r}
\label{eq:PSADG}
\end{equation}
as functions of the CP-violating phase $\Phi_S$. Using instead of ${\cal A}^{\mu\mu}_{\Delta\Gamma_s}$ the 
CP asymmetry 
${\cal S}_{\mu\mu}$ yields
\begin{equation}
|P|=\sqrt{\left(\frac{\sin\Phi_S+{\cal S}_{\mu\mu}}{\sin\Phi_S+\sin\Phi_P}\right)r},
\quad
|S|=\sqrt{\left(\frac{\sin\Phi_P-{\cal S}_{\mu\mu}}{\sin\Phi_P+\sin\Phi_S}\right)r}.
\label{eq:PSSmumu}
\end{equation}
The expression in Eq.~(\ref{eq:sinPhiP}) leaves us with a twofold ambiguity for $\varphi_P$ for every value of $\varphi_S$.
Information on the sign of ${\cal C}_{\mu\mu}$ allows us to determine the correct branch and thus obtain a single solution 
for $\varphi_P$ as a function of $\varphi_S$. However, both branches have the same dependence of $|P|$ and $|S|$ 
on $\varphi_S$, so a single solution for $|P|$ and $|S|$ as a function of $\varphi_S$ can be obtained even when 
no information on the sign of ${\cal C}_{\mu\mu}$ is available. In the flowchart in Fig.~\ref{fig:strategy_general}, we
illustrate this general method for analyzing the observables provided by the $B^0_s\to\mu^+\mu^-$ decay, and 
we will provide an example of this formalism in the next subsection.

\subsection{Discussion and Illustration}
\boldmath
\subsubsection{Vanishing Mixing-Induced CP Violation}
\unboldmath
An interesting situation arises for ${\cal S}_{\mu\mu}=0$. Although one may naively conclude that the
CP-violating phases take then simply trivial values, this is actually not the case because of the structure of the 
expression in Eq.~(\ref{S-expr}). In fact, we obtain the following extremal values on the circle with radius 
$\sqrt{r}$ in the $|P|$--$|S|$ plane:
\begin{equation}
|P_\pm|=\sqrt{\left(\frac{1\mp{\cal A}^{\mu\mu}_{\Delta\Gamma_s}}{2}\right)r},
\quad
|S_\pm|=\sqrt{\left(\frac{1\pm{\cal A}^{\mu\mu}_{\Delta\Gamma_s}}{2}\right)r},
\end{equation}
where the region between these points can be accessed by varying $\Phi_S$. 
In the case of ${\cal A}^{\mu\mu}_{\Delta\Gamma_s}=\pm1$, we have
\begin{equation}
|S|=0, \quad |P|=\sqrt{r}, \quad \sin\Phi_P=0, 
\end{equation}
yielding ${\cal A}^{\mu\mu}_{\Delta\Gamma_s}=+\cos\Phi_P=\pm1$, or
\begin{equation}
|P|=0, \quad |S|=\sqrt{r}, \quad \sin\Phi_S=0, 
\end{equation}
yielding ${\cal A}^{\mu\mu}_{\Delta\Gamma_s}=-\cos\Phi_S=\pm1$. For
$|{\cal A}^{\mu\mu}_{\Delta\Gamma_s}|<1$, we get
\begin{equation}
\left.\frac{|S|}{|P|}\right.=\sqrt{\frac{(1-{\cal A}^{\mu\mu}_{\Delta\Gamma_s})(1+
{\cal A}^{\mu\mu}_{\Delta\Gamma_s})}{1+2\,{\cal A}^{\mu\mu}_{\Delta\Gamma_s}\cos\Phi_S+
({\cal A}^{\mu\mu}_{\Delta\Gamma_s})^2}}\,.
\end{equation}
A particularly interesting situation arises for ${\cal A}^{\mu\mu}_{\Delta\Gamma_s}=0$, corresponding to
the following point in the $|P|$--$|S|$ plane:
\begin{equation}
|P|=|S|=\sqrt{\frac{r}{2}}\,.
\end{equation}

\boldmath
\subsubsection{Sizeable Mixing-Induced CP Violation}\label{sssec:genex}
\unboldmath
Let us now turn to mixing-induced CP violation in $B^0_s\to\mu^+\mu^-$, and discuss a scenario with a large
value of ${\cal S}_{\mu\mu}$, which requires significant CP-violating phases originating from physics beyond the 
SM. In order to illustrate this situation and the formalism discussed in Subsection~\ref{ssec:theo-des}, we 
consider an example which is characterized by 
\begin{equation}\label{S-input}
|S|= 0.30, \quad \varphi_S = 20^\circ.
\end{equation}
Assuming furthermore
\begin{equation}
 \varphi_P = 30^\circ,
\end{equation}
the central value of the observable $ \overline{R}$ in Eq.~(\ref{eq:Rbar}) yields
\begin{equation}
|P|= 0.89.
\end{equation}
These values of $|P|$ and $|S|$ fall well within the currently allowed region in the $|P|$--$|S|$ plane shown 
in Fig.~\ref{fig:absSabsP_ex}. We obtain the following set of observables:
\begin{equation}
\overline{R}=0.84, \quad {\cal A}^{\mu\mu}_{\Delta\Gamma_s}= 0.37, \quad {\cal S}_{\mu\mu} = 0.71,  
\quad {\cal C}_{\mu\mu} = 0.60,
\end{equation}
and assume that they were measured at a future experiment.

Let us now illustrate how we may obtain insights into NP effects using these observables. The deviation of 
${\cal A}^{\mu\mu}_{\Delta\Gamma_s}$ from the SM prediction $+1$ would indicate NP effects. Having the
measured ${\cal A}^{\mu\mu}_{\Delta\Gamma_s}$ at hand, we may use Eq.~(\ref{rr-def}) to convert 
$\overline{R}$ into $r$, yielding
\begin{equation}
r = 0.87.
\end{equation}
Moreover, the precision of the measured $B^0_s\to\mu^+\mu^-$ branching ratio will then have significantly increased (see 
Section~\ref{sec:Exp} for a more detailed discussion), allowing us to reduce the width of the circular band in 
Fig.~\ref{fig:absSabsP_ex}. However, without any information on ${\cal S}_{\mu\mu}$, we could not narrow down 
further $|S|$ and $|P|$ in a model-independent way, i.e.\ we would still be left with the whole circular region, 
and could in particular not establish a non-vanishing scalar contribution~$S$.

\begin{center}
\begin{figure}
\begin{center}
\includegraphics[width=0.55\textwidth]{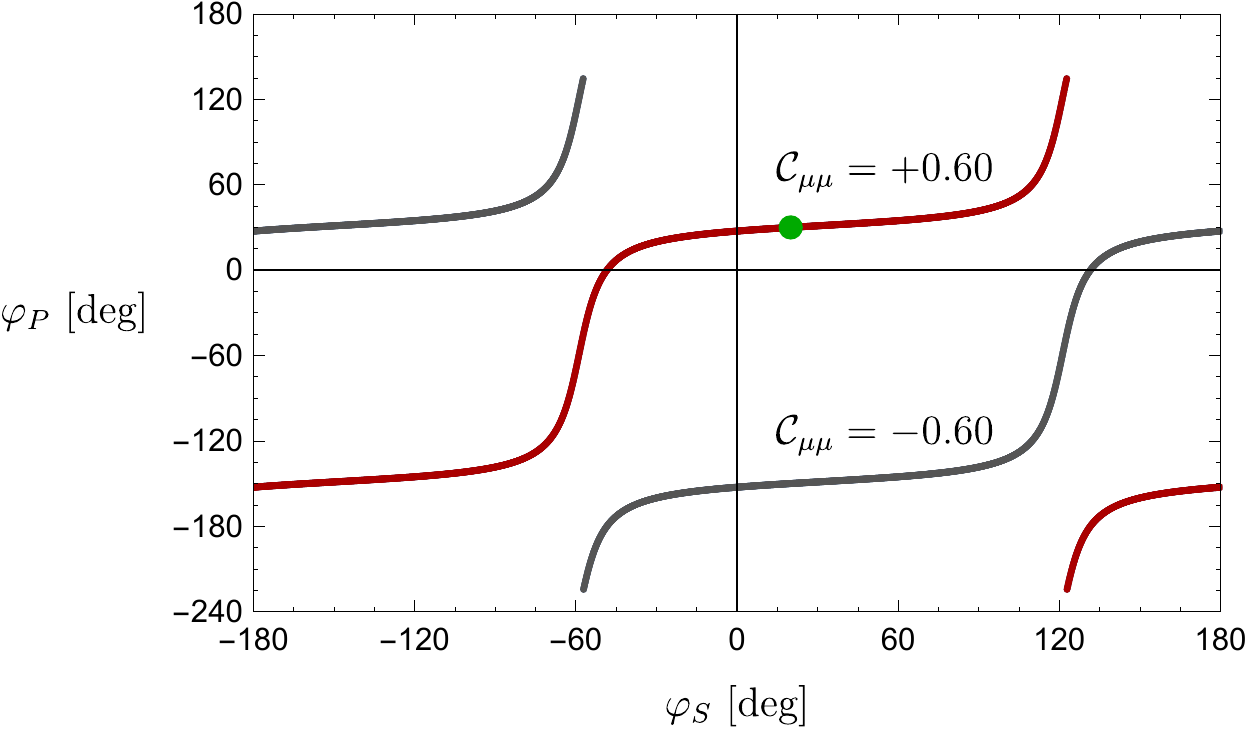}
\caption{Correlation between $\varphi_P$  and $\varphi_S$ for  ${\cal A}^{\mu\mu}_{\Delta\Gamma_s}=0.37$ and 
${\cal S}_{\mu\mu}=0.71$. The red and grey curves correspond to ${\cal C}_{\mu\mu}>0$ and ${\cal C}_{\mu\mu}<0$, 
respectively. The green dot marks the input parameters in Eq.~(\ref{S-input}).\label{fig:PhiPvsPhiS}}
\end{center}
\end{figure}
\end{center}

\begin{center}
\begin{figure}
\begin{center}
\includegraphics[width=0.44\textwidth]{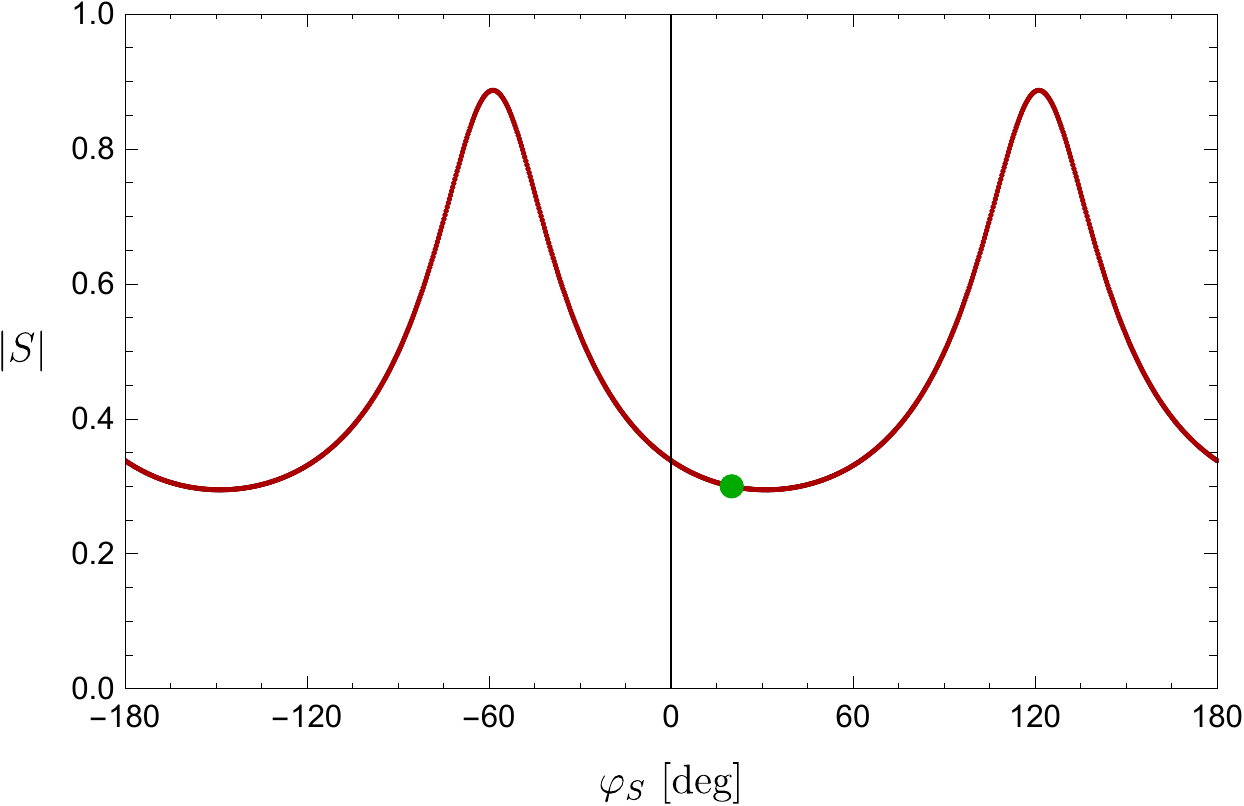}
\includegraphics[width=0.44\textwidth]{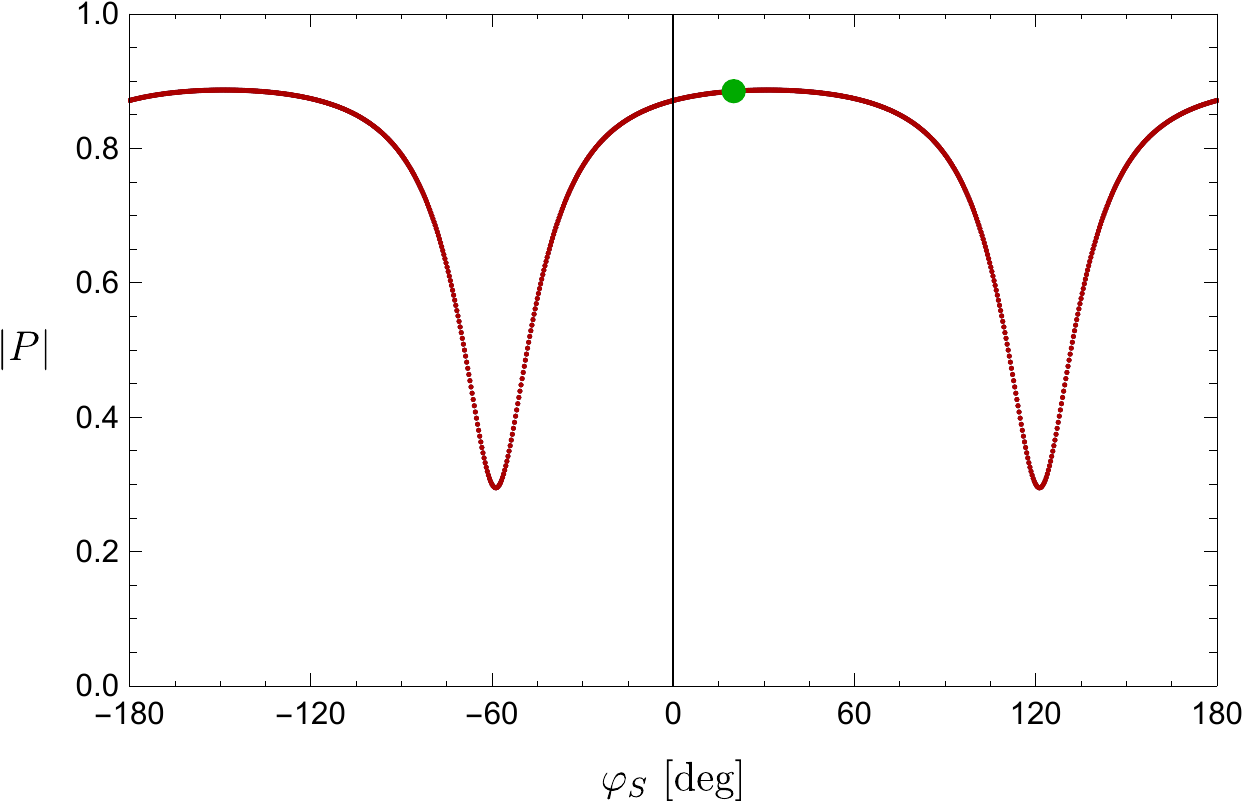}
\caption{The coefficients $|S|$ and $|P|$ determined as functions of $\varphi_S$ for the example discussed 
in the text. The corresponding input parameters are marked by the green dot.
\label{fig:SphiS_ex-PphiS_ex}}
\end{center}
\end{figure}
\end{center}

The measurement of the observable ${\cal S}_{\mu\mu}$ different from zero would signal new sources of CP violation. 
Using then Eq.~(\ref{eq:sinPhiP}), we could determine $\varphi_P$ as a function of $\varphi_S$, as illustrated in 
Fig.~\ref{fig:PhiPvsPhiS}. The information on the sign of ${\cal C}_{\mu\mu}$ would allow us to resolve the ambiguity, 
as indicated in the figure. Note that the points $(\varphi_S,\varphi_P)=(0^\circ,0^\circ)$ and $(180^\circ,180^\circ)$ 
would be excluded through the contours. Using Eqs.~(\ref{eq:PSADG}) or (\ref{eq:PSSmumu}), we obtain $|S|$ and 
$|P|$ as functions of $\varphi_S$, as shown in Fig.~\ref{fig:SphiS_ex-PphiS_ex}. Here, information about the sign 
of ${\cal C}_{\mu\mu}$ plays no further role. Interestingly, we would now be able to put a lower bound on $|S|$, i.e.\ 
could conclude that we have new scalar contributions. We insist on the fact that in order to obtain this highly 
non-trivial information, a measurement of the CP asymmetry ${\cal S}_{\mu\mu}$ is required.

Although we can only determine the $B^0_s\to\mu^+\mu^-$ parameters as functions of $\varphi_S$, this analysis 
would have profound implications, establishing in particular new scalar and pseudo-scalar contributions with 
CP-violating phases. In order to obtain further insights, more information is needed and assumptions 
about short-distance coefficients have to be made.

\boldmath
\section{Relations Between (Pseudo)-Scalar Coefficients}\label{sec:CP-rel}
\unboldmath
\subsection{General Framework}\label{ssec:gen-frame}
The effects of new particles enter the coefficients in Eqs.~(\ref{P-def}) and (\ref{S-def}) through the short-distance
coefficients $C_P$, $C_P'$ and $C_S$, $C_S'$, which describe new pseudo-scalar and scalar contributions,
respectively, and $C_{10}$, $C_{10}'$. As the constraints from the ATLAS and CMS experiments at the LHC for 
direct searches of new particles support the picture of a NP scale $\Lambda_{\rm NP}$ which is much larger than 
the electroweak scale $\Lambda_{\rm EW}$, the corresponding NP effects can be 
described in a model-independent way through an effective Lagrangian where
the heavy degrees of freedom, i.e.\ the NP particles, have been integrated out at $\Lambda_{\rm NP}$. If we require 
then invariance under the SM gauge group $SU(2)_L\times U(1)_Y$ for the renormalization group evolution
between $\Lambda_{\rm NP}$ and $\Lambda_{\rm EW}$, a ``SM Effective Field Theory" (SMEFT) can be set 
up \cite{BW,GIMR} and matched to the effective Hamiltonian in Eq.~(\ref{Heff}) describing  $B^0_s\to\mu^+\mu^-$ 
decays. Following these lines and applying the machinery of effective quantum field theory, the following relations 
among the corresponding short-distance coefficients can be derived \cite{AGMC}:
\begin{equation}\label{rel-PS}
C_P=-C_S
\end{equation}
\begin{equation}\label{rel-PS-prime}
C_P'=C_S'.
\end{equation}
A further application of these relations -- assuming no new sources of CP violation -- 
can be found in Ref.~\cite{ANS}, while a fit of data to the SMEFT scenario with complex coefficients
was performed in Ref.~\cite{BBJ}. For a discussion within specific models, see Ref.~\cite{BFGK}.

In this section, we explore the implication of Eqs.~(\ref{rel-PS}) and (\ref{rel-PS-prime}) for the general analysis 
of CP violation discussed in Section~\ref{sec:CP-gen}. To this end, we express the relevant quantities in terms of the 
scalar short-distance coefficients
\begin{equation}
C_S\equiv |C_S|e^{i\tilde\varphi_S}, \quad
C_S'\equiv |C_S'|e^{i\tilde\varphi_S'},
\end{equation}
which yields
\begin{equation}\label{conv-1}
P\equiv |P|e^{i\varphi_P}=|P|\cos\varphi_P + i |P| \sin\varphi_P=
{\cal C}_{10}-\frac{1}{w}\left[\frac{1+|x|e^{i\Delta}}{1-|x|e^{i\Delta}}\right]|S|e^{i\varphi_S}
\end{equation}
with
\begin{equation}
w\equiv\sqrt{1-\frac{4m_\mu^2}{M_{B_s}^2}}, \qquad {\cal C}_{10} \equiv \frac{C_{10} - C'_{10}}{C_{10}^{\rm SM}}
\end{equation}
and
\begin{equation}
x\equiv|x|e^{i\Delta}\equiv\left|\frac{C_S'}{C_S}\right|e^{i(\tilde\varphi_S'-\tilde\varphi_S)}.
\end{equation}
We will refer to this notation as the SMEFT parametrization. It it useful to write Eq.~(\ref{conv-1}) in the following form:
\begin{equation}\label{gen-rel-PS}
wP+\left[\frac{1+|x|e^{i\Delta}}{1-|x|e^{i\Delta}}\right]S = w \, {\cal C}_{10}.
\end{equation}
In the Appendix, we present expressions that allow us to obtain the $B^0_s\to\mu^+\mu^-$ observables in terms of the parametrization introduced above. As $P$ requires input for $C_{10}$, $C_{10}'$, we shall now first discuss these coefficients.

\boldmath
\subsection{Closer Look at $C_{10}$ and $C'_{10}$}\label{sec:C10}
\unboldmath
The Wilson coefficients $C_{10}$ and $C_{10}'$ enter in $P$ through the following combination:
\begin{equation}\label{eq:CalC10}
{\cal C}_{10}\equiv |{\cal C}_{10}|e^{i\varphi_{10}} \equiv \frac{C_{10} - C'_{10}}{C_{10}^{\rm SM}}
=1+{\cal C}_{10}^{\rm NP},
\end{equation}
where $\varphi_{10}$ is a CP-violating phase and
\begin{equation}
{\cal C}_{10}^{\rm NP}\equiv|{\cal C}_{10}^{\rm NP}|e^{i\varphi_{10}^{\rm NP}}
=\frac{C_{10}^{\rm NP} - C'_{10}}{C_{10}^{\rm SM}}
\end{equation}
parametrizes NP effects.
The relations
\begin{equation}
|{\cal C}_{10}|=\sqrt{1+2|{\cal C}_{10}^{\rm NP}|\cos\varphi_{10}^{\rm NP}+|{\cal C}_{10}^{\rm NP}|^2},
\end{equation}
\begin{equation}
|{\cal C}_{10}|\cos\varphi_{10}  =  1+|{\cal C}_{10}^{\rm NP}|\cos\varphi_{10}^{\rm NP}, \quad
|{\cal C}_{10}|\sin\varphi_{10} = |{\cal C}_{10}^{\rm NP}|\sin\varphi_{10}^{\rm NP},
\end{equation}
\begin{equation}\label{tanphi10}
\tan\varphi_{10} =\frac{|{\cal C}_{10}^{\rm NP}|\sin\varphi_{10}^{\rm NP}}{1+
|{\cal C}_{10}^{\rm NP}|\cos\varphi_{10}^{\rm NP}}
\end{equation}
allow us to express ${\cal C}_{10}$ in terms of the -- in general -- complex NP coefficient ${\cal C}_{10}^{\rm NP}$.

In order to reveal the substructure of $P$, information on ${\cal C}_{10}$ is required. In specific models, we 
may calculate ${\cal C}_{10}^{\rm NP}$ (see, for instance, Ref.~\cite{BFGK}). Alternatively, using experimental 
data for $B\to K^{(*)}\ell^+\ell^-$ decays, we may determine $C_{10} - C'_{10}$ from experiment (see Ref.~\cite{ANSS}
and references therein). In practice, the corresponding 
NP contributions are extracted through involved global fits to sets of large numbers of observables. We use the 
results from Ref.~\cite{ANSS}, where different scenarios for NP in real Wilson coefficients are discussed. 
Considering NP in individual Wilson coefficients, the authors find that the data is best explained by a contribution 
to the short-distance coefficient $C_9$ of the four-fermion operator 
$O_9=(\bar s \gamma_\mu P_L b) (\bar\mu\gamma^\mu\mu)$, which does not contribute to 
$B_s^0 \to \mu^+\mu^-$, yielding ${\cal C}_{10}^\text{NP} = 0$ and thus ${\cal C}_{10} = 1$. However, a similarly 
good fit is obtained by assuming the relation $C_9^\text{NP} = -C_{10}^\text{NP}$ for real coefficients, which appears 
in models with new particles that couple only to left-handed leptons. In this case, we find
\begin{equation}\label{C10-range}
{\cal C}_{10}^{\rm NP}=-0.16^{+0.04}_{-0.04},
\end{equation}
where the minus sign follows from $C_{10}^\text{SM}$ taking a negative value, as given in Eq.~(\ref{C10-SM}),
resulting in 
\begin{equation}\label{C10-range-Fit}
{\cal C}_{10} = 0.84_{-0.04}^{+0.04}.
\end{equation}

In Ref.~\cite{ANSS}, CP-violating phases are neglected. However, the short-distance coefficients are in general 
complex, and the phases can be included in the fit. In Ref.\ \cite{Altmannshofer:2014rta}, such an analysis is 
performed. The results are presented as 2D confidence contours in the complex plane of the coefficients $C_{10}$
and $C'_{10}$. To probe for the possible size of $\varphi_{10}$ and $ |\mathcal{C}_{10}|$, 
we assume that $C'_{10}=0$ and convert the $1\sigma$ allowed regions for the complex Wilson coefficient 
$C_{10}$ shown in Ref.~\cite{Altmannshofer:2014rta} into $\mathcal{C}_{10}$ using Eq. (\ref{eq:CalC10}),
yielding
\begin{eqnarray}
-40^{\circ} <  \varphi_{10} < -14^{\circ} \quad& \lor& \quad 14^{\circ} <  \varphi_{10} < 40^{\circ},\\
0.79< &|\mathcal{C}_{10}|&<0.98.
\end{eqnarray}
Due to the structure of Eq.~(\ref{tanphi10}), we obtain a rather constrained range for the CP-violating phase
$ \varphi_{10}$. It is also interesting to note that the range for the absolute value $|\mathcal{C}_{10}|$ is 
consistent with the result in Eq.~(\ref{C10-range-Fit}). 
   
In the future, analyses of CP-violating effects in 
$B\to K^{(*)}\ell^+\ell^-$ and $B_s\to \phi\mu^+\mu^-$ observables, as introduced in 
Refs.~\cite{CP-SL-rare-1,CP-SL-rare-2}, will allow us to get a much sharper picture of $|{\cal C}_{10}|$ and
a possible complex phase $\varphi_{10}$. It would be very useful to add the complex coefficient ${\cal C}_{10}$ as a 
default output to the corresponding sophisticated fits to the semileptonic rare $B_{(s)}$ decay data. 

For the numerical illustrations below, we will either use the range in Eq.~(\ref{C10-range-Fit}) for real 
Wilson coefficients $C_{10}$ and $C_{10}'$, or we will consider the case $|{\cal C}_{10}|=1$, 
$\varphi_{10} = 0^\circ$, where NP effects would enter exclusively through (pseudo)-scalar contributions.

An interesting situation arises if we consider a scenario where NP effects enter only through ${\cal C}_{10}$, with
vanishing coefficients $C_P$, $C_P'$ and $C_S$, $C_S'$, yielding $P={\cal C}_{10}$ and $S=0$. Specific examples 
are given by models with extra $Z'$ bosons (see, for instance, Ref.\ \cite{BFGK}) and scenarios with modified 
$Z$ couplings (such as in models with vector-like quarks \cite{BBCJ}). We would then have the simple 
expressions
\begin{equation}
{\cal A}^{\mu\mu}_{\Delta\Gamma_s}=\cos(2\varphi_{10}-\phi_s^{\rm NP}), \quad 
{\cal S}_{\mu\mu} =\sin(2\varphi_{10}-\phi_s^{\rm NP}) 
\label{eq:ObsCsCp0}
\end{equation}
with ${\cal C}_{\mu\mu} = 0$. Consequently, the observables would lie on a circle with radius one in the
${\cal A}^{\mu\mu}_{\Delta\Gamma_s}$--${\cal S}_{\mu\mu}$ plane.

\boldmath
\subsection{Extraction of $|x|$ and $\Delta$}\label{ssec:extractionxDelta}
\unboldmath
Applying the method presented in Subsection~\ref{ssec:theo-des}, we may determine $|S|$, $|P|$ and 
$\varphi_P$ from the $B^0_s\to\mu^+\mu^-$ observables as functions $\varphi_S$. Using Eq.~(\ref{conv-1}), 
we may convert these parameters into the ratio $x$ of the -- in general -- complex scalar short-distance
coefficients:
\begin{equation}
|x|e^{i\Delta} = \frac{w(P-{\cal C}_{10})+S}{w(P-{\cal C}_{10})-S},
\end{equation}
with
\begin{equation}
|x|=\sqrt{\frac{w^2\left|P-{\cal C}_{10}\right|^2+\left|S\right|^2+
2 \,w\,\Re\left[(P^*-{\cal C}_{10}^*)S\right]}{w^2\left|P-{\cal C}_{10}\right|^2+\left|S\right|^2-
2 \,w\,\Re\left[(P^*-{\cal C}_{10}^*)S\right]}}
\end{equation}
and
\begin{equation}
\cos\Delta\propto w^2|P-{\cal C}_{10}|^2-|S|^2, \quad
\sin\Delta\propto 2 \,w\, \Im\left[(P^*-{\cal C}_{10}^*)S\right],
\end{equation}
yielding
\begin{equation}
\tan\Delta=\frac{2 \,w\, \Im\left[(P^*-{\cal C}_{10}^*)S\right]}{w^2|P-{\cal C}_{10}|^2-|S|^2}.
\end{equation}
The quantities entering these expression can be expressed in terms of the absolute values and phases of
the relevant complex coefficients as 
\begin{equation}
|P-{\cal C}_{10}|=\sqrt{|P|^2-2|P||{\cal C}_{10}|\cos(\varphi_{10}-\varphi_P)+|{\cal C}_{10}|^2}
\end{equation}
and
\begin{equation}
\Im\left[(P^*-{\cal C}_{10}^*)S\right]=|S|\Bigl[|P|\sin(\varphi_S-\varphi_P)-|{\cal C}_{10}|\sin(\varphi_S-\varphi_{10}) \Bigr]
\end{equation}
\begin{equation}
\Re\left[(P^*-{\cal C}_{10}^*)S\right]=|S|\Bigl[|P|\cos(\varphi_S-\varphi_P)-|{\cal C}_{10}|\cos(\varphi_S-\varphi_{10}) \Bigr].
\end{equation}

\begin{figure}[t] 
   \centering
   \includegraphics[width=5.0in]{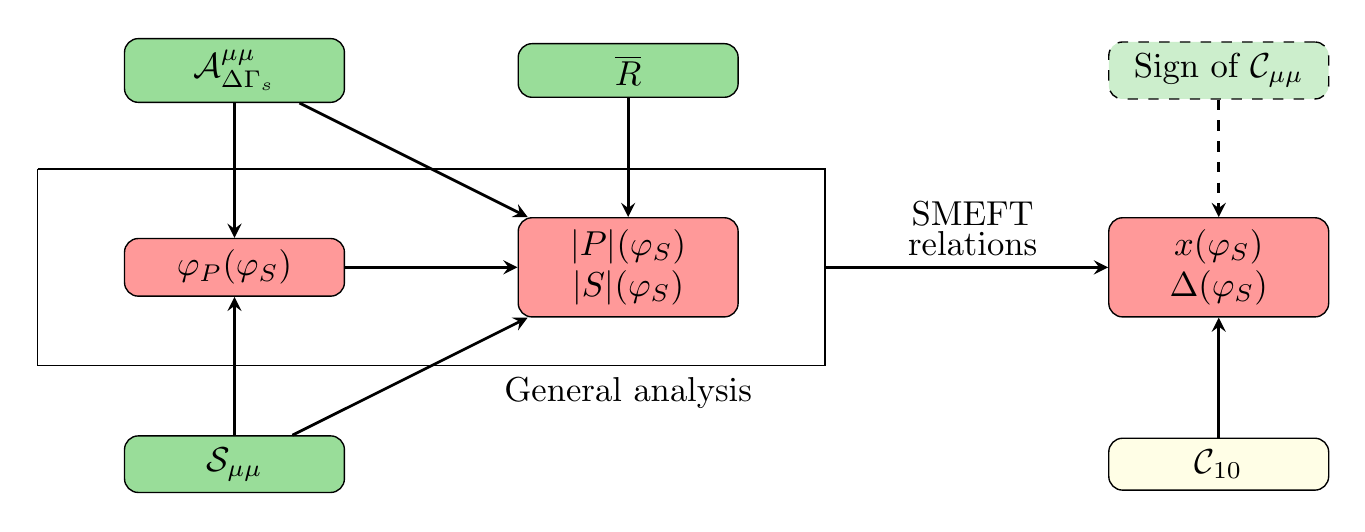} 
   \caption{Flowchart to illustrate the use of the SMEFT relations in Subsection~\ref{ssec:gen-frame} for the analysis of the
   $B^0_s\to\mu^+\mu^-$ observables as described in the text.}\label{fig:strategy_general_relations}
\end{figure}

\begin{figure}
   \centering
   \includegraphics[width=2.8in]{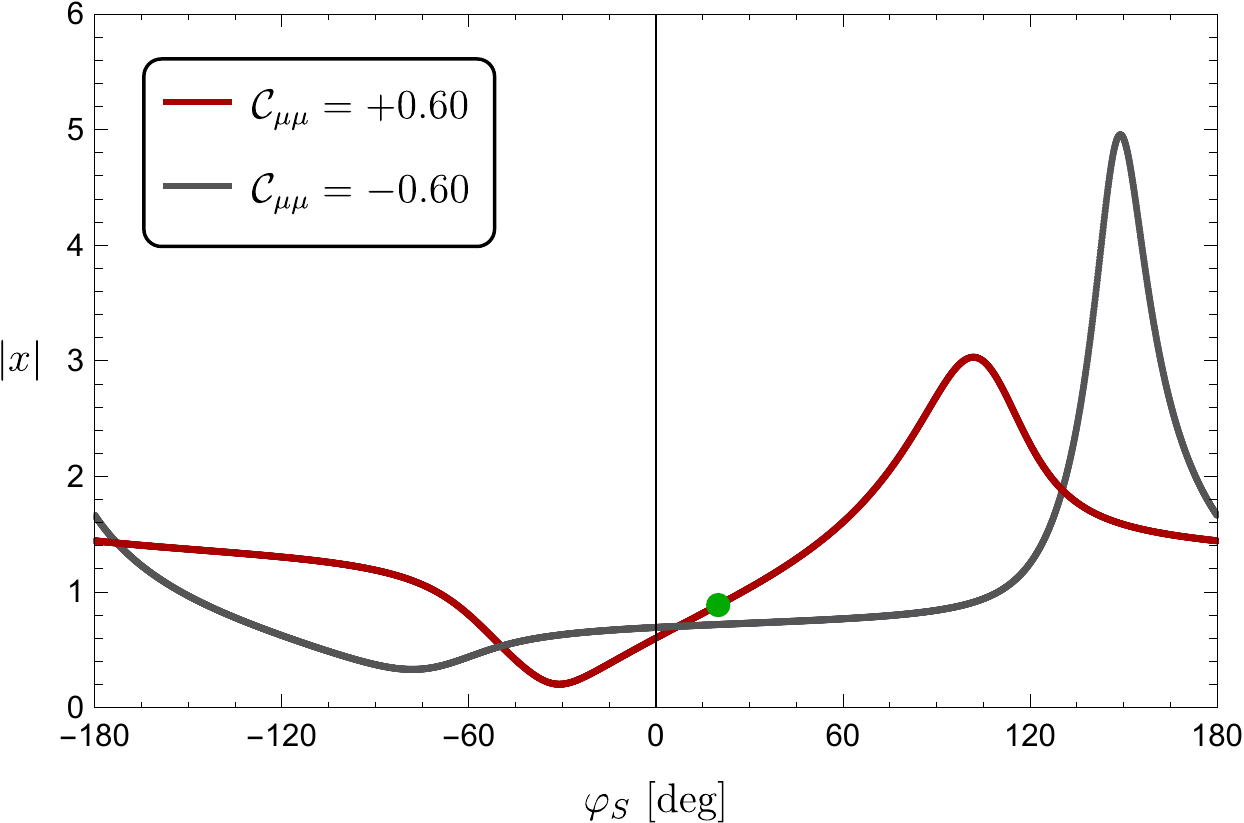} 
    \includegraphics[width=3.0in]{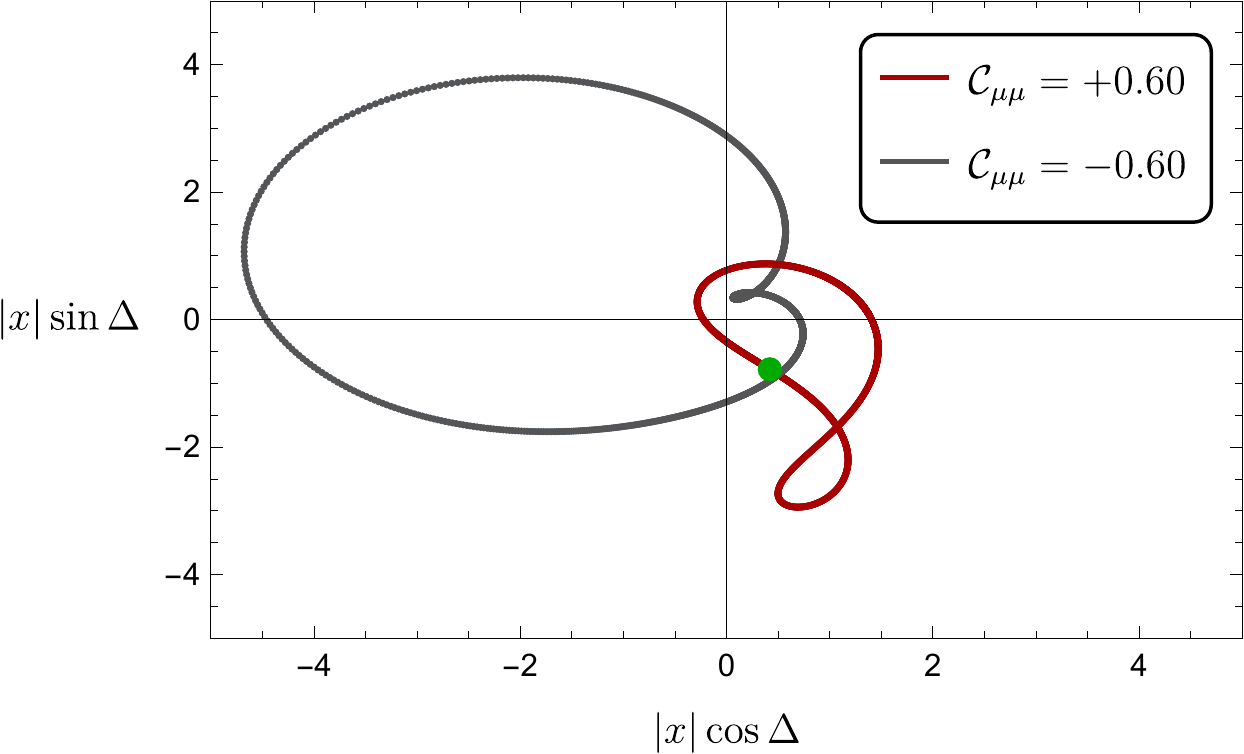} 
   \caption{Implementation of Fig.~\ref{fig:strategy_general_relations} for the example in Subsection~\ref{sssec:genex}, 
   corresponding to Eq.~(\ref{X-Del-Illu}), which is illustrated by the green dots. In the left panel, we give the resulting 
   dependence of $|x|$ on $\varphi_S$, while in the right panel, we show $|x|e^{i\Delta}$ in the complex plane. 
   The grey contours could be excluded through sign information for the observable 
   ${\cal C}_{\mu\mu}$.}\label{fig:X-Del-Applic}
\end{figure}

It is instructive to consider the example in Subsection~\ref{sssec:genex}, where $|S|=0.30$ and $\varphi_S=20^\circ$. 
Using the expressions given above, we can convert the corresponding values of $|P|=0.89$ and $\varphi_P=30^\circ$ 
into \begin{equation}\label{X-Del-Illu}
|x| = 0.89, \qquad \Delta = -62^\circ,
\end{equation}
where we have assumed no NP in ${\cal C}_{10}$, so $|{\cal C}_{10}| = 1$ and $\varphi_{10} = 0^\circ$.
In Fig.~\ref{fig:strategy_general_relations}, we give a flowchart for this strategy, and show in Fig.~\ref{fig:X-Del-Applic}
the situation corresponding to Eq.~(\ref{X-Del-Illu}). Using information on the sign of ${\cal C}_{\mu\mu}$,  we would
only be left with the red contours. We observe that $|x|e^{i\Delta}$ could be constrained in a very non-trivial way. 
The resulting contours depend strongly on the associated $B^0_s\to\mu^+\mu^-$ decay observables. 

In order to constrain the parameters more stringently, it is useful to make assumptions about scenarios, 
as we will illustrate in the next section. Following these lines, we may rule out a given scenario or confirm it, allowing 
us then to extract the corresponding parameters. By the time we may have measurements of CP violation in 
$B^0_s\to\mu^+\mu^-$ available, we should have a much better picture of the physics beyond the SM, thanks 
to the interplay between model building and data coming both from the high-energy and the high-precision frontiers. 
In particular, we should then also have some preferred scenarios, including specific patterns for the CP-violating
phases, which could be confronted with experimental data and the new strategies presented in this paper.

\subsection{Illustration}\label{ssec:illu}
As experimental data have already constrained the NP contribution $\phi_s^{\rm NP}$ to the
$B^0_s$--$\bar B^0_s$ mixing phase to be tiny, as given in 
Eq.~(\ref{phis-NP}), we may simplify the discussion by neglecting this quantity. Moreover, for the decay 
$B^0_s\to\mu^+\mu^-$, we have with excellent precision $w=1$. Let us now illustrate the formalism and
strategy discussed above through various examples. Here we shall choose values for the input parameters
to calculate the decay observables. Assuming then that these quantities have been measured at the future
LHC upgrade(s), we discuss the pictures emerging from the strategy discussed above. For simplicity, 
we do not consider experimental aspects in this section but will illustrate scenarios assuming uncertainties 
of future measurements in Section~\ref{sec:Exp}.

\boldmath
\subsubsection{$x=0$ and $|x|\to\infty$}
\label{sec:x0xinfty}
\unboldmath
The case $x=0$, which corresponds to $C_S'=C_P'=0$, is frequently considered in the literature for vanishing 
CP-violating phases (see, for instance, Ref.\ \cite{ANS}). It is interesting to note that the relation in 
Eq.~(\ref{gen-rel-PS}) gives
\begin{equation}\label{Rel-p}
 wP+S=w \, {\cal C}_{10},
\end{equation}
which reduces to $P+S=1$ for $w=1$ and ${\cal C}_{10}=1$. 
Allowing for possible CP violation, using the expressions in the Appendix we obtain 
\begin{equation}\label{rx0-calc}
r|_{x=0}=|{\cal C}_{10}|^2-2\cos(\varphi_{10}-\varphi_S)|{\cal C}_{10}||S|+2|S|^2
\end{equation}
as well as
\begin{equation}\label{ADGx0-calc}
{\cal A}^{\mu\mu}_{\Delta\Gamma_s}|_{x=0}=
\frac{|{\cal C}_{10}|^2\cos2\varphi_{10}-2\cos(\varphi_{10}+\varphi_S)|{\cal C}_{10}||S|}{|{\cal C}_{10}|^2-
2\cos(\varphi_{10}-\varphi_S)|{\cal C}_{10}||S| + 2 |S|^2}
\end{equation}
\begin{equation}\label{Smumux0-calc}
{\cal S}_{\mu\mu}|_{x=0}=
\frac{|{\cal C}_{10}|^2\sin2\varphi_{10}-2\sin(\varphi_{10}+\varphi_S)|{\cal C}_{10}||S|}{|{\cal C}_{10}|^2-
2\cos(\varphi_{10}-\varphi_S)|{\cal C}_{10}||S| + 2 |S|^2}
\end{equation}
\begin{equation}\label{Cmumux0-calc}
{\cal C}_{\mu\mu}|_{x=0}=\frac{2|S|\left[|{\cal C}_{10}|\cos(\varphi_{10}-\varphi_S)-
|S|\right]}{|{\cal C}_{10}|^2-2\cos(\varphi_{10}-\varphi_S)|{\cal C}_{10}||S| + 2 |S|^2}.
\end{equation}

Using Eq.~(\ref{rr-def}), and substituting $r$ and ${\cal A}_{\Delta\Gamma_s}^{\mu\mu}$ according
to Eqs.~(\ref{rx0-calc})~and~(\ref{ADGx0-calc}), we may determine $|S|$ as a function of $\varphi_{10}-\varphi_S$
from the measured value of $\overline{R}$:
\begin{displaymath}
\hspace*{-2.0truecm}
|S| = \frac{|{\cal C}_{10}|}{2}\Bigg\{\left[\cos(\varphi_{10}-\varphi_S)+y_s\cos(\varphi_{10}+\varphi_S)\right]
\end{displaymath}
\begin{equation}
\pm\sqrt{\left[\cos(\varphi_{10}-\varphi_S)+y_s\cos(\varphi_{10}+\varphi_S)\right]^2-
2\left[1+y_s\cos2\varphi_{10}-\frac{\overline{R}}{|{\cal C}_{10}|^2}(1+y_s)\right]}\Bigg\}.
\label{eq:S0}
\end{equation}
Note that the discriminant must have a value greater than or equal to zero, which implies the following upper bound:
\begin{equation}
|{\cal C}_{10}| \leq \sqrt{\left(\frac{2}{1-y_s}\right)\overline{R}}.
\end{equation}
The current experimental value of $\overline{R}$ in Eq.~(\ref{eq:Rbar}) yields
\begin{equation}
|{\cal C}_{10}| \leq 1.3 \pm 0.1,
\end{equation}
which is obviously consistent with ${\cal C}_{10}=1$.

\begin{center}
\begin{figure}
\includegraphics[width=0.5\textwidth]{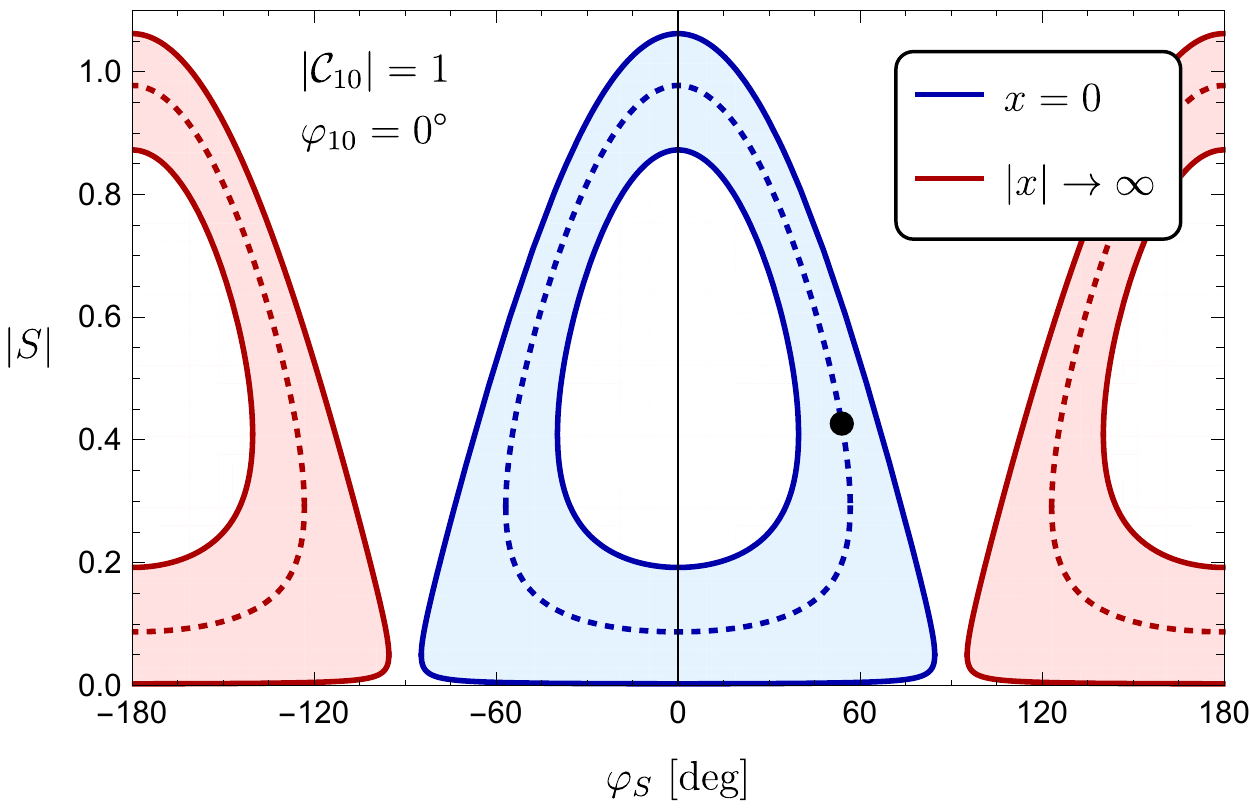}
\includegraphics[width=0.5\textwidth]{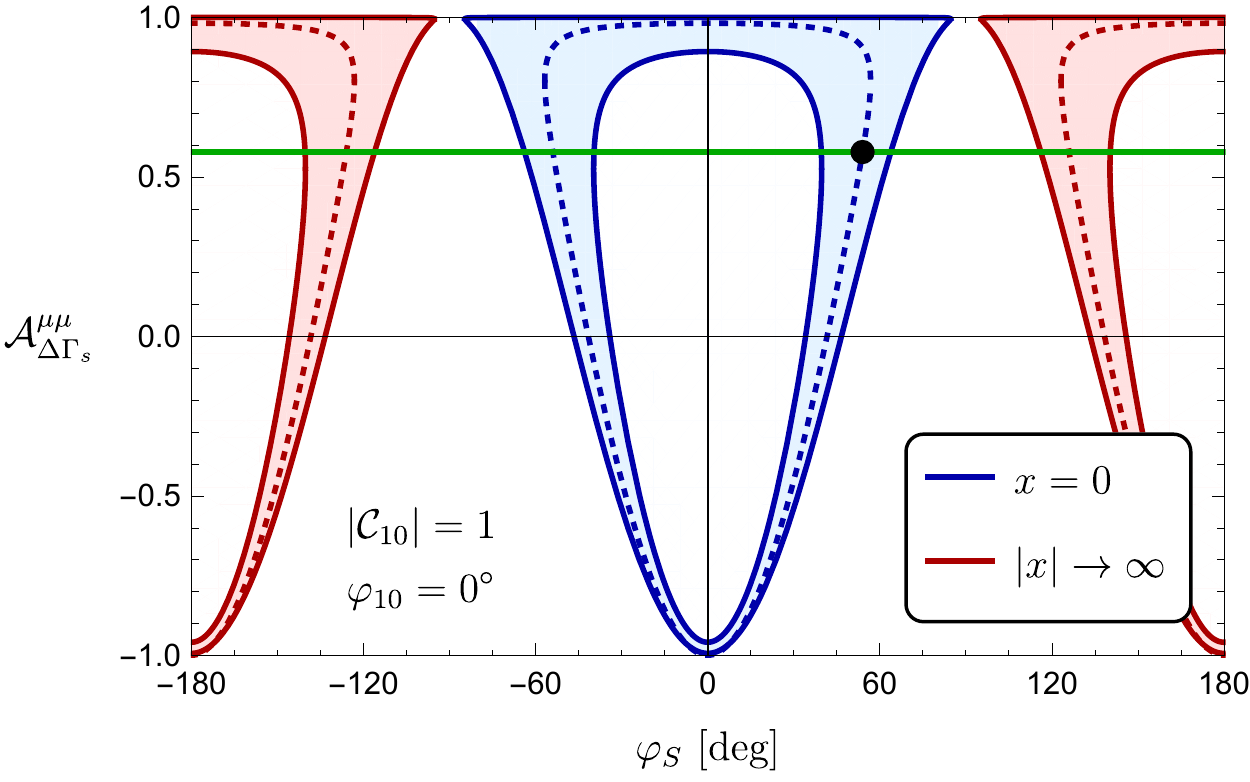}
\includegraphics[width=0.5\textwidth]{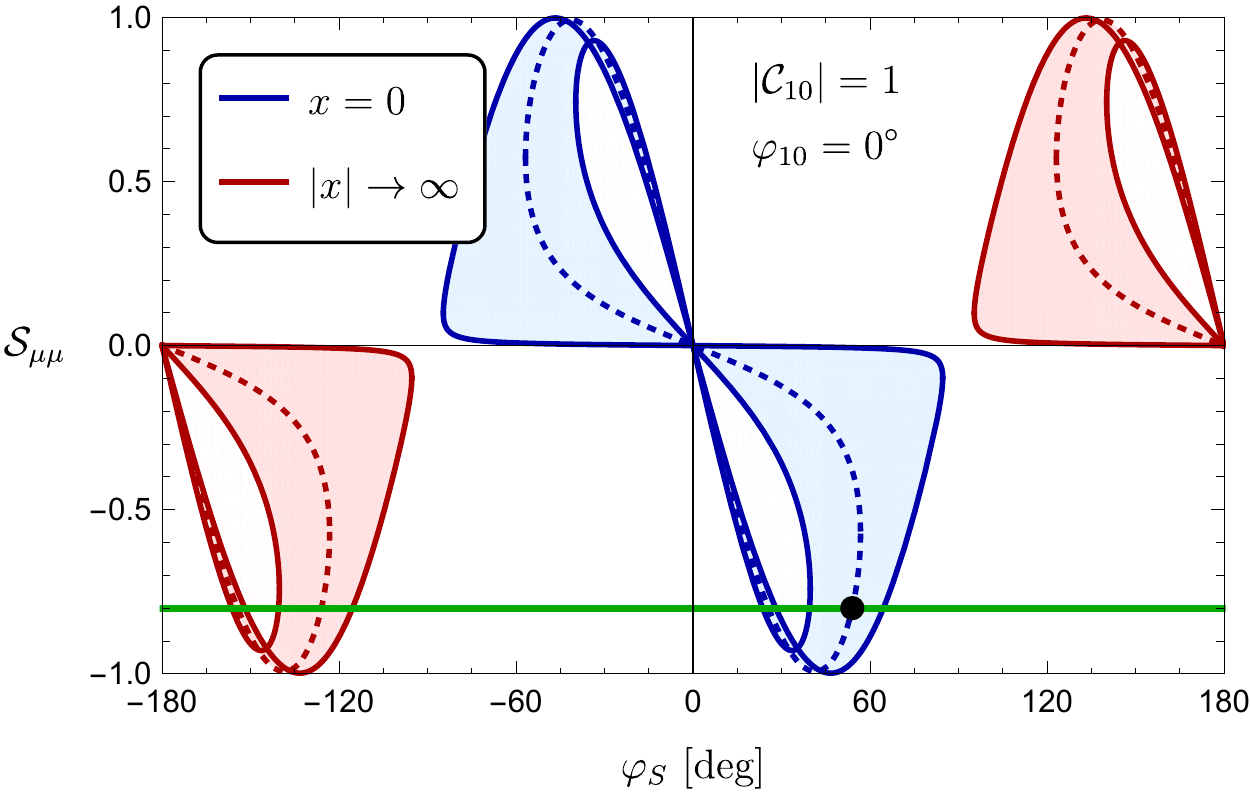}
\includegraphics[width=0.5\textwidth]{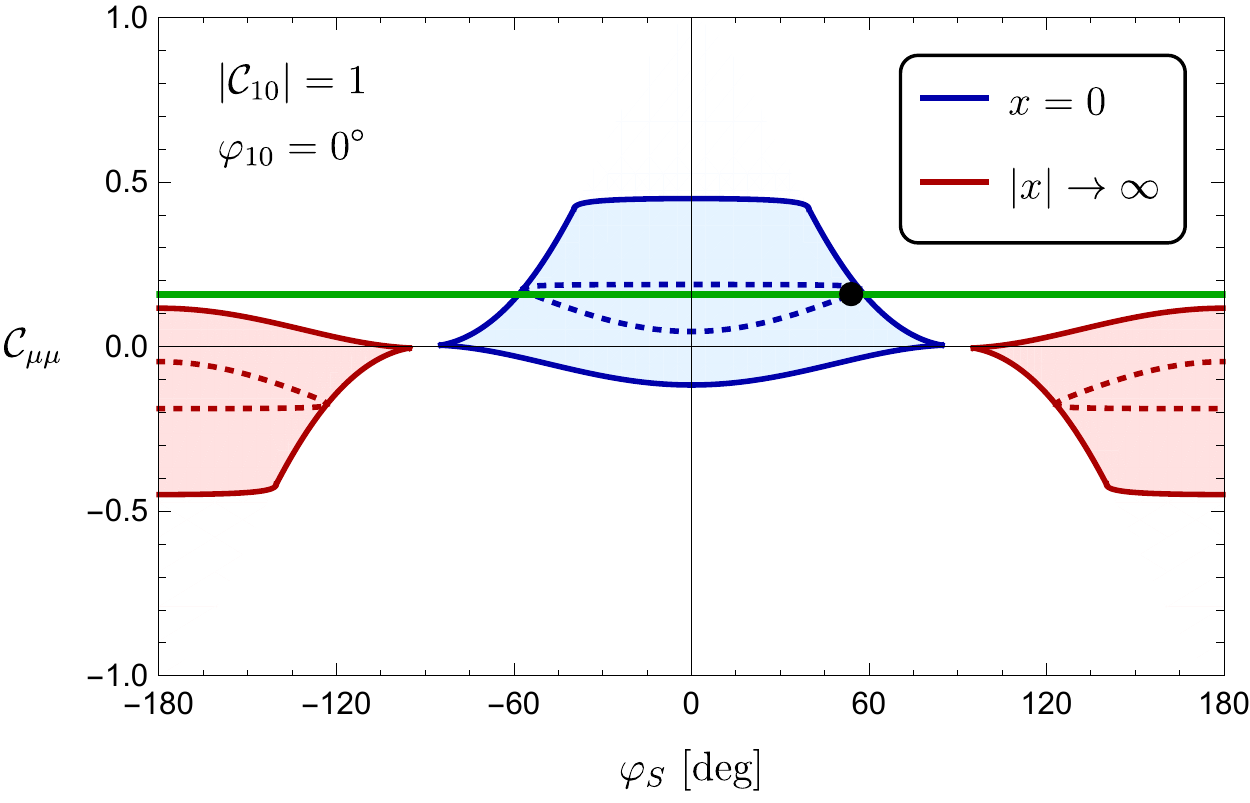}
\caption{Functional dependences between $|S|$, ${\cal A}_{\Delta\Gamma_s}^{\mu\mu}$,
${\cal S}_{\mu\mu}$, ${\cal C}_{\mu\mu}$ and the CP-violating phase $\varphi_S$ for 
$|\mathcal{C}_{10}|=1, \  \varphi_{10}=0^\circ$. 
The blue and red contours correspond to the scenarios $x=0$ and $|x|\rightarrow \infty$, respectively.
The allowed regions are determined within the  $1\,\sigma$ range for $\overline{R}$ given in Eq.\ (\ref{eq:Rbar}), 
where the dashed curve is associated with the central value for this observable. Notice that for each value of 
$\varphi_S$, we have in general two possible solutions for the observables, leading to closed loops in the parameter 
space. The black dot refers to the input parameters of the scenario in Eq.\ (\ref{eq:example0infinity}), whereas the green line shows the value of the observables in Eq.~(\ref{eq:obsx0}).\label{fig:absSvsPhisC1}}
\end{figure}
\end{center}

\begin{center}
\begin{figure}
\includegraphics[width=0.5\textwidth]{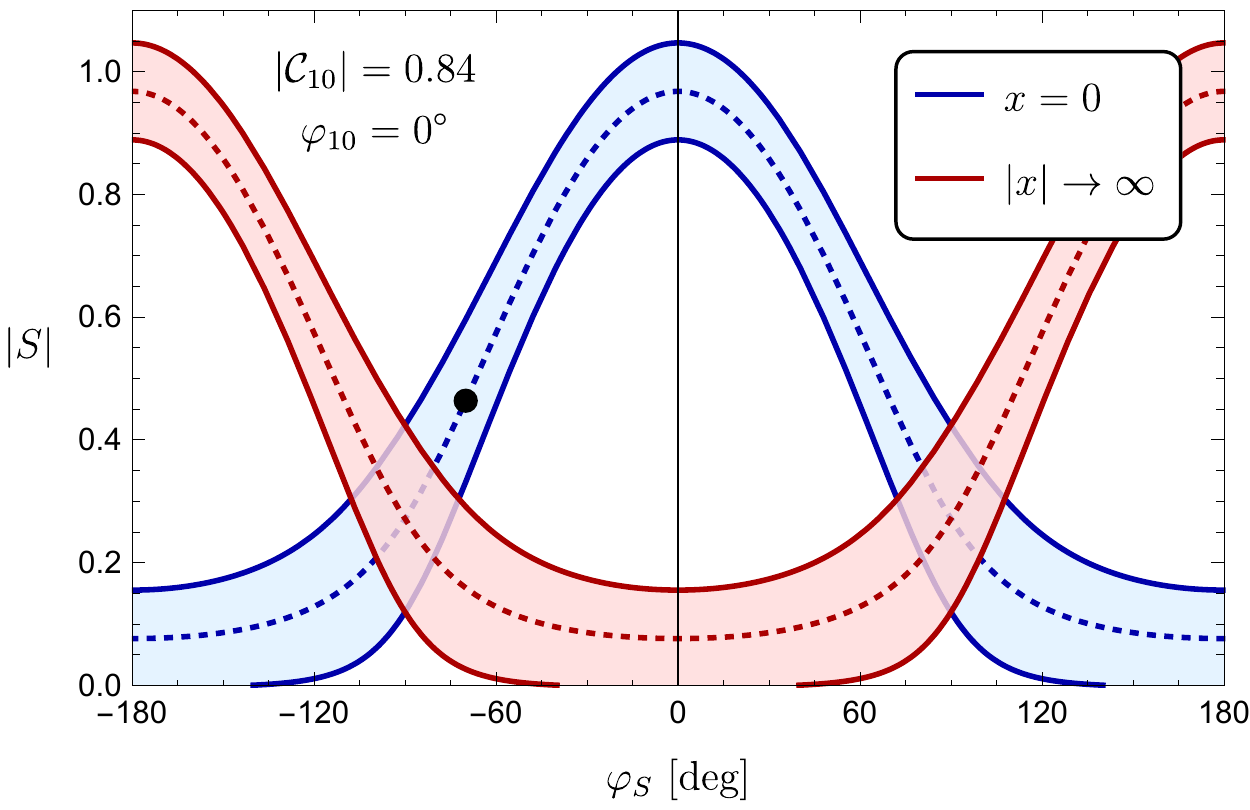}
\includegraphics[width=0.5\textwidth]{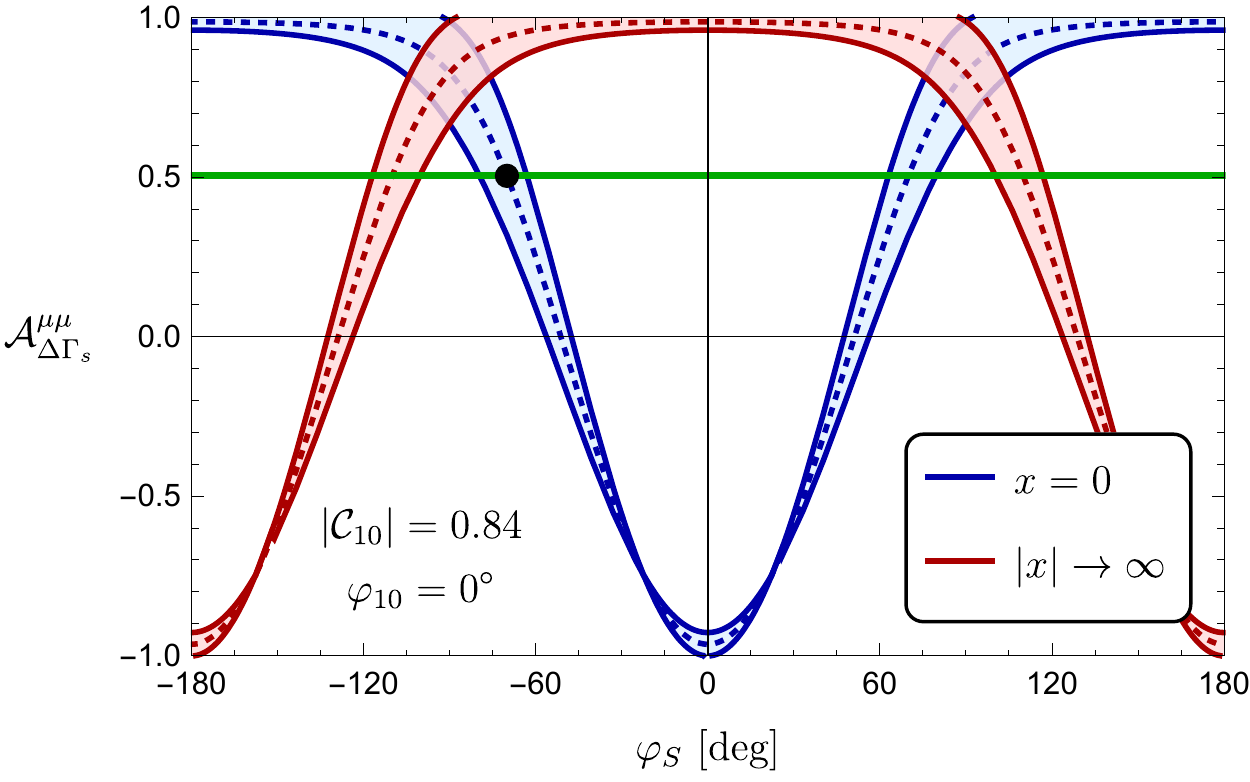}
\includegraphics[width=0.5\textwidth]{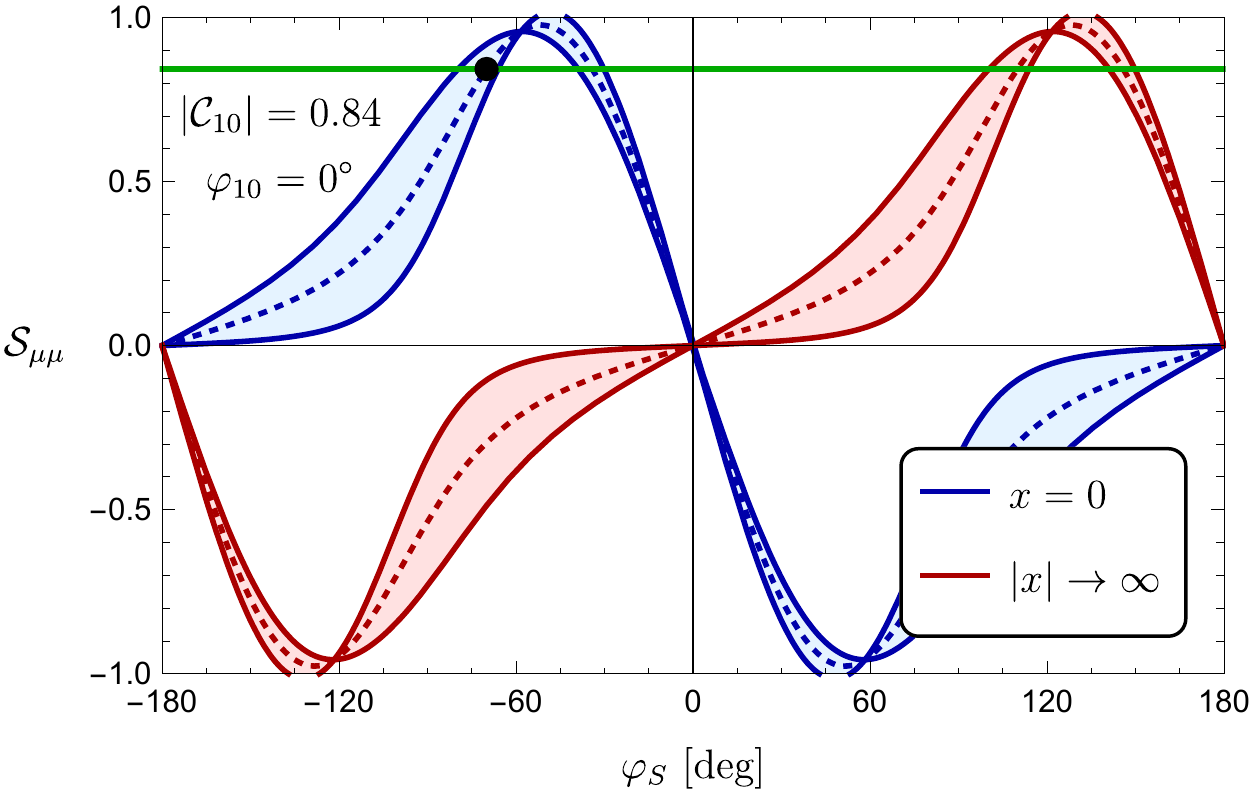}
\includegraphics[width=0.5\textwidth]{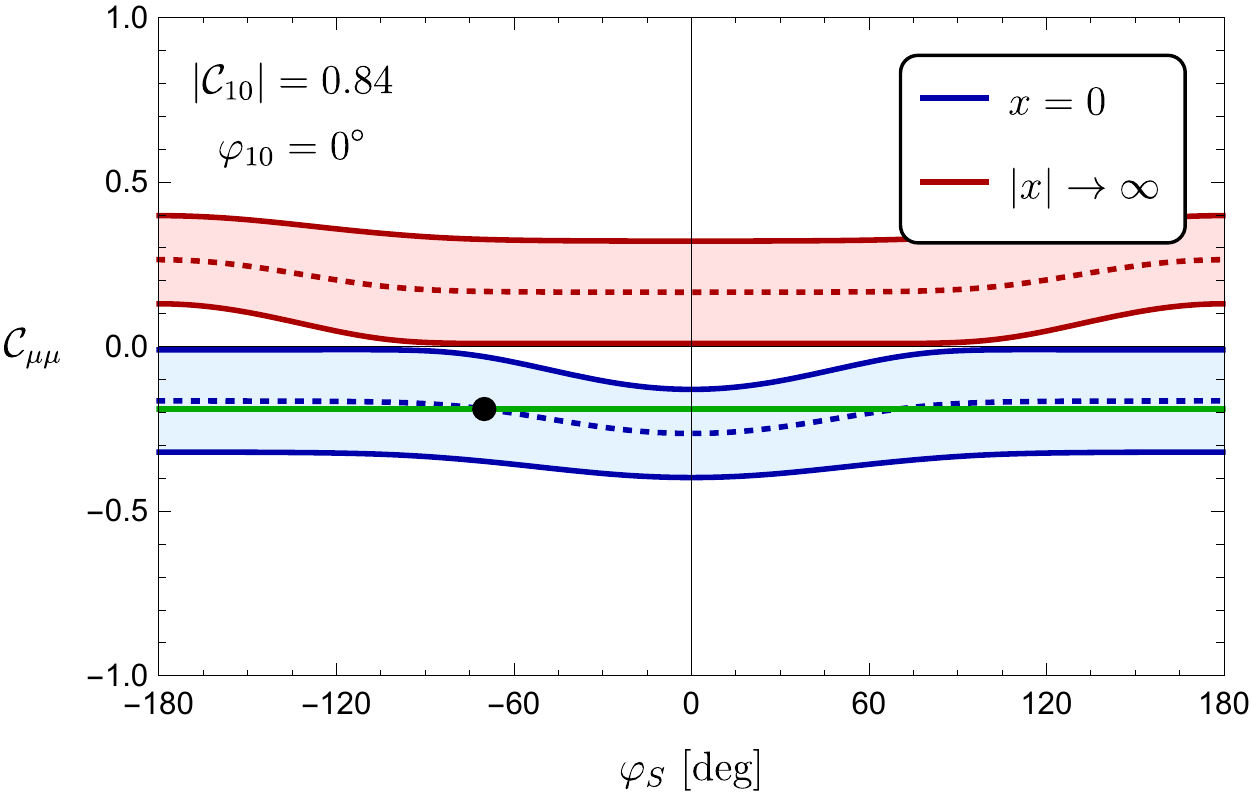}
\caption{
Functional dependences between $|S|$, ${\cal A}_{\Delta\Gamma_s}^{\mu\mu}$,
${\cal S}_{\mu\mu}$, ${\cal C}_{\mu\mu}$ and the CP-violating phase
$\varphi_S$ for $|\mathcal{C}_{10}|=0.84$ and $\varphi_{10}=0^\circ$. The blue and red contours
correspond to the scenarios $x=0$ and $|x|\rightarrow \infty$, respectively. The allowed regions are
determined within the  $1\,\sigma$ range for $\overline{R}$ given in Eq.\ (\ref{eq:Rbar}), where
the dashed curve is associated with the central value for this observable.
The black dot refers to the input parameters of the scenario in Eq.\ (\ref{eq:example0infinity2}), whereas the green line shows the value of the observables in Eq.~(\ref{eq:obsx02}).
\label{fig:absSvsPhisC84}}
\end{figure}
\end{center}

\vspace*{-2.0truecm}

The number of allowed solutions for a given angle $\varphi_S$ depends on the value of the Wilson coefficient 
${\cal C}_{10}$.  In order illustrate this feature, we consider two scenarios for $\mathcal{C}_{10}$. Let us 
first assume that there is a vanishing NP contribution $\mathcal{C}^{\rm NP}_{10}=0$, which yields
$|{\cal C}_{10}|=1, \  \varphi_{10}=0^\circ$. In this case,  Eq.~(\ref{eq:S0}) results in two solutions for $|S|$ 
as a function of $\varphi_S$, 
as can be seen in the top-left plot in Fig.\ \ref{fig:absSvsPhisC1}. Using Eqs.\ (\ref{ADGx0-calc}), (\ref{Smumux0-calc}) 
and (\ref{Cmumux0-calc}), we can determine the observables $\mathcal{A}^{\mu\mu}_{\Delta \Gamma_s}$, 
$\mathcal{S}^{\mu\mu}$ and $\mathcal{C}_{\mu\mu}$ as functions of $\varphi_S$, respectively, as shown in
Fig.~\ref{fig:absSvsPhisC1}. In particular, once $\mathcal{A}^{\mu\mu}_{\Delta \Gamma_s}$ has been measured,
the value of $\mathcal{S}_{\mu\mu}$ can be predicted. Should this CP asymmetry be measured correspondingly, 
this scenario would be confirmed, allowing us to determine the corresponding NP parameters. On the other hand,
should the measurement of $\mathcal{S}_{\mu\mu}$ be in conflict with the prediction, the NP scenario would be
ruled out by experimental data.

\begin{figure}[t] 
   \centering
   \includegraphics[width=5.5in]{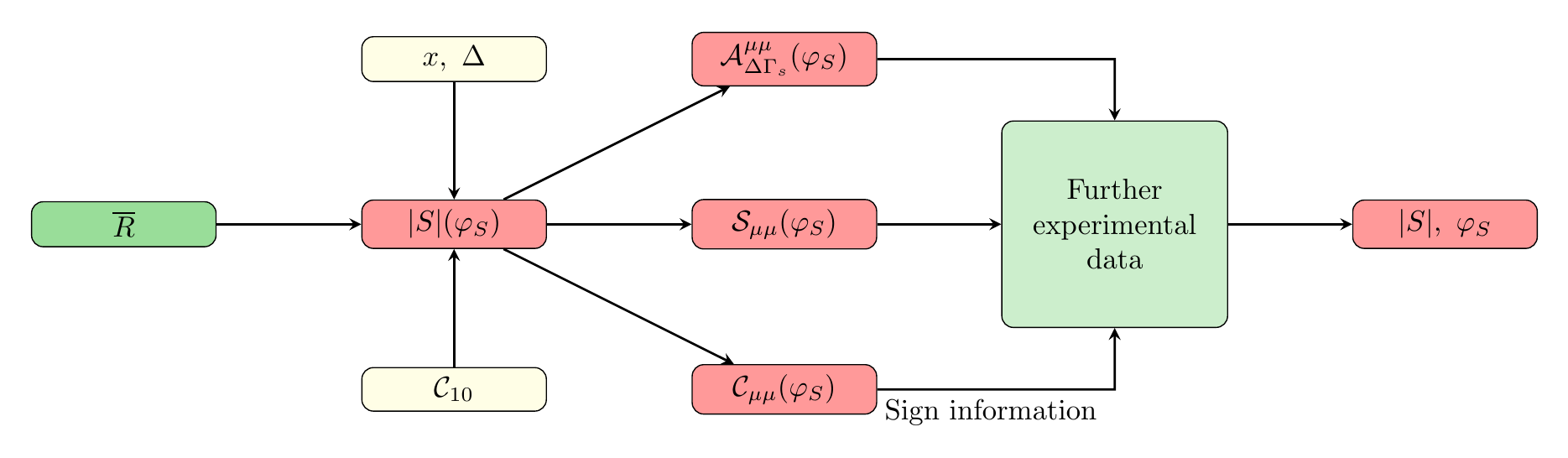} 
   \caption{Flowchart to illustrate the use of the relations in Subsection~\ref{ssec:gen-frame} with information on
   ${\cal C}_{10}$ to convert the measured value of $\overline{R}$ into predictions of the $B^0_s\to\mu^+\mu^-$ 
   observables. Once these are measured in accordance with the pattern characterizing the NP scenario, 
   $|S|$ and $\varphi_S$ can be extracted from the data.}\label{fig:Strategy_relations}
\end{figure}

Let us now consider a scenario with NP contributions to ${\cal C}_{10}$. If we follow the analysis of Ref.\ \cite{ANSS}
and use the central value of ${\cal C}_{10}$ in Eq.~(\ref{C10-range-Fit}), we 
obtain the functional dependence of $|S|$ and the corresponding observables on $\varphi_S$ shown 
Fig.~\ref{fig:absSvsPhisC84}. Interestingly, for a given value of $\varphi_S$, Eq.~(\ref{eq:S0}) gives now a single 
solution for $|S|$. Consequently, unlike their counterparts in Fig.~\ref{fig:absSvsPhisC1}, the contours no longer form 
closed loops, thereby indicating that the degeneracy with respect to $\varphi_S$ has disappeared. In Fig.~\ref{fig:Strategy_relations}, we illustrate this strategy, which is actually 
more general, i.e.\ does not only apply to the case of $x=0$.

\newpage

A closer look at the expressions in the Appendix shows that the case of $x=0$ is connected with 
$|x|\to \infty$, where the scalar and pseudo-scalar coefficients $C_S$ and $C_P$ vanish while $C_P'=C_S'$ 
takes a non-vanishing value. The expression in Eq.~(\ref{gen-rel-PS}) takes then the form
\begin{equation}\label{Rel-m}
 wP - S=w \, {\cal C}_{10},
\end{equation}
which reduces to $P-S=1$ for $w=1$ and ${\cal C}_{10}=1$. For the observables $r$ as well as 
${\cal A}^{\mu\mu}_{\Delta\Gamma_s}$ and ${\cal S}_{\mu\mu}$, we have the symmetry relation 
\begin{equation}
\varphi_S\to \pi+ \varphi_S,
\label{eq:symmetry_trans}
\end{equation}
which is equivalent to $|S|\to -|S|$, and yields
\begin{equation} \label{rxinf-calc}
r|_{|x|\to \infty}=|{\cal C}_{10}|^2+2\cos(\varphi_{10}-\varphi_S)|{\cal C}_{10}||S|+2|S|^2
\end{equation}
\begin{equation} \label{ADGxinf-calc}
{\cal A}^{\mu\mu}_{\Delta\Gamma_s}|_{|x|\to \infty}=
\frac{|{\cal C}_{10}|^2\cos2\varphi_{10}+2\cos(\varphi_{10}+\varphi_S)|{\cal C}_{10}||S|}{|{\cal C}_{10}|^2+
2\cos(\varphi_{10}-\varphi_S)|{\cal C}_{10}||S| + 2 |S|^2}
\end{equation}
\begin{equation}
\label{eq:SmumuInfty}
{\cal S}_{\mu\mu}|_{|x|\to \infty}=
\frac{|{\cal C}_{10}|^2\sin2\varphi_{10}+2\sin(\varphi_{10}+\varphi_S)|{\cal C}_{10}||S|}{|{\cal C}_{10}|^2+
2\cos(\varphi_{10}-\varphi_S)|{\cal C}_{10}||S| + 2 |S|^2}.
\end{equation}
In the case of ${\cal C}_{\mu\mu}$, the symmetry is broken by an overall minus sign:
\begin{equation} \label{Cmumuxinf-calc}
{\cal C}_{\mu\mu}|_{|x|\to \infty}=\frac{2|S|\left[|{\cal C}_{10}|\cos(\varphi_{10}-\varphi_S)+
|S|\right]}{|{\cal C}_{10}|^2+2\cos(\varphi_{10}-\varphi_S)|{\cal C}_{10}||S| + 2 |S|^2}.  
\end{equation}
More explicitly, we have
\begin{eqnarray} \label{eq:x0xinf-symm-r-ADG-smumu}
r|_{x=0}(\varphi_S + \pi)&=&r|_{|x|\rightarrow \infty}(\varphi_S)\nonumber \\
{\cal A}^{\mu\mu}_{\Delta\Gamma_s}|_{x=0}(\varphi_S + \pi) 
& =&{\cal A}^{\mu\mu}_{\Delta\Gamma_s}|_{|x|\rightarrow \infty}(\varphi_S) \\
{\cal S}_{\mu\mu}|_{x=0}(\varphi_S + \pi)&=&{\cal S}_{\mu\mu}|_{|x|\rightarrow \infty}(\varphi_S),\nonumber
\end{eqnarray}
while
\begin{eqnarray} \label{eq:x0xinf-symm-cmumu}
{\cal C}_{\mu\mu}|_{x=0}(\varphi_S+\pi)&=&-{\cal C}_{\mu\mu}|_{|x|\rightarrow \infty}(\varphi_S).
\end{eqnarray}
As we will see below, this feature has interesting phenomenological implications.

In order to illustrate the expressions given above, we consider two examples with different values of
the coefficient $\mathcal{C}_{10}$:\\

\noindent
{\bf Example (a):}\\

\noindent
We first assume a situation with vanishing NP contributions ${\cal C}_{10}^{\rm NP}=0$, and employ the
following setup:
\begin{equation}
\label{eq:example0infinity}
\overline{R}=0.84\pm 0.16, \quad  x=0, \quad \varphi_S=54^{\circ}, \quad |\mathcal{C}_{10}|=1, 
\quad \varphi_{10}=0^{\circ}.
\end{equation}
Using Eq.\ (\ref{eq:S0}), we determine $|S|$ as a function of $\varphi_S$. As discussed above, for
$|\mathcal{C}_{10}|=1, \  \varphi_{10}=0^\circ$ and the central value of $\overline{R}$ in 
Eq.\ (\ref{eq:example0infinity}), we obtain
a twofold solution. For the sake of illustration, we consider only the solution with the plus sign in front of the square 
root, yielding
\begin{equation}
|S|=0.43.
\end{equation}
With the help of Eq.~(\ref{conv-1}), we may now calculate 
\begin{equation}
|P|=0.82, \quad \varphi_P=-25^{\circ}.
\end{equation}
The corresponding values for the observables read as follows:
\begin{equation}\label{eq:obsx0}
{\cal A}^{\mu\mu}_{\Delta\Gamma_s}=0.58 , 
\quad {\cal S}_{\mu\mu} = -0.80,  \quad {\cal C}_{\mu\mu} = 0.16.
\end{equation}

Let us now assume that these observables have been measured, and discuss how we may then -- 
with the help of the strategy discussed above -- reveal the dynamics of the $B^0_s\to\mu^+\mu^-$ decay
and distinguish between the $x=0$ and $|x|\rightarrow \infty$ cases:
\begin{itemize}
\item It is plausible to expect that $\mathcal{A}^{\mu\mu}_{\Delta \Gamma_s}$ is the next observable to be measured. 
With the help of the top-right plot in Fig.~\ref{fig:absSvsPhisC1}, we identify four possible values for $\varphi_S$ 
which are compatible with the ``experimental" result of $\mathcal{A}^{\mu\mu}_{\Delta \Gamma_s}=0.58$ in 
Eq.\ (\ref{eq:obsx0}): $\varphi^{(1)}_S=-126^{\circ}$, $\varphi^{(2)}_S=-54^{\circ}$, $\varphi^{(3)}_S=54^{\circ}$ and 
$\varphi^{(4)}_S=126^{\circ}$.

\item We may now predict the observable $\mathcal{S}_{\mu\mu}$. Using the bottom-left plot in 
Fig.\ \ref{fig:absSvsPhisC1} or the expressions in  Eqs.\ (\ref{Smumux0-calc}) and (\ref{eq:SmumuInfty}), we obtain
$\mathcal{S}_{\mu\mu}=-0.80$ for $\varphi^{(1)}_S=-126^{\circ}$  (branch $|x|\rightarrow \infty$)  and 
$\varphi^{(3)}_S=54^{\circ}$ (branch $x=0$). Moreover, we find $\mathcal{S}_{\mu\mu}=0.80$ for 
$\varphi^{(2)}_S=-54^{\circ}$ (branch $x=0$) and $\varphi^{(4)}_S=126^{\circ}$ (branch $|x|\rightarrow \infty$).

\item The measurement $\mathcal{S}_{\mu\mu}=-0.80$ would then allow us to narrow down the four solutions 
for $\varphi_S$ to only two at $\varphi^{(1)}_S=-126^{\circ}$ and $\varphi^{(3)}_S=54^{\circ}$, corresponding 
to $|x|\rightarrow \infty$ and $x=0$, respectively. It should be emphasized that both solutions would be valid at this
stage of the analysis, i.e.\ we would have confirmed a CP-violating NP scenario with either $|x|\rightarrow \infty$ 
or $x=0$.

\item This ambiguity can be resolved through information on the sign of $\mathcal{C}_{\mu\mu}$,
which is given by $\mathcal{C}_{\mu\mu}=-0.16$ and $\mathcal{C}_{\mu\mu}=+0.16$ for $|x|\rightarrow \infty$ 
and $x=0$, respectively, as can be seen in Fig. \ref{fig:absSvsPhisC1}. Consequently, the fact that 
$\mathcal{C}_{\mu\mu}$ breaks the symmetry in Eq.~(\ref{eq:symmetry_trans}) gives us a powerful tool 
to distinguish between $x=0$ and $|x|\rightarrow \infty$.
\end{itemize}

\noindent
{\bf Example (b):}\\

\noindent
Now we have a look at a scenario with NP contributions to $\mathcal{C}_{10}$, which is characterized as follows:
\begin{equation}
\label{eq:example0infinity2}
\overline{R}=0.84\pm 0.16, \quad x=0, \quad \varphi_S=-70^{\circ}, \quad |\mathcal{C}_{10}|=0.84, 
\quad \varphi_{10}=0^{\circ}.
\end{equation}
Here the value of ${\cal C}_{10}$ follows from Eq.\ (\ref{C10-range-Fit}), and is discussed in more 
detail in Subsection~\ref{sec:C10}. In contrast to Example (a), we obtain now a single solution for $|S|$ from 
Eq.\ (\ref{eq:S0}), which is given by
\begin{equation}
|S|=0.46.
\end{equation}
Using Eq.\ (\ref{conv-1}), we find
\begin{equation}
|P|=0.81, \quad \varphi_P=33^{\circ},
\end{equation}
resulting in the following values of the observables:
\begin{equation}\label{eq:obsx02}
{\cal A}^{\mu\mu}_{\Delta\Gamma_s}=0.50 , 
\quad {\cal S}_{\mu\mu} = 0.84,  \quad {\cal C}_{\mu\mu} = -0.19.
\end{equation}
In analogy to Example (a), using the plots in Fig.~\ref{fig:absSvsPhisC84}, we may again show the
compatibility of the ``measured" observables with the scenario $x=0$, and rule out the case of 
$|x|\rightarrow \infty$ through the sign of the ${\cal C}_{\mu\mu}$ asymmetry. For the convenience of the reader, we summarize the main features of these examples in Table~\ref{tab:Tab1}.

\begin{center}
\begin{figure}
\begin{center}
\includegraphics[width=0.5\textwidth]{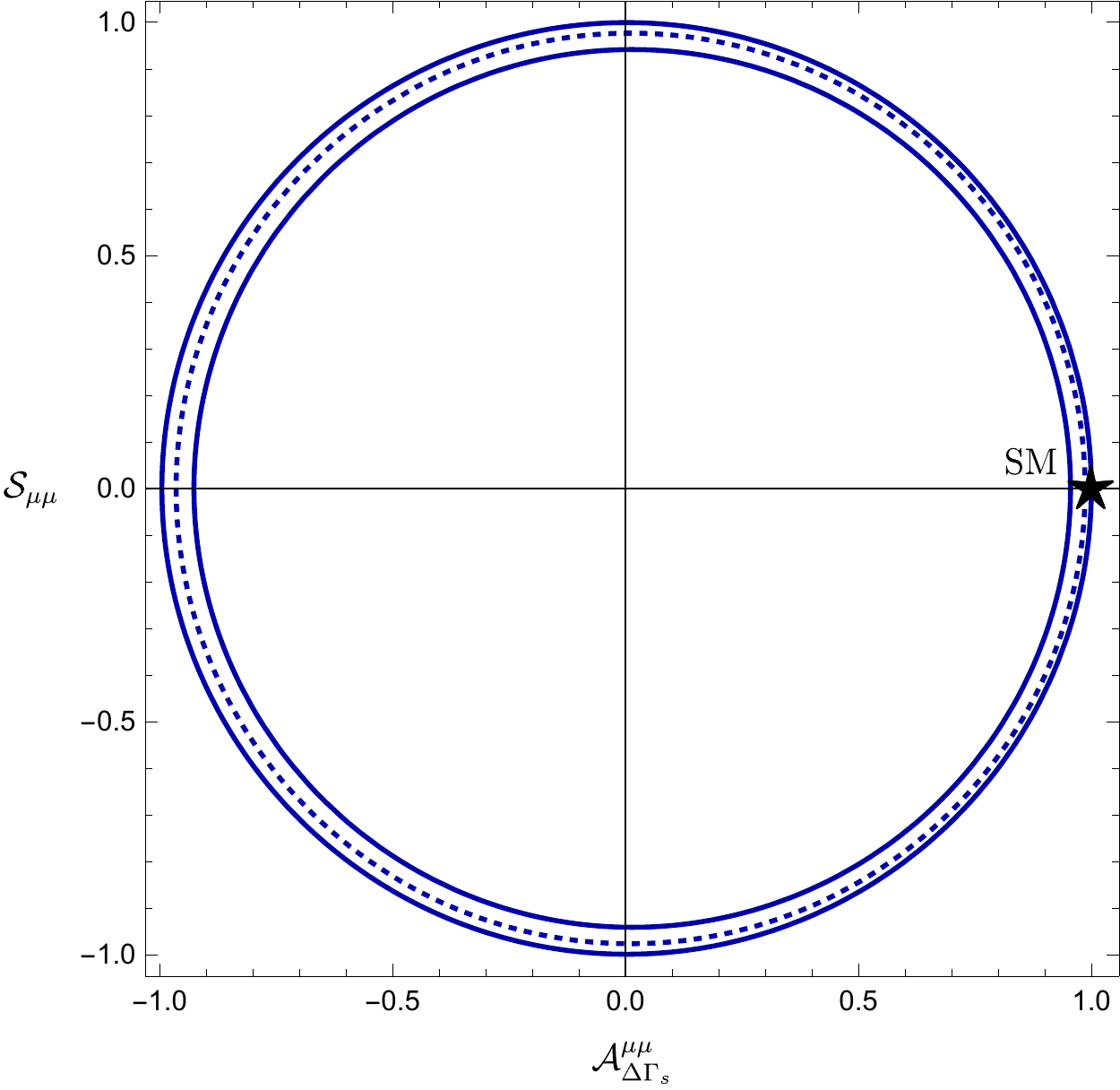}
\caption{Allowed region in the ${\cal A}_{\Delta\Gamma_s}^{\mu\mu}$--${\cal S}_{\mu\mu}$ plane following from 
the current experimental value of $\overline{R}$ for $x=0$; the same correlation is obtained for $|x|\rightarrow \infty$. 
The circular region corresponds to the $1\sigma$ uncertainty of $\overline{R}$ in 
Eq.~(\ref{eq:Rbar}). The black star indicates the SM point.
\label{fig:aDGsmumu_corr-x0-xinf}}
\end{center}
\end{figure}
\end{center}

In Fig.~\ref{fig:aDGsmumu_corr-x0-xinf}, we show the correlation between ${\cal A}^{\mu\mu}_{\Delta\Gamma_s}$ 
and ${\cal S}_{\mu\mu}$ through the CP-violating phase $\varphi_S$. It should be noted that the corresponding 
regions for  $|\mathcal{C}_{10}|=0.84, \  \varphi_{10} = 0^\circ$ and $|\mathcal{C}_{10}|=1, \  \varphi_{10}=0^\circ$ 
do not differ substantially and are included in a 
single plot. Due to the symmetry transformation in Eq.\ (\ref{eq:symmetry_trans}), the scenarios $x=0$ and 
$|x|\rightarrow\infty$ cover the same region once we make a scan over the full range of $\varphi_S$. The allowed
region in Fig.~\ref{fig:aDGsmumu_corr-x0-xinf} exhibits the following interesting features:
\begin{enumerate}

\item  The currently available measurement of $\overline{R}$ implies a remarkably constrained circular region in the 
${\cal A}_{\Delta\Gamma_s}^{\mu\mu}$--${\cal S}_{\mu\mu}$ plane for CP-violating NP scenarios characterized 
by $x=0$ and $|x|\rightarrow \infty$. 

\item A future measurement of the observable combination ${\cal A}^{\mu\mu}_{\Delta\Gamma_s}$ and 
${\cal S}_{\mu\mu}$ lying outside the allowed region would rule out the $x=0$ and $|x|\rightarrow \infty$
scenarios.  

\item The allowed region in the ${\cal A}_{\Delta\Gamma_s}^{\mu\mu}$--${\cal S}_{\mu\mu}$ plane is close
to the unit circle.  Consequently, due to Eq.\ (\ref{CP-rel}), the observable $\mathcal{C}_{\mu\mu}$ 
is constrained to take a smallish value.

\item The allowed region is similar to the one arising for the scenario described in Section~\ref{sec:C10}. 
While here $\varphi_{10}=0^{\circ}$ would imply the SM results ${\cal A}^{\mu\mu}_{\Delta\Gamma_s}=1$ 
and ${\cal S}_{\mu\mu}=0$, in the case of $x=0$ or $|x|\rightarrow \infty$ we may still deviate substantially 
from the SM even in spite of having a vanishing phase $\varphi_{10}$.

\end{enumerate}

In a complementary way, if we can obtain the value of the phase $\varphi_S$ from external information
or theoretical considerations, we will be able to predict the observables ${\cal A}^{\mu\mu}_{\Delta\Gamma_s}$ 
and ${\cal S}_{\mu\mu}$ compatible with vanishing short distance contributions  $C_{P,S}$ or  $C'_{P,S}$.  
Strong deviations from these determinations will indicate that the corresponding scenarios are not realized in Nature. A discussion of NP scenarios characterized by the relations $P\pm S=1$ (see Eqs.\ (\ref{Rel-p}) and (\ref{Rel-m})) 
can be found in Ref.~\cite{BFGK}.

\begin{table}
\begin{center}
\begin{tabular}{ |c|c|c|}
\hline 
\multicolumn{3}{|c|}{  \textbf{Example (a)}} \\
\multicolumn{3}{|c|}{ $|\mathcal{C}_{10}|=1,$ 
\quad $\varphi_{10}=0^{\circ}$}\\
\multicolumn{3}{|c|}{ $\overline{R}=0.84\pm 0.16,$ \quad  $\mathcal{A}^{\mu\mu}_{\Delta \Gamma_s}=0.58,$ \quad 
$\mathcal{S}_{\mu\mu}=-0.80,$\quad $\mathcal{C}_{\mu\mu}=0.16$}\\
\hline
\hline
 Observables & Solutions & Scenario\\
 \hline 
\multirow{ 2}{*}{ $\mathcal{A}^{\mu\mu}_{\Delta \Gamma_s}$} & $\varphi^{(1)}_S=-126^{\circ}$,\quad $\varphi^{(4)}_S=126^{\circ}$& 
$|x|\rightarrow \infty$
\quad $(C_S=C_P=0)$ \\ 
\cline{2-3}
 &$\varphi^{(2)}_S=-54^{\circ}$, \quad $\varphi^{(3)}_S=54^{\circ}$   & $x=0$ \quad $(C'_S=C'_P=0)$\\
\hline
\multirow{ 2}{*}{$\mathcal{A}^{\mu\mu}_{\Delta \Gamma_s}$, \quad 
$\mathcal{S}_{\mu\mu}$} &  $\varphi^{(1)}_S=-126^{\circ}$
&$|x|\rightarrow \infty$ \quad $(C_S=C_P=0)$\\
\cline{2-3}
 & $\varphi^{(3)}_S=54^{\circ}$    & $x=0$ \quad $(C'_S=C'_P=0)$\\
\hline
$\mathcal{A}^{\mu\mu}_{\Delta \Gamma_s}$, \quad $\mathcal{S}_{\mu\mu}$, \quad $\mathcal{C}_{\mu\mu}$ 
&$\varphi^{(3)}_S=54^{\circ}$
    & $x=0$ \quad $(C'_S=C'_P=0)$\\
\hline
\hline
\hline
\multicolumn{3}{|c|}{  \textbf{Example (b)}} \\
\multicolumn{3}{|c|}{ $|\mathcal{C}_{10}|=0.84,$
\quad $\varphi_{10}=0^{\circ}$}\\
\multicolumn{3}{|c|}{ $\overline{R}=0.84\pm 0.16,$ \quad  $\mathcal{A}^{\mu\mu}_{\Delta \Gamma_s}=0.50,$ \quad
$\mathcal{S}_{\mu\mu}=0.84,$\quad $\mathcal{C}_{\mu\mu}=-0.19$}\\
\hline
\hline
\multirow{ 2}{*}{ $\mathcal{A}^{\mu\mu}_{\Delta \Gamma_s}$} & $\varphi^{(1)}_S=-110^{\circ}$,\quad $\varphi^{(4)}_S=110^{\circ}$&
$|x|\rightarrow \infty$
\quad $(C_S=C_P=0)$ \\
\cline{2-3}
 &$\varphi^{(2)}_S=-70^{\circ}$, \quad $\varphi^{(3)}_S=70^{\circ}$   & $x=0$ \quad $(C'_S=C'_P=0)$\\
\hline
\multirow{ 2}{*}{$\mathcal{A}^{\mu\mu}_{\Delta \Gamma_s}$, \quad
$\mathcal{S}_{\mu\mu}$} &  $\varphi^{(4)}_S=110^{\circ}$
&$|x|\rightarrow \infty$ \quad $(C_S=C_P=0)$\\
\cline{2-3}
 & $\varphi^{(2)}_S=-70^{\circ}$    & $x=0$ \quad $(C'_S=C'_P=0)$\\
\hline
$\mathcal{A}^{\mu\mu}_{\Delta \Gamma_s}$, \quad $\mathcal{S}_{\mu\mu}$, \quad $\mathcal{C}_{\mu\mu}$
&$\varphi^{(2)}_S=-70^{\circ}$
    & $x=0$ \quad $(C'_S=C'_P=0)$\\
 \hline
\end{tabular}
\end{center}
\caption{
Summary of the strategy followed in Examples (a) and (b) to disentangle the scenario $x=0$ from $|x|\rightarrow \infty$ and determine the value of $\varphi_S$.}
\label{tab:Tab1}
\end{table}

\boldmath
\subsubsection{$\Delta=0^\circ$}
\label{sec:Delta0}
\unboldmath
Another interesting case arises if $C_S'$ and $C_S$ have the same CP-violating phases, i.e.\ 
$\Delta=0^\circ$, which yields
\begin{equation}\label{r-Del0}
r|_{\Delta=0^\circ}=|{\cal C}_{10}|^2-2\left(\frac{1+|x|}{1-|x|}\right)\cos(\varphi_{10}-\varphi_S)|{\cal C}_{10}||S|
+2\left[\frac{1+|x|^2}{(1-|x|)^2}\right]|S|^2
\end{equation}
\begin{equation}
{\cal A}^{\mu\mu}_{\Delta\Gamma_s}|_{\Delta=0^\circ}=
\frac{(1-|x|)^2|{\cal C}_{10}|^2\cos2\varphi_{10}-2\left(1-|x|^2\right)
\cos(\varphi_{10}+\varphi_S)|{\cal C}_{10}||S| +4 |x||S|^2\cos2\varphi_S}{(1-|x|)^2|{\cal C}_{10}|^2-2\left(1-|x|^2\right)
\cos(\varphi_{10}-\varphi_S)|{\cal C}_{10}||S| + 2 (1+|x|^2)|S|^2}
\label{eq:ADGx1}
\end{equation}
\begin{equation}
{\cal S}_{\mu\mu}|_{\Delta=0^\circ}=
\frac{(1-|x|)^2|{\cal C}_{10}|^2\sin2\varphi_{10}-2\left(1-|x|^2\right)
\sin(\varphi_{10}+\varphi_S)|{\cal C}_{10}||S| +4 |x||S|^2\sin2\varphi_S}{(1-|x|)^2|{\cal C}_{10}|^2-2\left(1-|x|^2\right)
\cos(\varphi_{10}-\varphi_S)|{\cal C}_{10}||S| + 2 (1+|x|^2)|S|^2}
\label{eq:Smumux1}
\end{equation}
\begin{equation}
{\cal C}_{\mu\mu}|_{\Delta=0^\circ}=\frac{2|S|\left[(1-|x|)^2|{\cal C}_{10}|\cos(\varphi_{10}-\varphi_S)-
(1-|x|^2)|S|\right]}{(1-|x|)^2|{\cal C}_{10}|^2-2\left(1-|x|^2\right)
\cos(\varphi_{10}-\varphi_S)|{\cal C}_{10}||S| + 2 (1+|x|^2)|S|^2}.
\label{eq:Cmumux1}
\end{equation}
In analogy to the scenarios $x=0$ and $|x|\to \infty$ discussed in Subsection~\ref{sec:x0xinfty}, the expressions in 
Eqs.\ (\ref{r-Del0})--(\ref{eq:Smumux1}) are invariant under the symmetry 
transformation
\begin{equation}\label{sym-Del-0}
|x| \to 1/|x|, \quad \varphi_S \to \varphi_S +\pi,
\end{equation}
leading to
\begin{eqnarray} \label{eq:Del-0-symm-r-ADG-smumu}
r|_{\Delta=0^\circ}(|x|,\varphi_S)&=&r|_{\Delta=0^\circ}(1/|x|,\varphi_S+\pi) \nonumber \\
{\cal A}^{\mu\mu}_{\Delta\Gamma_s}|_{\Delta=0^\circ}(|x|,\varphi_S)&=&
{\cal A}^{\mu\mu}_{\Delta\Gamma_s}|_{\Delta=0^\circ}(1/|x|,\varphi_S+\pi)\nonumber\\
{\cal S}_{\mu\mu}|_{\Delta=0^\circ}(|x|,\varphi_S)&=&{\cal S}_{\mu\mu}|_{\Delta=0^\circ}(1/|x|,\varphi_S+\pi),
\end{eqnarray}
while the symmetry is again broken by the observable ${\cal C_{\mu\mu}}$ through an overall sign change:
\begin{eqnarray}
{\cal C}_{\mu\mu}|_{\Delta=0^\circ}(|x|,\varphi_S)&=&-{\cal C}_{\mu\mu}|_{\Delta=0^\circ}(1/|x|,\varphi_S+\pi).
\end{eqnarray}

The three observables $r$, $\mathcal{A}^{\mu\mu}_{\Delta \Gamma_s}$ and $\mathcal{S}_{\mu\mu}$  in
Eqs.\ (\ref{r-Del0})--(\ref{eq:Smumux1}) depend on the three unknowns $x$,  $|S|$ and $\varphi_S$. Consequently,
 if the observables are measured, we may determine these parameters. The twofold ambiguity following from 
 the symmetry transformation in Eq.~(\ref{sym-Del-0}) can be resolved through the measurement of the sign of
 ${\cal C_{\mu\mu}}$. Unfortunately, in view of the highly non-linear structure of the equations, we cannot give
simple analytic solutions. However, the parameters can be determined numerically. In Section~\ref{sec:Exp}, 
 we will illustrate this determination through fits to scenarios of future measurements.

\begin{center}
\begin{figure}
\includegraphics[width=0.5\textwidth]{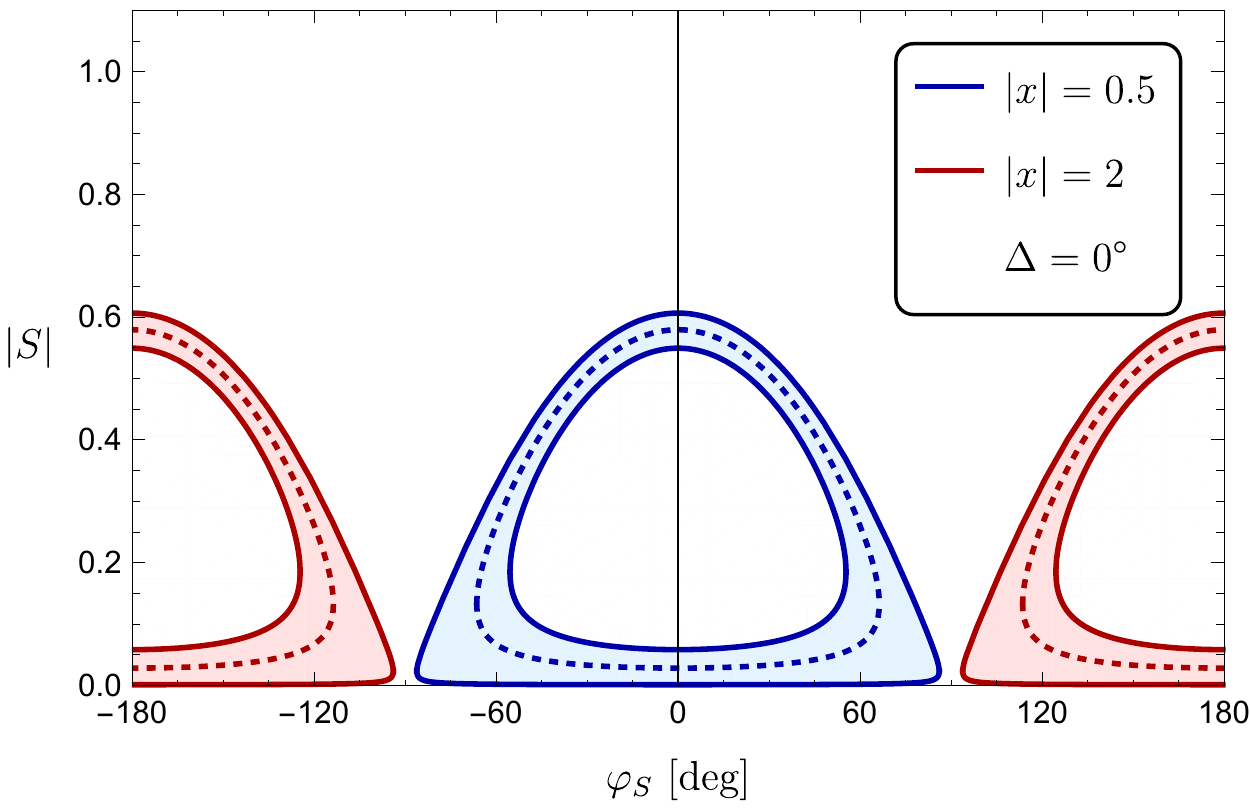}
\includegraphics[width=0.5\textwidth]{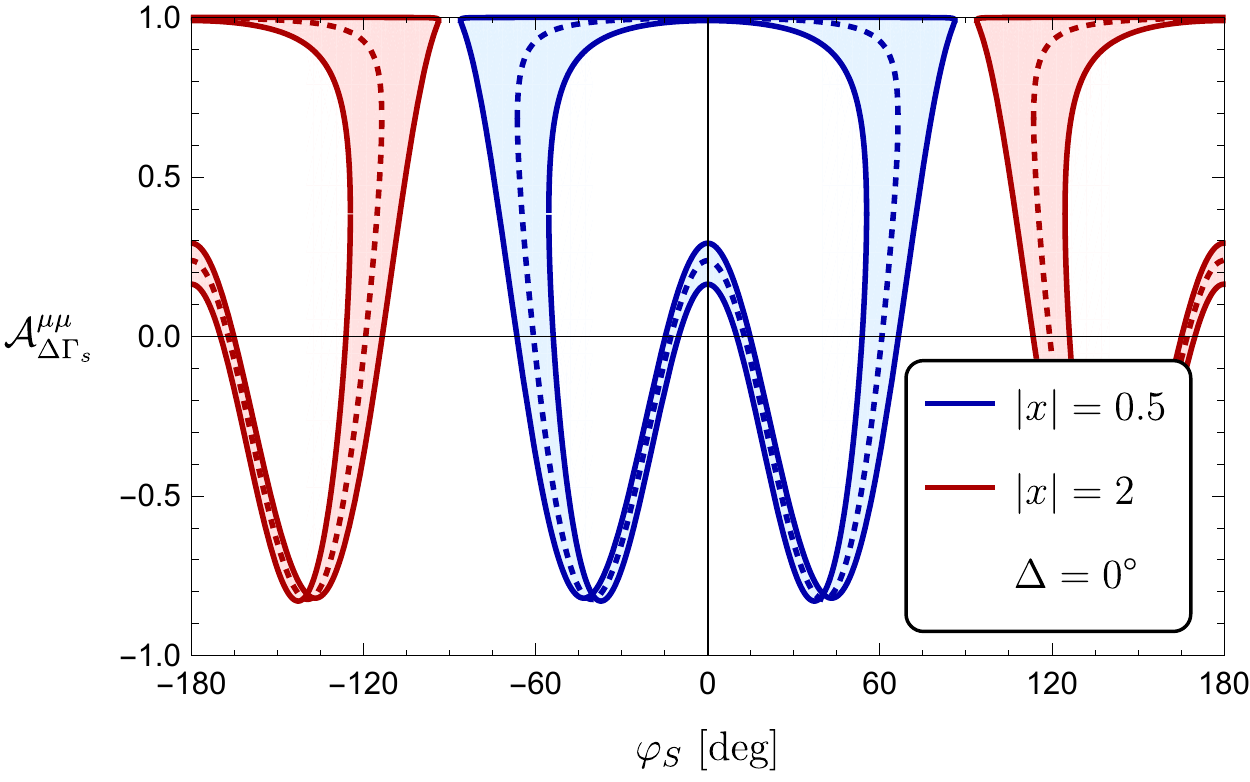}
\includegraphics[width=0.5\textwidth]{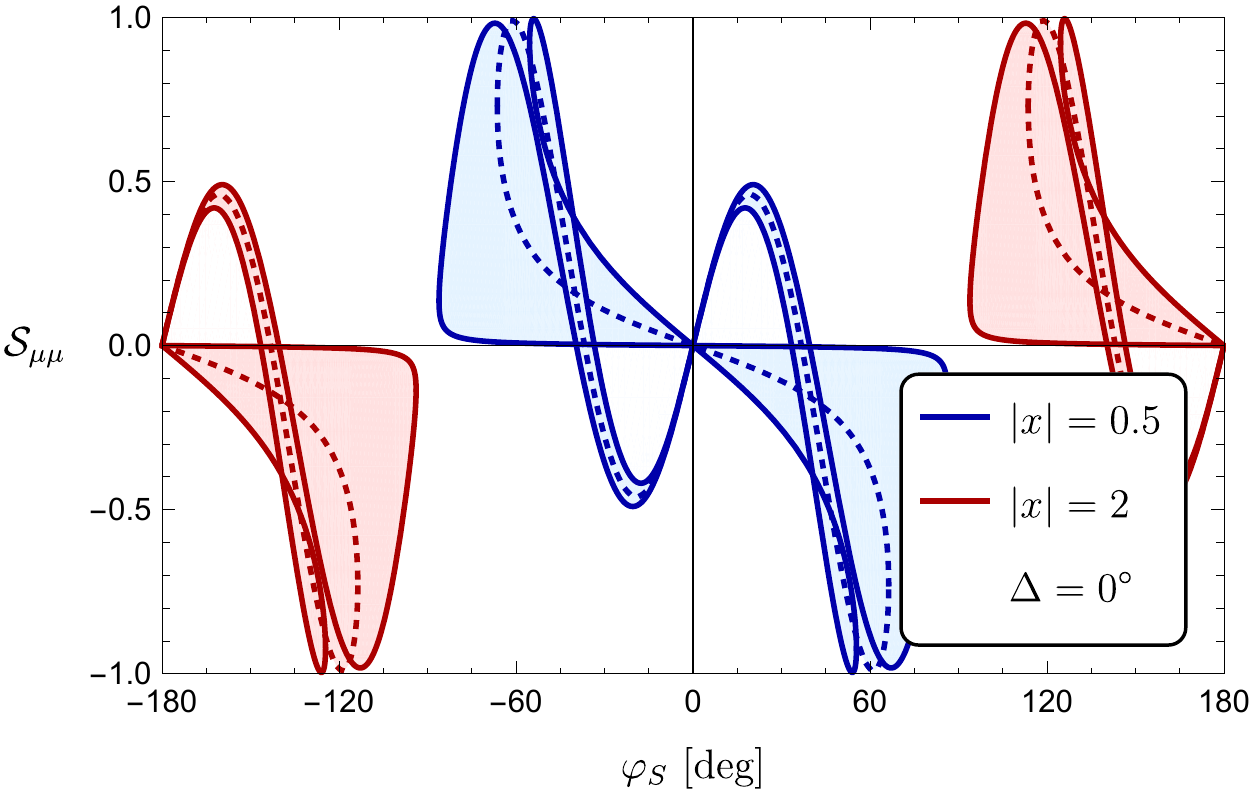}
\includegraphics[width=0.5\textwidth]{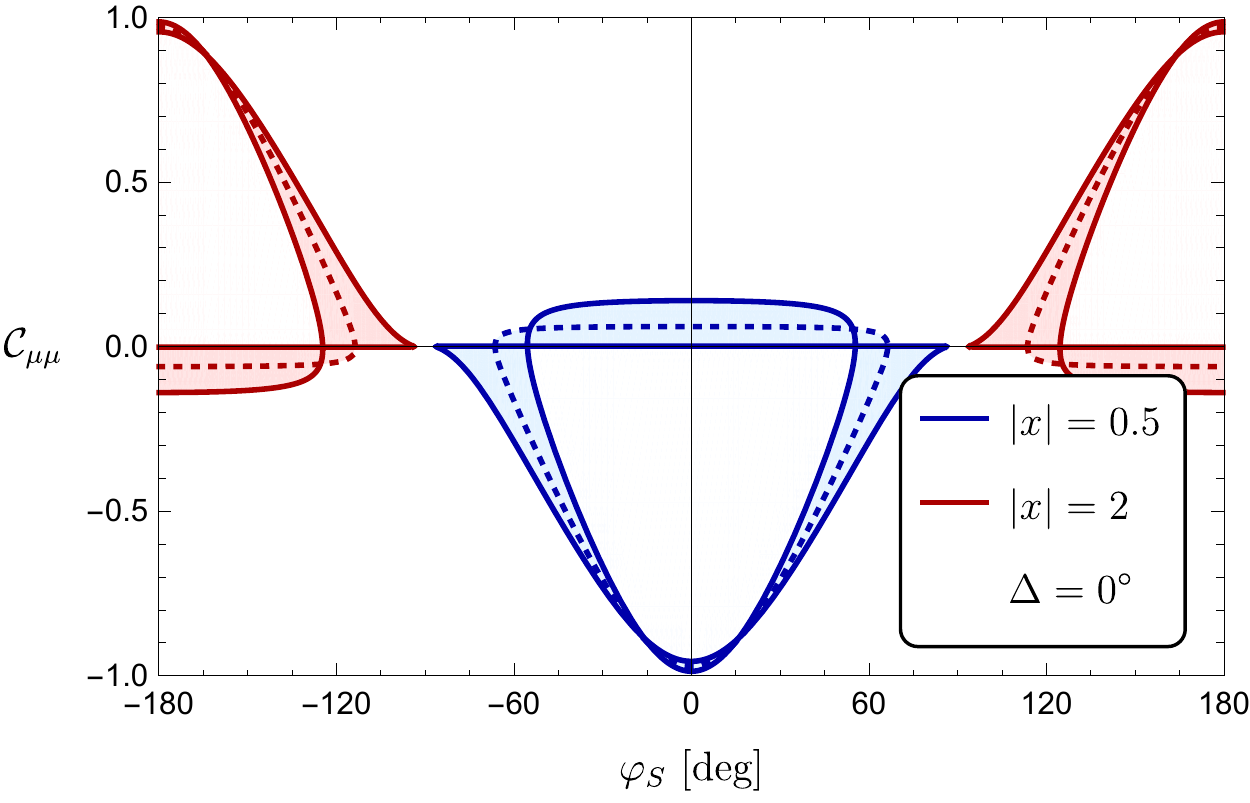}
\caption{Functional dependences between $|S|$, ${\cal A}_{\Delta\Gamma_s}^{\mu\mu}$,
${\cal S}_{\mu\mu}$, ${\cal C}_{\mu\mu}$ and the CP-violating phase $\varphi_S$ for 
$|\mathcal{C}_{10}|=1, \  \varphi_{10}=0^\circ$ and $\Delta=0^\circ$. 
The blue and red contours correspond to the scenarios $|x|=0.5$ and the associated value $|x| = 2$, respectively.
The allowed regions are determined within the  $1\,\sigma$ range for $\overline{R}$ given in Eq.~(\ref{eq:Rbar}), 
where the dashed curve corresponds to the central value for this observable. Notice that for each value of 
$\varphi_S$ we have in general two possible solutions for the observables, leading to closed loops in the parameter 
space. \label{fig:asymmvsPhis_Delta0_x05}}
\end{figure}
\end{center}

Alternatively, we can apply the strategy depicted in the flowchart in Fig.~\ref{fig:Strategy_relations}. We start with the 
experimental value of $\overline{R}$ given in Eq.~(\ref{eq:Rbar}). Furthermore, we assume that 
$|x|=0.5$ and $|{\cal C}_{10}| = 1, \  \varphi_{10}=0^\circ$. This allows us to solve for $|S|$ as a function 
of $\varphi_S$, and to subsequently 
determine ${\cal A}_{\Delta\Gamma_s}^{\mu\mu}$, ${\cal S}_{\mu\mu}$ and ${\cal C}_{\mu\mu}$ as functions 
of $\varphi_S$. The results are shown as the blue contours in Fig.~\ref{fig:asymmvsPhis_Delta0_x05}. Here, also the 
symmetric situation with $|x|=2$ is shown in red, illustrating nicely how ${\cal C}_{\mu\mu}$ breaks the symmetry.

\begin{figure}
   \centering
   \includegraphics[width=2.6in]{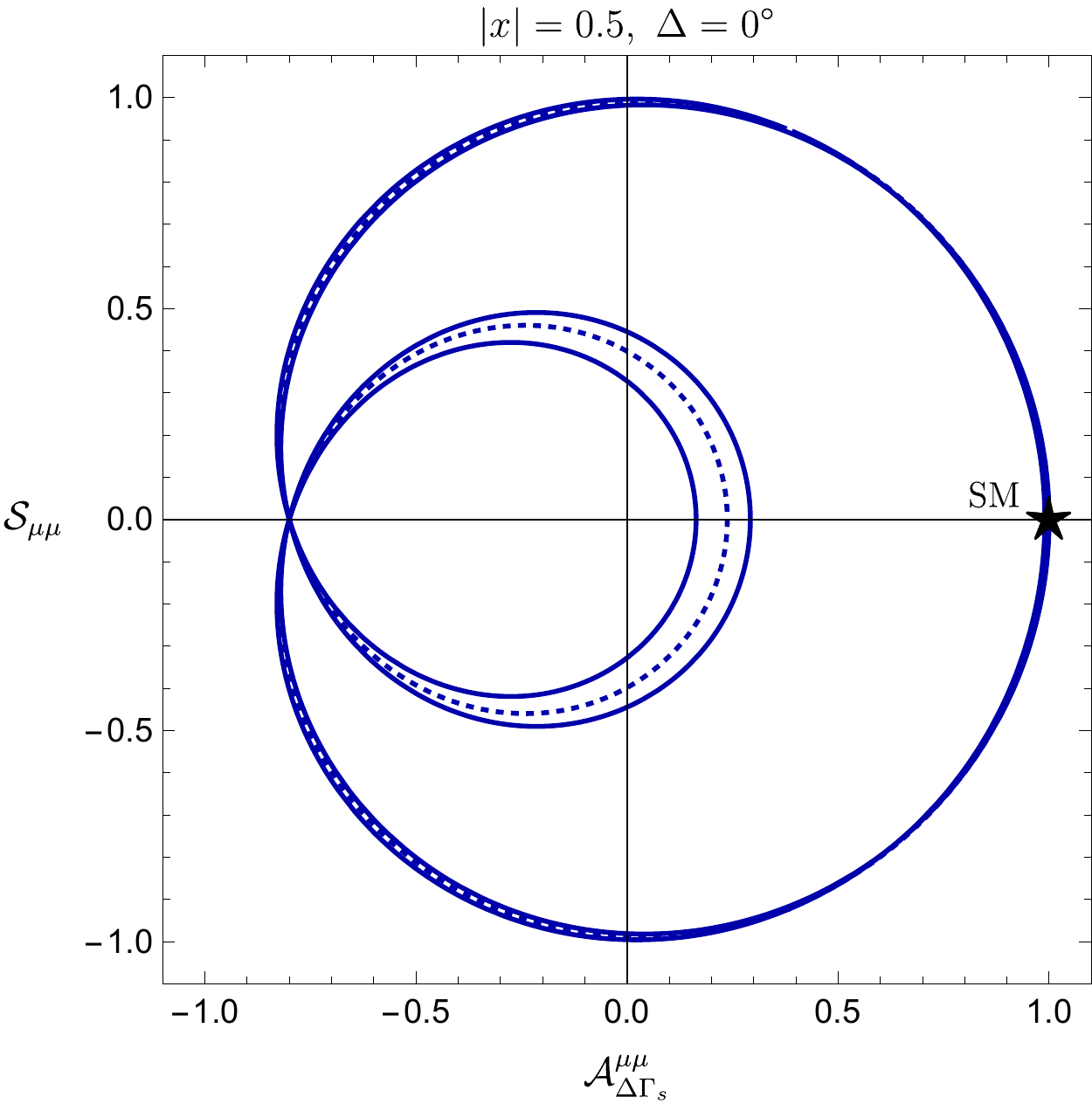} 
   \includegraphics[width=2.6in]{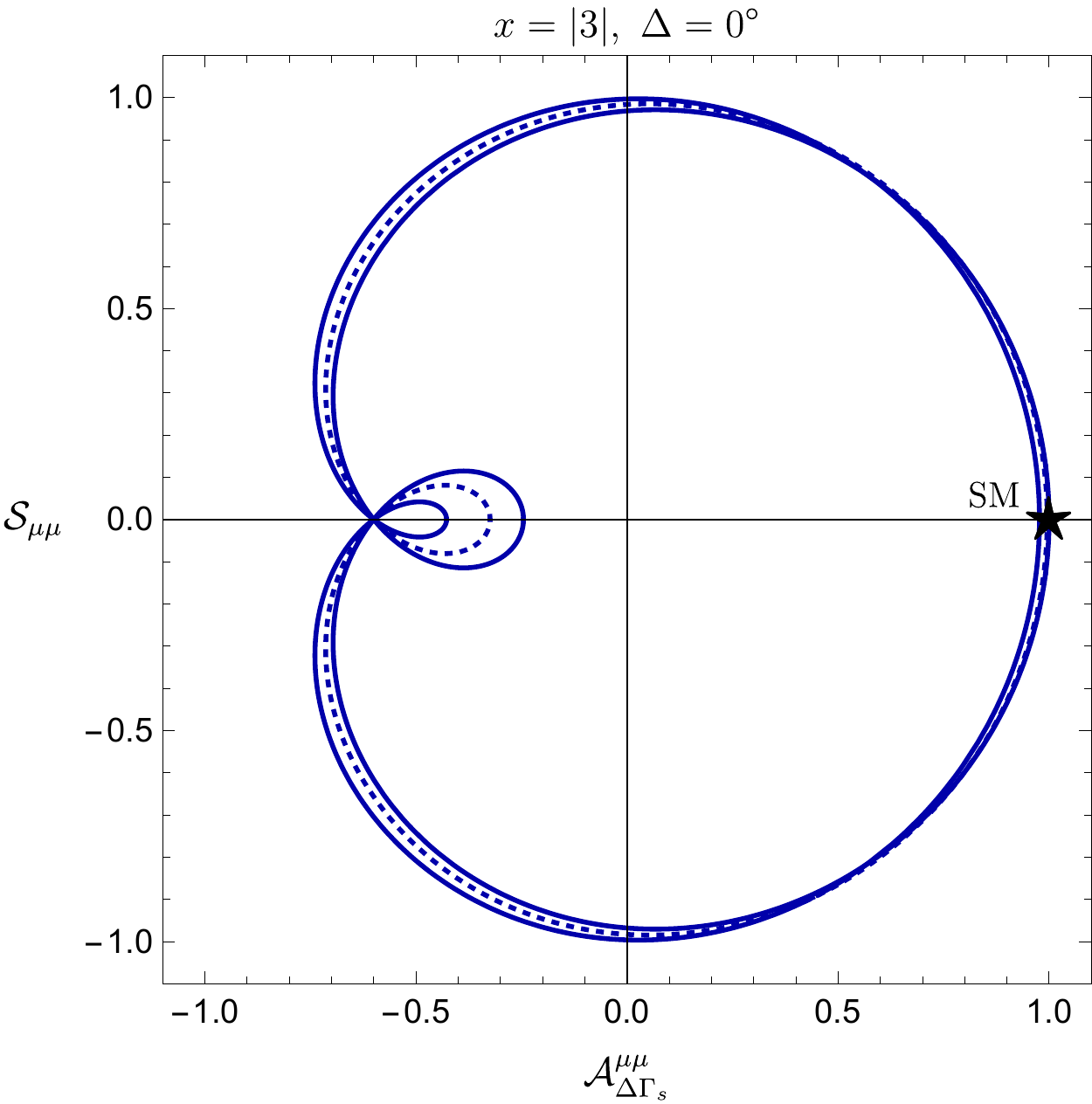} 
   \caption{Correlations between ${\cal A}^{\mu\mu}_{\Delta\Gamma_s}$ and ${\cal S}_{\rm \mu\mu}$ 
   in the case of $\Delta=0^\circ$ for $|x|=0.5$ and $|x|=3$ in the left and right panels, respectively. The 
   region corresponds to the $1\sigma$ uncertainty of $\overline{R}$ in Eq.~(\ref{eq:Rbar}). The black star 
   indicates the SM point.} 
   \label{fig:x-contours}
\end{figure}

Finally, in Fig.~\ref{fig:x-contours}, we show the correlation between ${\cal A}^{\mu\mu}_{\Delta\Gamma_s}$ 
and ${\cal S}_{\rm \mu\mu}$ for $|x|=0.5$ and $|x| = 3$. Contrary to the situation for $x=0, \  |x|\to\infty$, we 
are not constrained to a contour close to the unit circle, but can also obtain values in the interior region. 
For the scenario $|x|=3$, the relations $|S|$, ${\cal A}_{\Delta\Gamma_s}^{\mu\mu}$, ${\cal S}_{\mu\mu}$ and 
${\cal C}_{\mu\mu}$ as functions of $\varphi_S$ are similar to the ones 
shown in Fig.~\ref{fig:asymmvsPhis_Delta0_x05}.

In the expression for $r$ given in Eq.~(\ref{r-Del0}), a pole seems to arise for $|x|=1$, which corresponds to
\begin{equation}\label{CSCS-p}
C_S=C'_S.
\end{equation}
However, this is a spurious divergence, which is cancelled by the $C_S-C_S'$ term 
in the expression for $S$ in Eq.~(\ref{S-def}), implying
\begin{equation}
S|_{|x|=1, \  \Delta=0^\circ}=0.
\label{eq:Sx1Delta0}
\end{equation}
Using the relations in Eqs.~(\ref{rel-PS}) and (\ref{rel-PS-prime}), we obtain 
\begin{equation}
C_P'=C_S'=C_S=-C_P.
\end{equation}
Consequently, Eq.~(\ref{P-def}) yields
\begin{eqnarray}
\hspace*{-0.5truecm}|P|e^{i\varphi_P}|_{|x|=1, \  \Delta=0^\circ}=
\mathcal{C}_{10} - 
\frac{M^2_{B_s}}{m_{\mu}}\Bigl(\frac{m_b}{m_b + m_s}\Bigl)\frac{C_S}{C^{\rm SM}_{10}} \nonumber \\
= |\mathcal{C}_{10}|e^{i\varphi_{10}} - \frac{M^2_{B_s}}{m_{\mu}}\Bigl(\frac{m_b}{m_b + m_s}\Bigl)
\frac{|C_S|}{C^{\rm SM}_{10}}e^{i\tilde{\varphi}_S}
\end{eqnarray}
and shows that also the divergence in Eq.~(\ref{conv-1}) for $|x|=1, \  \Delta=0^\circ$ is spurious. 
If we neglect, for simplicity, again
the tiny NP contribution $\phi_s^{\rm NP}$ to the $B^0_s$--$\bar B^0_s$ mixing phase, we obtain
\begin{eqnarray}
r|_{|x|=1, \  \Delta=0^\circ}&=&|P|^2\\
\mathcal{A}^{\mu\mu}_{\Delta \Gamma_s}|_{|x|=1, \  \Delta=0^\circ}&=&\cos(2\varphi_P)\\
\mathcal{S}_{\mu\mu}|_{|x|=1, \  \Delta=0^\circ}&=&\sin(2\varphi_P)\\
\mathcal{C}_{\mu\mu}|_{|x|=1, \  \Delta=0^\circ}&=&0.
\label{eq:obs_x_1_delta0}
\end{eqnarray}
For a discussion of NP models describing this situation, see Ref.~\cite{BFGK}. Obviously, also extensions of the SM 
with scalars, which couple in a left-right-symmetric way to quarks (see the operators in Eq.~(\ref{ops}) and the
relations in Eq.~(\ref{CSCS-p})), fall into this category.

\begin{table}[t]
\begin{center}
\begin{tabular}{ |c|ll|ll|c|}
\hline
Subsection&\multicolumn{2}{|c|}{SMEFT} & \multicolumn{2}{|c|}{Standard}  & \multicolumn{1}{|c|}{Extra}\\
&\multicolumn{2}{|c|}{Parameterization}&\multicolumn{2}{|c|}{Parameterization} &\multicolumn{1}{|c|}{Assumptions}\\
\hline
\multirow{2}{*}{\ref{sec:x0xinfty}}&\multicolumn{2}{|c|}{$x=0$}                 & \multicolumn{2}{|c|}{$C'_S=C'_P=0$} & 
\multirow{2}{*}{$|\mathcal{C}_{10}|=1$,\quad $|\mathcal{C}_{10}|=0.84$} \\
\cline{2-5}
&\multicolumn{2}{|c|}{$|x|\rightarrow \infty$} & \multicolumn{2}{|c|}{$C_S=-C_P=0$}  &    \\
\hline
\multirow{2}{*}{\ref{sec:Delta0}}&$\Delta=0^{\circ}$,& $|x|=0.5$ &$\tilde{\varphi}_S=\tilde{\varphi}'_S,$ & $|C_S|=2|C'_S|$  & \multirow{2}{*}{$|\mathcal{C}_{10}|=1$}  \\
\cline{2-5}
&$\Delta=0^{\circ}$,& $|x|=2$ &$\tilde{\varphi}_S=\tilde{\varphi}'_S,$ & $|C_S|=0.5|C'_S|$  &  \\
\hline
\ref{sec:Delta180}&$\Delta=180^{\circ}$,& $|x|=1$ &$C_S=-C'_S$, & $C_P=C'_P$  &   \\
\hline
\end{tabular}
\end{center}
\caption{Summary of the scenarios described in Subsection~\ref{ssec:illu}. In all the cases we have assumed $\varphi_{10}=0^{\circ}$.}
\label{tab:Tab2}
\end{table}

As in the case given by Eq.~(\ref{eq:ObsCsCp0}), the correlation between the observables 
$\mathcal{A}^{\mu\mu}_{\Delta \Gamma_s}$ and $\mathcal{S}_{\mu\mu}$ describes a circle with radius one. 
The overall phase $\varphi_P$ includes effects from the, in general, complex quantities
 $\mathcal{C}_{10}$ and $C_S$. This is particularly
interesting if future measurements reveal $(\mathcal{A}_{\Delta\Gamma_s}^{\mu\mu})^2+(\mathcal{S}_{\mu\mu})^2$
compatible with the unit circle and if we have bounds available on the phase  $\varphi_{10}$ from other processes. Then,
results incompatible with Eq.\ (\ref{eq:ObsCsCp0}) will indicate the potential presence of a scalar or 
pseudo-scalar contribution.

\boldmath
\subsubsection{$\Delta=180^\circ$}
\label{sec:Delta180}
\unboldmath
In the case of $\Delta=180^\circ$, we obtain the following expressions for the $B^0_s\to\mu^+\mu^-$ observables:
\begin{equation}\label{r-Del180}
r|_{\Delta=180^\circ}=|{\cal C}_{10}|^2-2\left(\frac{1-|x|}{1+|x|}\right)\cos(\varphi_{10}-\varphi_S)|{\cal C}_{10}||S|
+2\left[\frac{1+|x|^2}{(1+|x|)^2}\right]|S|^2
\end{equation}
\begin{equation}
{\cal A}^{\mu\mu}_{\Delta\Gamma_s}|_{\Delta=180^\circ}=
\frac{(1+|x|)^2|{\cal C}_{10}|^2\cos2\varphi_{10}-2\left(1-|x|^2\right)
\cos(\varphi_{10}+\varphi_S)|{\cal C}_{10}||S| - 4 |x||S|^2\cos2\varphi_S}{(1+|x|)^2|{\cal C}_{10}|^2-2\left(1-|x|^2\right)
\cos(\varphi_{10}-\varphi_S)|{\cal C}_{10}||S| + 2 (1+|x|^2)|S|^2}
\label{eq:ADGx1-Del180}
\end{equation}
\begin{equation}
{\cal S}_{\mu\mu}|_{\Delta=180^\circ}=
\frac{(1+|x|)^2|{\cal C}_{10}|^2\sin2\varphi_{10}-2\left(1-|x|^2\right)
\sin(\varphi_{10}+\varphi_S)|{\cal C}_{10}||S| - 4 |x||S|^2\sin2\varphi_S}{(1+|x|)^2|{\cal C}_{10}|^2-2\left(1-|x|^2\right)
\cos(\varphi_{10}-\varphi_S)|{\cal C}_{10}||S| + 2 (1+|x|^2)|S|^2}
\label{eq:Smumux1-Del180}
\end{equation}
\begin{equation}
{\cal C}_{\mu\mu}|_{\Delta=180^\circ}=\frac{2|S|\left[(1+|x|)^2|{\cal C}_{10}|\cos(\varphi_{10}-\varphi_S)-
(1-|x|^2)|S|\right]}{(1+|x|)^2|{\cal C}_{10}|^2-2\left(1-|x|^2\right)
\cos(\varphi_{10}-\varphi_S)|{\cal C}_{10}||S| + 2 (1+|x|^2)|S|^2}.
\label{eq:Cmumux1-Del180}
\end{equation}
These equations could be solved numerically to determine $|x|$, $|S|$ and $\varphi_S$, in analogy to the
discussion of $\Delta=0^\circ$.

It is interesting to have a closer look at $x=-1$, i.e.\ $|x|=1$ and $\Delta=180^\circ$.
In terms of the short-distance coefficients, this case corresponds to
\begin{equation}\label{eq:condition180}
C_S=-C'_S.
\end{equation}
Using the relations in Eqs.\ (\ref{rel-PS}) and (\ref{rel-PS-prime}), we obtain furthermore
\begin{equation}
C_P=C'_P,
\end{equation}
implying
\begin{equation}
P=\mathcal{C}_{10}.
\end{equation}
Using Eqs.\ (\ref{R-expr}) and (\ref{y-pm}), we obtain
\begin{equation}\label{eq:S180}
|S|^2=\frac{\Bigl(1+y_s\Bigl)\overline{R}-\Bigl[1+y_s \cos(2\varphi_{10})\Bigl]|\mathcal{C}_{10}|^2}{1-
y_s \cos(2\varphi_{S})}.
\end{equation}
Special care should be paid when using Eq.\ (\ref{eq:S180}), since the expression on the right-hand side has
to be greater than or equal to zero. This feature implies the following upper bound:
\begin{equation}
 |\mathcal{C}_{10}|\leq \sqrt{\left (\frac{1+y_s}{1-y_s}\right) \overline{R}},
\end{equation}
where we have used that $1+ y_s \cos(2\varphi_{10})\geq1-y_s$, with $y_s$ given in Eq.~(\ref{ys-exp}). With the current 
experimental value of $\overline{R}$ in Eq.~(\ref{eq:Rbar}), the corresponding bound is given by
\begin{equation}
 |\mathcal{C}_{10}| \leq 0.98\pm0.09.
\end{equation}

The different scenarios described in the previous subsections are presented in Table \ref{tab:Tab2}, where
we show the connection between the standard parametrization used for the short distance contributions and the SMEFT
one introduced in Sec. \ref{ssec:gen-frame}.

\boldmath
\section{Experimental Aspects}\label{sec:Exp}
\unboldmath
Up to now we have not considered experimental uncertainties in the observables 
$\mathcal{A}^{\mu\mu}_{\Delta \Gamma_s}$,  
${\cal S}_{\mu\mu}$ and $\mathcal{C}_{\mu\mu}$ when studying the different scenarios. Nevertheless, 
we would like to demonstrate the potential for the determination of the underlying parameters at future experiments.
Since the asymmetries are not independent, due to the relation in Eq.~(\ref{CP-rel}), it is not possible to determine all four parameters $|S|$, $\varphi_S$, $|x|$ and $\Delta$. However, as discussed in Subsection~\ref{ssec:extractionxDelta}, we expect to have a better picture of physics beyond the SM by the time the CP asymmetries of $B^0_s\to\mu^+\mu^-$ have been measured.
Therefore, we consider some of the examples discussed in Subsection~\ref{ssec:illu}, which correspond to specific values of $|x|$ or $\Delta$. We assume uncertainties for the observables, allowing us to extract the NP parameters through fits.

Unless  specified otherwise, within this section we use a future measurement of  
\begin{equation}\label{eq:rBarUp}
\overline{R} = 0.84 \pm 0.09,
\end{equation}
where we have assumed a relative uncertainty of $10 \%$ for $\overline{\mathcal{B}}(B_s \to \mu^+\mu^-)$, which is 
achievable at the LHCb upgrade \cite{LHCbup}, while keeping the current central value fixed. Notice that the 
relative uncertainty  in our ``measurement'' for $\bar{R}$ in Eq.~(\ref{eq:rBarUp}) leads to a $2\sigma$ tension with 
the SM. Thus the statistical significance will not be high enough to claim for the discovery of NP effects.
The major limiting factor of the precision is the ratio $f_d/f_s$ of the fragmentation functions of the $B^0_d$ and 
$B^0_s$ mesons \cite{FST}, which is required for normalization purposes. To the best of our knowledge, no 
information about the expected precision of future measurements of 
${\cal A}_{\Delta\Gamma_s}^{\mu\mu}$, ${\cal S}_{\mu\mu}$ and ${\cal C}_{\mu\mu}$ is available. The key 
question we want to address is the precision of the measurement of these observables that is required to 
establish in particular new (pseudo)-scalar contributions at the $5\,\sigma$ confidence level.

\boldmath
\subsection{$x=0$ and $|x|\to\infty$} \label{ssec:Expx0xinf}
\unboldmath
To begin with, we evaluate the impact of experimental errors for the observables in Example (a) of
Subsection~\ref{sec:x0xinfty}, corresponding to a scenario where $x = 0$.
An absolute uncertainty of $\pm 0.2$ for the asymmetries leads to the 
following set of observables:
\begin{equation} \label{eq:obsx0fit}
{\cal A}_{\Delta\Gamma_s}^{\mu\mu} = 0.58 \pm 0.20, 
\quad {\cal S}_{\mu\mu} = -0.80 \pm 0.20, \quad {\cal C}_{\mu\mu} = 0.16 \pm 0.20.
\end{equation}
In such a situation, ${\cal S}_{\mu\mu}$ would indicate CP-violating NP effects at the $4 \sigma$ level, while 
${\cal A}_{\Delta\Gamma_s}^{\mu\mu}$ and ${\cal C}_{\mu\mu}$ would deviate from the SM picture at the $2\sigma$ 
and $1 \sigma$ levels, respectively. Let us assume that the values above have been measured at a future experiment, 
and that there are strong reasons to consider models characterized by $x=0$. We will now illustrate through a $\chi^2$ 
fit how well we can reveal the underlying decay dynamics.

\begin{figure}
   \centering
   \includegraphics[width=0.45\textwidth]{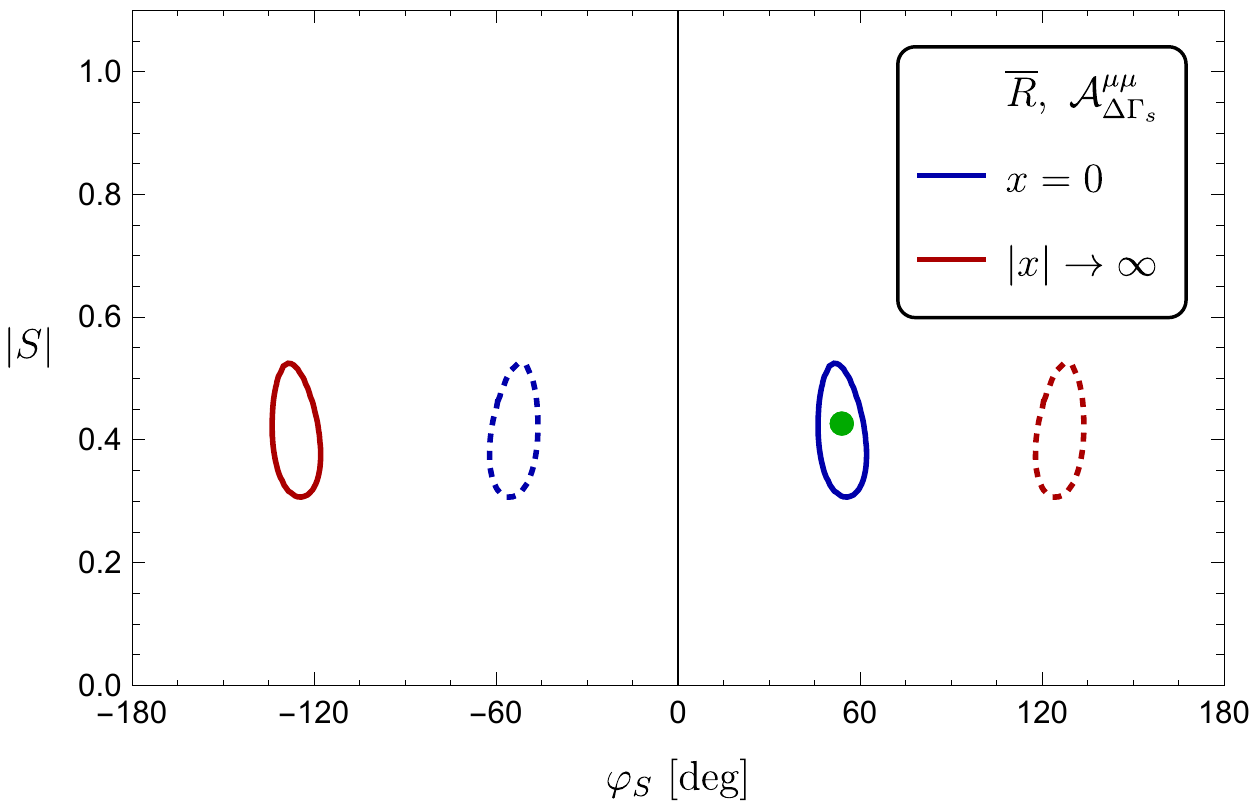} 
   \includegraphics[width=0.45\textwidth]{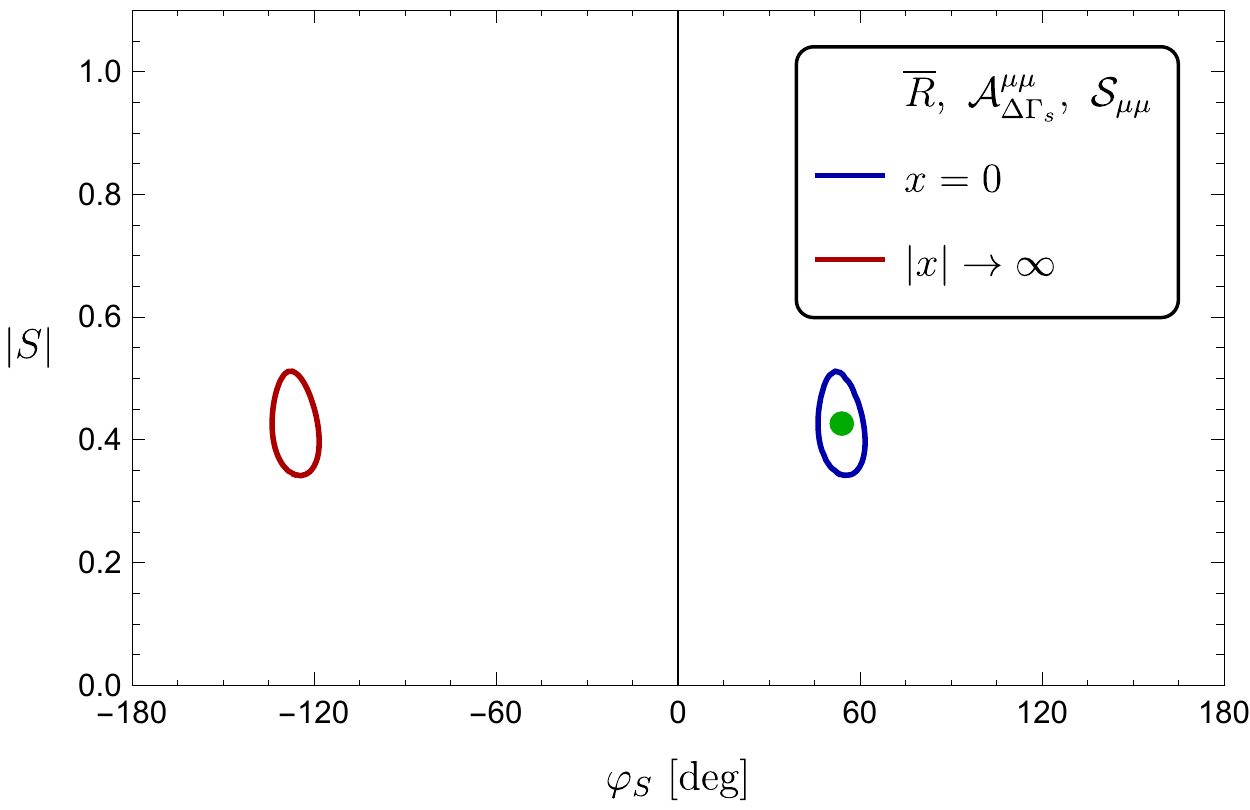} 
   \caption{Illustration of the determination of $|S|$ and $\varphi_S$ from the observables in Eq.~(\ref{eq:obsx0fit})
   for a scenario with $x=0$, which is degenerate with $|x|\to\infty$. The contours correspond to the $1\sigma$ allowed 
   regions obtained from $\chi^2$ fits. In the left panel, we show the result of the fit to only $\overline{R}$ and 
   ${\cal A}_{\Delta\Gamma_s}^{\mu\mu}$, while in the right panel we have also included ${\cal S}_{\mu\mu}$.
   The blue contours were obtained by assuming $x=0$, whereas the red contours follow for $|x|\to\infty$. 
   A measurement of the sign of ${\cal C}_{\mu\mu}$ would allow us to distinguish these cases, excluding the
   $|x|\to\infty$ scenario.} 
   \label{fig:fitx0xinf}
\end{figure}

Let us first obtain the regions allowed for $|S|$ and $\varphi_S$ if we only include $\overline{R}$ 
and ${\cal A}_{\Delta\Gamma_s}^{\mu\mu}$ in the statistical analysis. Thus, using the expression in 
Eqs.~(\ref{rx0-calc})~and~(\ref{ADGx0-calc}) with the ``data'' in Eqs.~(\ref{eq:rBarUp})~and~(\ref{eq:obsx0fit}), 
we perform a $\chi^2$ fit to these two observables and obtain the blue contours in the left panel of Fig.~\ref{fig:fitx0xinf}, 
which correspond to $1\sigma$ allowed regions. We indicate the input parameters used to determine our observables in 
Eq.~(\ref{eq:obsx0fit}) with the green dot. This plot allows us to establish a non-zero value for $|S|$ at the $3\sigma$ 
level. If we include also the ``measurement'' for ${\cal S}_{\mu\mu}$ indicated in Eq.~(\ref{eq:obsx0fit}), along with 
Eq.~(\ref{Smumux0-calc}), and repeat the $\chi^2$ fit, we can eliminate the dashed contour in the left panel and 
obtain the right plot of Fig.~\ref{fig:fitx0xinf}.

As we have pointed out in Subsection~\ref{sec:x0xinfty}, there is a symmetry between $x=0$ and $|x|\to\infty$,
implying the same values of ${\cal A}_{\Delta\Gamma_s}^{\mu\mu}$ and ${\cal S}_{\mu\mu}$ for these two 
cases. Conversely, we could not distinguish $x=0$ and $|x|\to\infty$ at the phenomenological level having 
only measurements of these observables
available. Indeed, repeating the $\chi^2$ fits assuming $|x|\to\infty$ for the same set of input observables leads to 
the red contours in Fig.~\ref{fig:fitx0xinf}. Although we use $x=0$ as our favoured model in this illustration, it 
would certainly be desirable to rule out the degenerate $|x|\to\infty$ scenario. As ${\cal C}_{\mu\mu}$ breaks 
the symmetry by an 
overall minus sign, we could actually exclude the $|x|\to\infty$ case through experimental information on the sign 
on this CP asymmetry. If we add ${\cal C}_{\mu\mu}$ to the analysis, a solution only arises in case of the $x=0$ scenario, thereby singling out the blue contour.  We would then find
\begin{equation}
|S| = 0.43_{-0.08}^{+0.07}, \qquad \varphi_S = (54_{-7}^{+6})^\circ,
\end{equation}
where ${\cal C}_{\mu\mu}$ has a minor impact on the numerical values themselves, apart from excluding $|x|\to\infty$.
In this scenario, the assumed experimental uncertainties in Eq.~(\ref{eq:obsx0fit}) would allow us to establish
non-zero values of $|S|$ and $\varphi_S$ at the $5\sigma$ and $7\sigma$ levels, respectively, which would provide
highly non-trivial insights into the underlying dynamics.

\boldmath
\subsection{$\Delta=0^\circ$}
\unboldmath
Let us now have a closer look at another interesting scenario: $\Delta=0^\circ$, where $C_S'$ and $C_S$ have 
the same CP-violating phases. The expressions in Eqs.~(\ref{r-Del0})--(\ref{eq:Cmumux1}) form a system
of three independent equations which allows us to determine $|S|$, $\varphi_S$ and $|x|$. Due to the
highly non-linear structure of the mathematical expressions, we cannot provide analytical solutions in general. 
Instead we give an example of how to solve the system through a $\chi^2$ fit. We consider the input parameters
\begin{equation} \label{eq:inputDelta0fit1}
|x|=0.5, \qquad \varphi_S = 20^\circ, \qquad |{\cal C}_{10}| = 1, \qquad \varphi_{10} = 0^\circ,
\end{equation}
allowing us to determine the following solution for $|S|$, which is consistent with the current central value for 
$\overline{R}$ shown in  Eq.~(\ref{eq:Rbar}):
\begin{equation} \label{eq:inputDelta0fit2}
|S| = 0.55.
\end{equation}
If we use the previous numerical values in Eqs.~(\ref{eq:ADGx1})--(\ref{eq:Cmumux1}), our observables are
\begin{equation} \label{eq:obsDelta0fit}
{\cal A}_{\Delta\Gamma_s}^{\mu\mu} = -0.27 \pm 0.20, \quad 
{\cal S}_{\mu\mu} = 0.46 \pm 0.20, \quad {\cal C}_{\mu\mu} = -0.85 \pm 0.20,
\end{equation}
where we have considered the same absolute uncertainties as in Subsection~\ref{ssec:Expx0xinf}. Assuming 
that these observables have been measured correspondingly at a future experiment, 
${\cal A}_{\Delta\Gamma_s}^{\mu\mu}$ would indicate NP at the $6\sigma$ level, while ${\cal S}_{\mu\mu}$ and 
${\cal C}_{\mu\mu}$ would differ from the SM at the $2\sigma$ and $4\sigma$ levels, respectively. The latter 
observable would require a non-vanishing scalar contribution $S$. Performing a $\chi^2$ fit to these quantities, we 
can determine the underlying decay parameters $|x|$, $|S|$ and $\varphi_S$ simultaneously from the best fit point.

\begin{figure}
   \centering
   \includegraphics[width=0.45\textwidth]{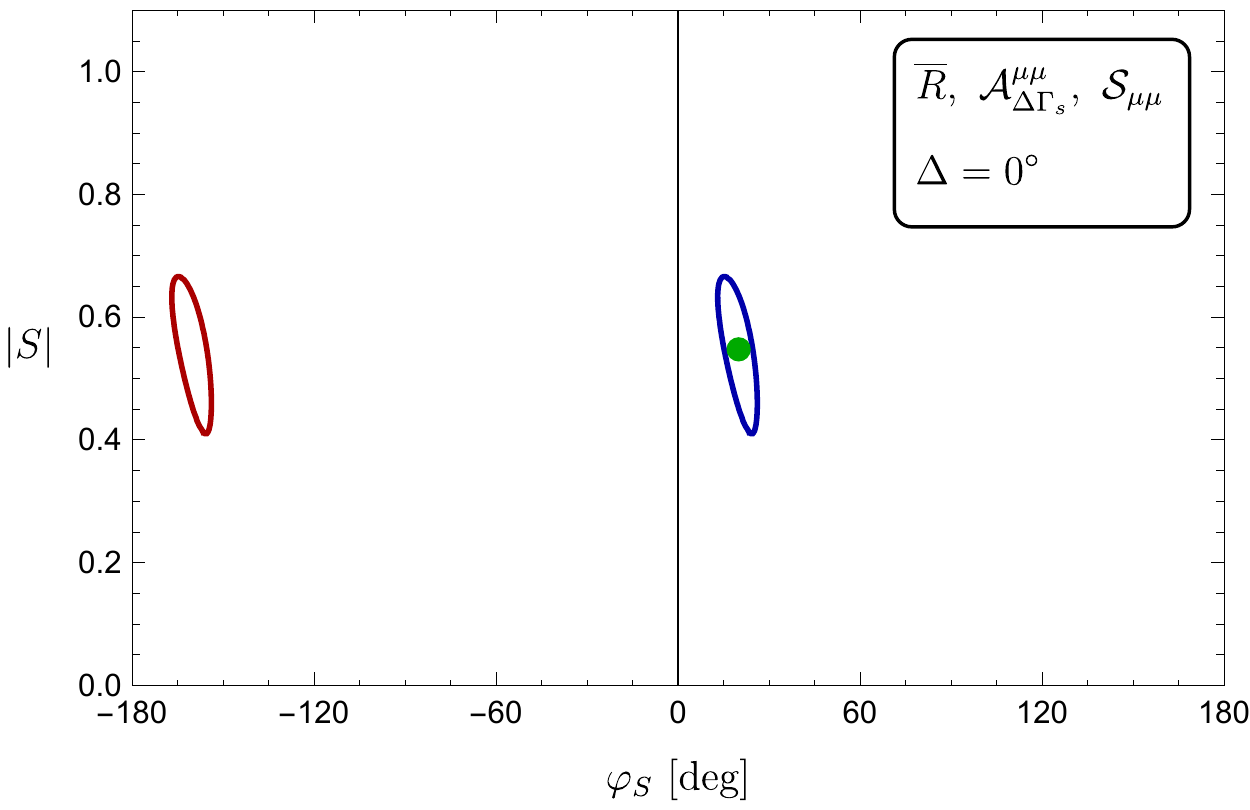} 
   \includegraphics[width=0.45\textwidth]{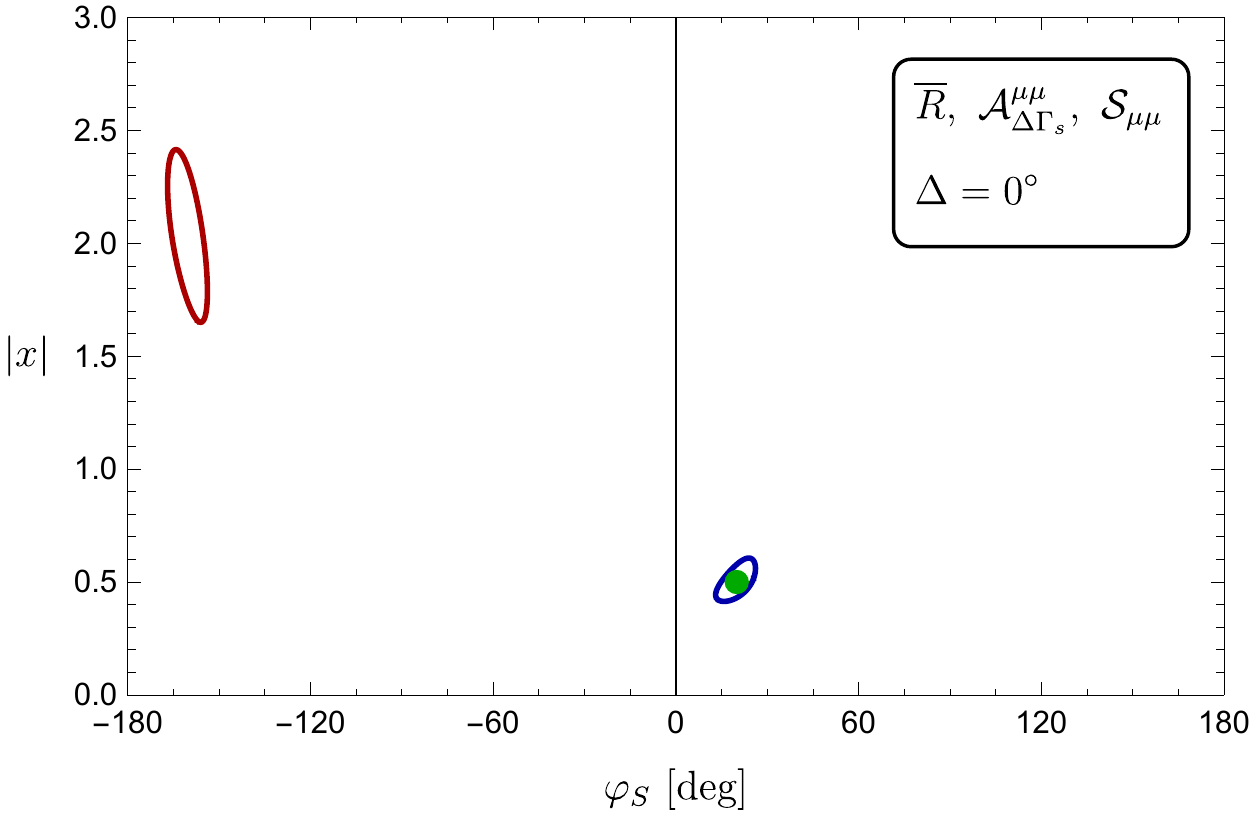} 
   \caption{Illustration of the determination of $|S|$, $|x|$ and $\varphi_S$ in the scenario where we assume 
   $\Delta=0^\circ$. The contours correspond to the $1\sigma$ allowed regions obtained by performing a $\chi^2$ 
   fit to $\overline{R}$, ${\cal A}_{\Delta\Gamma_s}^{\mu\mu}$ and ${\cal S}_{\mu\mu}$ given in Eq.~(\ref{eq:obsDelta0fit}). 
   We obtain two solutions, indicated in blue and red, as expected from the symmetry relations in 
   Eq.~(\ref{eq:Del-0-symm-r-ADG-smumu}). The green dot marks the input parameters given in 
   Eqs.~(\ref{eq:inputDelta0fit1})~and~(\ref{eq:inputDelta0fit2}).} 
   \label{fig:fitDelta0x05}
\end{figure}

We start our statistical analysis 
by considering only $\overline{R}$, ${\cal A}_{\Delta\Gamma_s}^{\mu\mu}$ and ${\cal S}_{\mu\mu}$. In the left and 
right panels of Fig.~\ref{fig:fitDelta0x05}, we show the corresponding $1\,\sigma$ confidence regions in the
$\varphi_S$--$|S|$ and $\varphi_S$--$|x|$ planes, respectively. We obtain two solutions, given by the blue and red 
contours, as we expect based on the symmetry relations in Eq.~(\ref{eq:Del-0-symm-r-ADG-smumu}). Our input
parameters are indicated by the green dot. Consequently, non-zero values of $|S|$ and $|x|$ at the 
$4\sigma$ and $6\sigma$ levels, respectively, could then be established.

\begin{figure}
   \centering
   \includegraphics[width=0.45\textwidth]{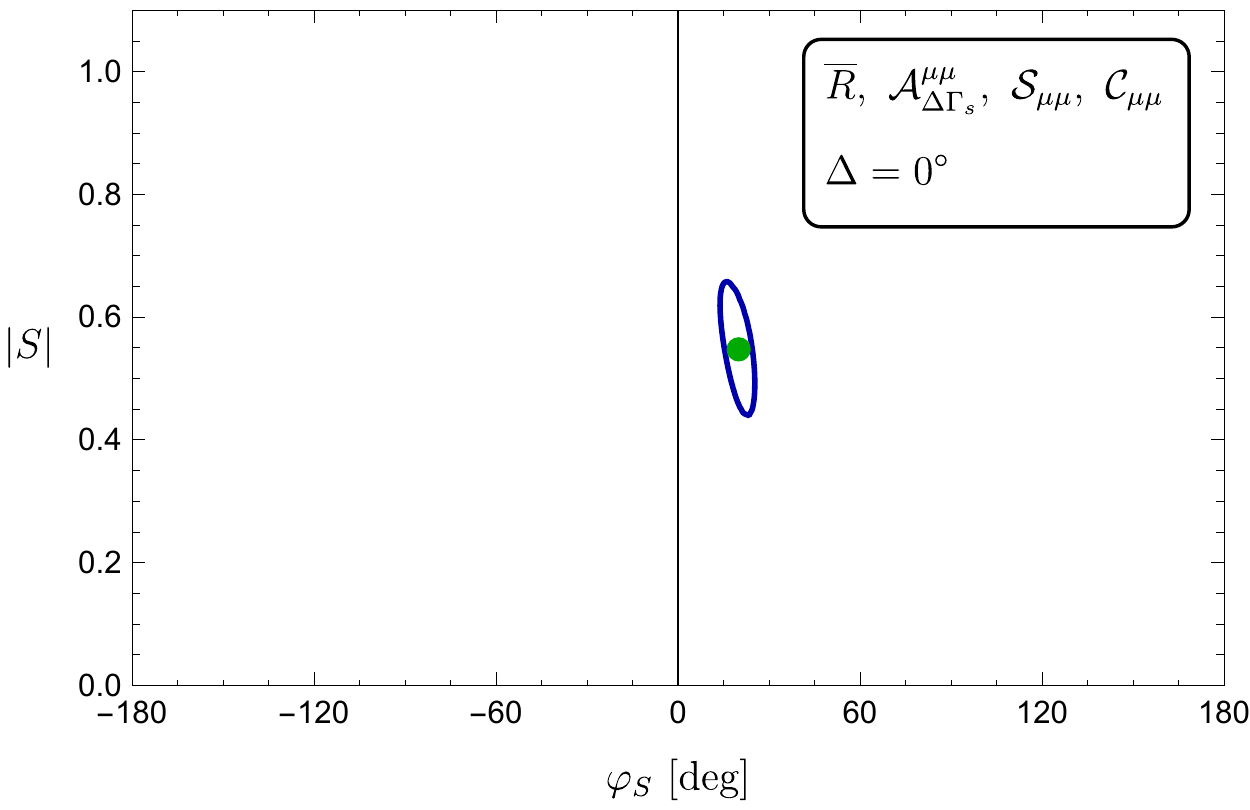} 
   \includegraphics[width=0.45\textwidth]{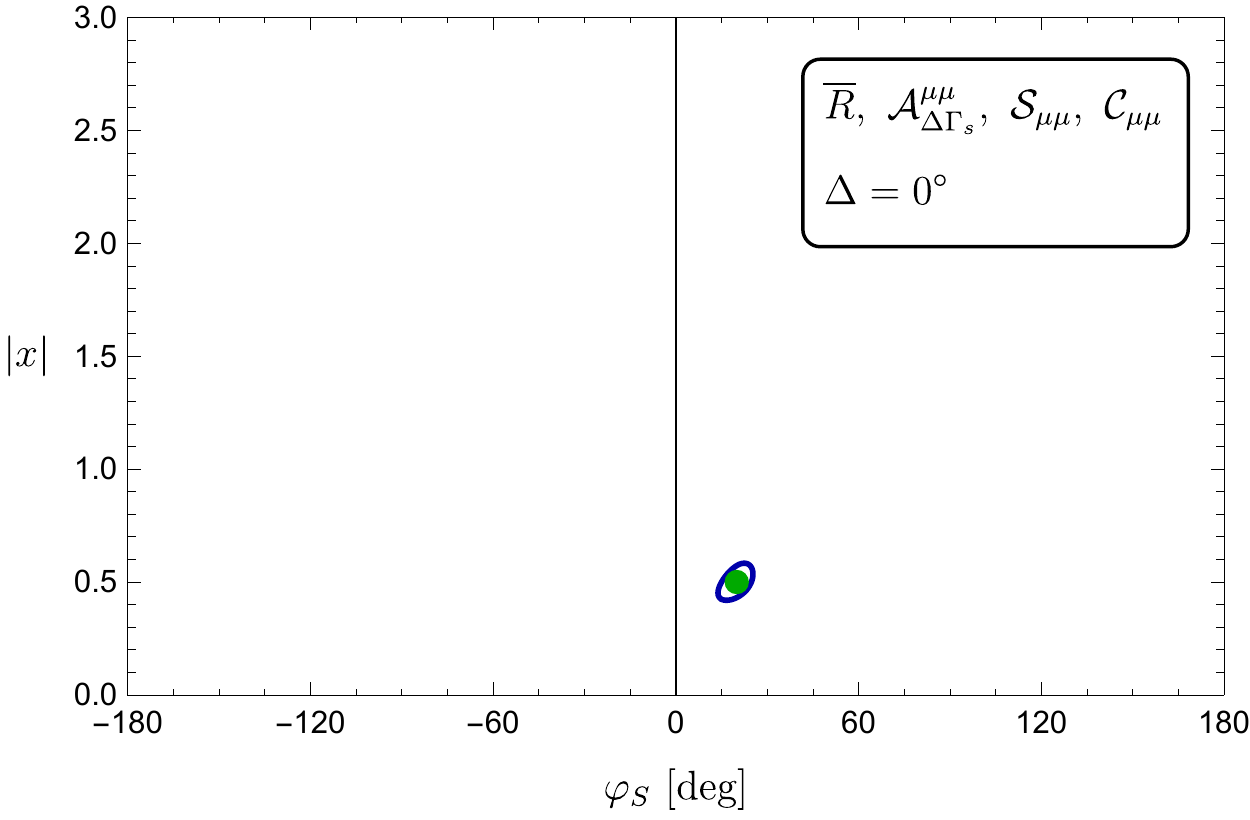} 
   \caption{Illustration of the determination of $|S|$, $|x|$ and $\varphi_S$ in a scenario where 
   we assume $\Delta=0^\circ$. The contours correspond to the $1\sigma$ allowed regions obtained by performing 
   a $\chi^2$ fit to $\overline{R}$, ${\cal A}_{\Delta\Gamma_s}^{\mu\mu}$, ${\cal S}_{\mu\mu}$ and 
   ${\cal C}_{\mu\mu}$ given in Eq.~(\ref{eq:obsDelta0fit}). The green dot marks the input parameters given 
   in Eqs.~(\ref{eq:inputDelta0fit1})~and~(\ref{eq:inputDelta0fit2}).} 
   \label{fig:fitwithCmumuDelta0x05}
\end{figure}

If we include also ${\cal C}_{\mu\mu}$ in the analysis, we can eliminate the solution corresponding 
to the red contours, since ${\cal C}_{\mu\mu}$ breaks the symmetry relation in Eq.~(\ref{sym-Del-0}) by an 
overall minus sign. The resulting $1\sigma$ regions are shown in Fig.~\ref{fig:fitwithCmumuDelta0x05}, corresponding
to the results
\begin{equation}
|S| = 0.55_{-0.10}^{+0.08}, \qquad \varphi_S=(20_{-5}^{+5})^\circ, \qquad |x| = 0.50_{-0.07}^{+0.07}.
\end{equation}
As a matter of fact, non-zero values of these parameters could be pinned down at the $5\sigma$, $4\sigma$ and 
$7\sigma$ levels, respectively.

In general, the precision for the CP asymmetries required to determine  $|S|$, $\varphi_S$ and $|x|$ with a 
given confidence level depends on the situation in parameter space. Moreover, we may end up with an ambiguity 
even after including ${\cal C}_{\mu\mu}$ in the $\chi^2$ fit. Nevertheless, this example nicely complements the one
in Subsection~\ref{ssec:Expx0xinf} and shows the potential of the CP asymmetries to determine the (pseudo)-scalar 
contributions, and even to discriminate between the corresponding primed and unprimed Wilson coefficients.

\boldmath
\section{Conclusions and Outlook}\label{sec:concl}
\unboldmath
The rare decay $B^0_s\to\mu^+\mu^-$ has been in the focus of particle physics for decades, offering one of the
theoretically cleanest probes for physics beyond the SM, in particular for new (pseudo)-scalar contributions, which
are still largely unconstrained. Finally, this channel could be observed by the CMS and LHCb collaborations and is 
now an experimentally well established process, exhibiting a branching ratio encoded in $\overline{R}$ in the ballpark 
of the SM. The observable ${\cal A}^{\mu\mu}_{\Delta\Gamma_s}$, which is accessible thanks to the decay width 
difference $\Delta\Gamma_s$ and requires an untagged -- but time-dependent -- analysis, will play an important role 
to shed light on possible NP contributions to $B^0_s\to\mu^+\mu^-$. In general, these effects involve also CP-violating 
phases, which are usually neglected in theoretical analyses for simplicity. 

In this paper, we have presented a comprehensive strategy for the future LHC upgrade(s), allowing us to reveal 
the presence of new sources of CP violation. The key role in this endeavour is played by the mixing-induced 
CP asymmetry ${\cal S}_{\mu\mu}$, which requires -- in contrast to ${\cal A}^{\mu\mu}_{\Delta\Gamma_s}$ -- also 
tagging information for the experimental analysis. Another observable, ${\cal C}_{\mu\mu}$, would become accessible 
if the helicity of the final-state muons could be determined; already sign information for this CP asymmetry would be 
very valuable information. These three observables do not depend on the decay constant $f_{B_s}$ and are 
not affected by theoretical uncertainties. 

Interestingly, the interplay of $\overline{R}$ with ${\cal A}^{\mu\mu}_{\Delta\Gamma_s}$ and ${\cal S}_{\mu\mu}$ 
allows us to establish new (pseudo)-scalar contributions and new sources of CP violation. In general, we can only
obtain constraints as we do not have sufficient independent observables to determine the short-distance coefficients
$|S|$, $|P|$ and their phases $\varphi_S$, $\varphi_P$. To obtain further insights, additional information
is required. This could either be obtained by assuming specific NP models, or in a model-independent way 
through relations between the short-distance coefficients $C_P^{(')}$, $C_S^{(')}$, which can be derived within the
SMEFT approach. We have followed the latter avenue, discussing a variety of scenarios to illustrate how the
corresponding parameters can be determined from the measured observables. 

Since the pseudo-scalar coefficient $P$ involves ${\cal C}_{10}$, we need information on this quantity. By the time
precise measurements of the observables ${\cal A}^{\mu\mu}_{\Delta\Gamma_s}$ and ${\cal S}_{\mu\mu}$ are
available, we expect to have a detailed picture of ${\cal C}_{10}$, following from analyses of semileptonic rare 
$B\to K^{(*)}\mu^+\mu^-$ and $B_s\to \phi\mu^+\mu^-$ decays. Current anomalies in the data for the former and 
$B\to K^{(*)} e^+e^-$ decays indicate NP effects in ${\cal C}_{10}$, which we have also considered in our explorations. It will be important to utilize CP violation in the corresponding observables in the future.

To the best of our knowledge, experimental feasibility studies for the measurement of ${\cal S}_{\mu\mu}$ at the
LHC upgrade(s) are not yet available. Performing fits to the observables for given future scenarios, we find that 
an absolute precision at the 0.2 level for ${\cal A}^{\mu\mu}_{\Delta\Gamma_s}$ and ${\cal S}_{\mu\mu}$ could have 
a dramatic impact on our search for new (pseudo)-scalar contributions in leptonic rare $B_s$ decays, allowing us to reveal the
underlying dynamics. We urge the LHC collaborations to add studies of CP violation in rare $B^0_s\to\ell^+\ell^-$
decays to their physics agenda for the long-term future and super-high-precision era of $B$ physics.


\section*{Acknowledgements}
This research has been supported by the Netherlands Foundation for Fundamental Research of Matter (FOM) 
programme 156, ``Higgs as Probe and Portal'', and by the National Organisation for Scientific Research (NWO).
D.G.E. acknowledges the support through a fellowship by the Universit\'e Paris-Sud and the hospitality by 
Nikhef and the Vrije Universiteit Amsterdam during her visit. We would like to thank Marcel Merk for useful discussions.


\appendix
\section*{Appendix} \label{sec:Obs}
In this appendix, we collect formulae which are useful for the analysis of $B_s^0 \to \mu^+ \mu^-$ within the SMEFT framework introduced in Subsection~\ref{ssec:gen-frame}. These expressions can be applied to any SMEFT scenario. In order to obtain the relevant observables in terms of the parameters $|S|$, $\varphi_S$, $|x|$ and $\Delta$, we write Eq.~(\ref{conv-1}) as
\begin{equation}
|P|\cos\varphi_P=|{\cal C}_{10}|\cos\varphi_{10} - \frac{1}{w}
\left[\frac{\left(1-|x|^2\right)\cos\varphi_S-2|x|\sin\Delta\sin\varphi_S}{1-2|x|\cos\Delta+|x|^2}\right]|S|\end{equation}
\begin{equation}
|P|\sin\varphi_P=|{\cal C}_{10}|\sin\varphi_{10} - \frac{1}{w}
\left[\frac{\left(1-|x|^2\right)\sin\varphi_S+2|x|\sin\Delta\cos\varphi_S}{1-2|x|\cos\Delta+|x|^2}\right]|S|,
\end{equation}
yielding
\begin{equation}
\tan\varphi_P=\frac{|{\cal C}_{10}|\sin\varphi_{10}-\left[ \left(1-|x|^2\right)\sin\varphi_S+
2|x|\sin\Delta\cos\varphi_S \right]G}{|{\cal C}_{10}|\cos\varphi_{10}-
\left[ \left(1-|x|^2\right)\cos\varphi_S-2|x|\sin\Delta\sin\varphi_S \right]G}
\end{equation}
with
\begin{equation}
G\equiv\frac{|S|}{w \left(1-2|x|\cos\Delta+|x|^2 \right)}\,.
\end{equation}
The scalar coefficient function is given as 
\begin{equation}
S \equiv |S|e^{i\varphi_S}= w \frac{M_{B_s}^2}{2m_\mu}\left(\frac{m_b}{m_b + m_s}\right) 
	\left( \frac{|C_{S}|}{C_{10}^{\rm SM}}\right)\left(1-|x|e^{i\Delta}\right) e^{i\tilde\varphi_S} 
\label{eq:Sofx}
\end{equation}
with
\begin{equation}
\tan\varphi_S=\frac{\left(1-|x|\cos\Delta\right)\sin\tilde\varphi_S-|x|\sin\Delta\cos\tilde\varphi_S}{\left(1-|x|\cos\Delta\right)
\cos\tilde\varphi_S+|x|\sin\Delta\sin\tilde\varphi_S}.
\end{equation}
As we noted in Eq.~(\ref{C10-neg}), $C_{10}^{\rm SM}$ is negative. 
We may also convert $\varphi_S$ into $\tilde\varphi_S$:
\begin{equation}
\cos\tilde\varphi_S\propto |x|\cos(\varphi_S-\Delta)-\cos\varphi_S, \quad
\sin\tilde\varphi_S\propto |x|\sin(\varphi_S-\Delta)-\sin\varphi_S,
\end{equation}
which yields
\begin{equation}
\tan\tilde\varphi_S=\frac{(1-|x|\cos\Delta)\sin\varphi_S+|x|\sin\Delta\cos\varphi_S}{(1-|x|\cos\Delta)\cos\varphi_S
-|x|\sin\Delta\sin\varphi_S}.
\end{equation}
Moreover, we have
\begin{equation}
|C_{S}|=\frac{1}{w} \frac{2m_\mu}{M_{B_s}^2}\left(\frac{m_b + m_s}{m_b}\right) 
	\frac{|C_{10}^{\rm SM}|}{\sqrt{1-2|x|\cos\Delta+|x|^2}} |S|.
\end{equation}

The observables in Eqs.~(\ref{ADG-expr}),  and (\ref{S-expr}) and (\ref{rr-def}) require the quantities
\begin{displaymath}
\hspace*{-2.0truecm}
|P|^2=|{\cal C}_{10}|^2-2\left[(1-|x|^2)\cos(\varphi_{10}-\varphi_S)+2|x|\sin\Delta\sin(\varphi_{10}-\varphi_S)\right]
|{\cal C}_{10}|G
\end{displaymath}
\begin{equation}
+\left[ \left(1-|x|^2\right)^2 + \left( 2 |x| \sin\Delta\right)^2 \right]G^2,
\end{equation}
\begin{displaymath}
\hspace*{-8.5truecm}|P|^2\cos2\varphi_P=|P|^2\left( \cos^2\varphi_P-\sin^2\varphi_P \right)
\end{displaymath}
\begin{equation}
=|{\cal C}_{10}|^2\cos2\varphi_{10}
-2\left[\left(1-|x|^2\right)\cos(\varphi_{10}+\varphi_S)-2|x|\sin\Delta\sin(\varphi_{10}+\varphi_S) \right]|{\cal C}_{10}|G
\end{equation}
\begin{displaymath}
+\left[\left\{\left(1-|x|^2\right)^2 - (2|x|\sin\Delta)^2\right\} \cos2\varphi_S - 4 |x| \left(1-|x|^2\right)\sin\Delta
\sin2\varphi_s \right]G^2,
\end{displaymath}
\begin{displaymath}
\hspace*{-8.5truecm}|P|^2\sin2\varphi_P=2|P|\sin\varphi_P|P|\cos\varphi_P=
\end{displaymath}
\begin{equation}
=|{\cal C}_{10}|^2\sin2\varphi_{10}
-2\left[\left(1-|x|^2\right)\sin(\varphi_{10}+\varphi_S)+2|x|\sin\Delta\cos(\varphi_{10}+\varphi_S) \right]|{\cal C}_{10}|G
\end{equation}
\begin{displaymath}
+\left[\left\{\left(1-|x|^2\right)^2 - (2|x|\sin\Delta)^2\right\} \sin 2\varphi_S + 4 |x| \left(1-|x|^2\right)\sin\Delta
\cos 2\varphi_s \right]G^2,
\end{displaymath}
while the CP asymmetry in Eq.~(\ref{Cobs}) involves
\begin{equation}
|P||S|\cos(\varphi_P-\varphi_S)=|S|\left[|{\cal C}_{10}|\cos(\varphi_{10}-\varphi_S) - \left(\frac{1-|x|^2}{1-2|x|\cos\Delta
+|x|^2}\right)\frac{|S|}{w}\right].
\end{equation}
In view of the complexity of the resulting general expressions, we refrain from listing them
for the observables $r$, ${\cal A}^{\mu\mu}_{\Delta\Gamma_s}$, ${\cal S}_{\mu\mu}$ and ${\cal C}_{\mu\mu}$. 
However, we have given formulae for specific examples discussed in Subsection~\ref{ssec:illu}.


%
%
%

%
%
%
\end{document}